\documentclass[reprint,superscriptaddress,aps,onecolumn,nofootinbib,notitlepage,showpacs]{revtex4-1}

\usepackage[intlimits]{amsmath}
\usepackage{scalerel}
\usepackage{stackengine,wasysym}
\usepackage{amssymb}
\usepackage[scr=boondox,scrscaled=1.03]{mathalfa}
\usepackage{placeins}
\usepackage{amsbsy}
\usepackage{bbold}
\usepackage[T1]{fontenc}
\usepackage[utf8]{inputenc}
\usepackage[portuguese,english]{babel}
\usepackage{multirow}
\usepackage{microtype}
\usepackage{amsfonts}
\usepackage{amssymb}
\usepackage{setspace}
\usepackage{graphicx}
\usepackage{wrapfig}
\usepackage{textcomp}         
\usepackage{array}
\usepackage{slashed}
\usepackage{url}
\usepackage[usenames,table]{xcolor}
\usepackage[colorlinks=true,linkcolor=blue,citecolor=blue,urlcolor=blue]{hyperref}
\usepackage{natbib}

\DeclareMathSizes{10.95}{10.95}{7}{7}
\setcitestyle{numbers,sort&compress}
 
\def\dj{d\kern-0.4em\char"16\kern-0.1em}

\def\be{\begin{equation}}
\def\ee{\end{equation}}
\def\bea{\begin{eqnarray}}
\def\eea{\end{eqnarray}}
\def\bfl{\begin{flushleft}}
\def\efl{\end{flushleft}}
\def\bfr{\begin{flushright}}
\def\efr{\end{flushright}}
\def\bc{\begin{center}}
\def\ec{\end{center}}
\def\ben{\begin{enumerate}}
\def\een{\end{enumerate}}
\def\bit{\begin{itemize}}
\def\eit{\end{itemize}}

\def\dzn{,\kern-0.1em,}

\def\graph{\includegraphics}
\def\tew{\textwidth}
\def\er{\eqref}

\def\d#1{{#1\kern-0.4em\char"16\kern-0.1em}}
\def\D#1{{\raise0.2ex\hbox{-}\kern-0.4em 31}}

\def\d{\mbox{d}}

\def\D{{\cal{D}}}

\newcommand {\apgt} {\ {\raise-.5ex\hbox{$\buildrel>\over\sim$}}\ }
\newcommand {\aplt} {\ {\raise-.5ex\hbox{$\buildrel<\over\sim$}}\ }

\begin{document}


\title{Probing the tensor structure of lattice \\ three-gluon vertex in Landau gauge}

\author{Milan Vujinovi\'c}
\affiliation{Institute of Physics, University of Graz, Universit\"atsplatz 5, 8010 Graz, Austria}
\affiliation{Instituto de Física de S\~ao Carlos, Universidade de S\~ao Paulo,\\
Caixa Postal 369, 13560-970 S\~ao Carlos, SP, Brasil}
\author{Tereza Mendes}
\affiliation{Instituto de Física de S\~ao Carlos, Universidade de S\~ao Paulo,\\
Caixa Postal 369, 13560-970 S\~ao Carlos, SP, Brasil}

\begin{abstract}

In this paper we test an approximate method that is often used in lattice studies of the Landau gauge three-gluon vertex.~The approximation consists
in describing the lattice correlator with tensor bases from the continuum theory.~With the help of vertex reconstruction, we show that this ``continuum''
approach may lead, for general kinematics, to significant errors in vertex tensor representations.~Such errors are highly unwelcome, as they can lead 
to wrong quantitative estimates for vertex form factors and related quantities of interest, like the three-gluon running coupling.~As a possible solution,
we demonstrate numerically and analytically that there exist special kinematic configurations for which the vertex tensor structures can be described
exactly on the lattice.~For these kinematics, the dimensionless tensor elements are equal to the continuum ones, regardless of the details of the lattice
implementation.~We ran our simulations for an $SU(2)$ gauge theory in two and three spacetime dimensions, with Wilson and $\mathcal{O}(a^2)$ tree-level
improved gauge actions.~Our results and conclusions can be straightforwardly generalised to higher dimensions and, with some precautions, to other lattice
correlators, like the ghost-gluon, quark-gluon and four-gluon vertices.     

\end{abstract}

\maketitle


\section{Introduction}

The primitively divergent vertex functions of quantum chromodynamics (QCD) and its quenched version, the pure Yang-Mills theory, have been the subject
of numerous non-perturbative investigations in the past two decades.~There are two main reasons why these objects attract considerable interest among 
researchers.~Firstly, by studying the vertices, and in particular their infrared (IR) properties, one might be able to learn something about confinement.~Of
particular interest in this regard are the Gribov-Zwanziger \cite{Gribov:1977wm, Zwanziger:1993dh, Zwanziger:2001kw, Zwanziger:2003cf} and Kugo-Ojima 
\cite{Kugo:1979gm} confinement scenarios, and their relation to the IR behaviour of the ghost and gluon propagators.~The second reason to study the vertex
functions has to do with the functional bound state calculations, for which these quantities constitute a key component, see e.\,g.~\cite{Vujinovic:2014ioa, 
Sanchis-Alepuz:2015qra, Williams:2015cvx, Binosi:2016rxz, Eichmann:2016yit, Rodriguez-Quintero:2018wma} and references therein.

The non-perturbative methods that have been used to study the vertices can roughly be divided into two main categories.~The first includes functional techniques
like the Dyson-Schwinger equations (DSEs) (see e.\,g.~\cite{Schleifenbaum:2004id, Kellermann:2008iw, Alkofer:2008dt, Huber:2012kd, Huber:2012zj, Aguilar:2013xqa,
Aguilar:2013vaa, Blum:2014gna, Eichmann:2014xya, Cyrol:2014kca, Vujinovic:2014ioa, Binosi:2014kka, Aguilar:2014lha, Sanchis-Alepuz:2015qra, Williams:2015cvx, 
Binosi:2016wcx, Aguilar:2016lbe, Huber:2017txg, Oliveira:2018fkj, Aguilar:2018csq}\,), functional renormalisation group (FRG) \cite{Pawlowski:2005xe,Mitter:2014wpa, 
Cyrol:2016tym, Cyrol:2017ewj, Corell:2018yil}, modified perturbation theory \cite{Pelaez:2013cpa,Pelaez:2015tba}, and others.~The second consists of various lattice
formulations and the corresponding Monte Carlo (MC) simulations, see e.\,g.~\cite{Parrinello:1994wd, Alles:1996ka, Boucaud:1998bq, Skullerud:2002ge, Skullerud:2003qu,
Cucchieri:2004sq, Cucchieri:2006tf, Ilgenfritz:2006he, Maas:2007uv, Cucchieri:2008qm, Maas:2011se, Boucaud:2013jwa, Duarte:2016jhj, Athenodorou:2016oyh, Sternbeck:2016ltn,
Boucaud:2017obn, Sternbeck:2017ntv}.~Both of these groups of approaches have their particular strengths and weaknesses.~In the case of Monte Carlo investigations, there
is an issue related to the tensor representations of lattice vertices, which we would like to address in detail in this paper. 

In most lattice studies of 3-point vertices, authors use the corresponding tensor elements from the continuum theory \cite{Parrinello:1994wd, Alles:1996ka, Boucaud:1998bq,
Skullerud:2002ge, Skullerud:2003qu, Boucaud:2013jwa, Duarte:2016jhj, Athenodorou:2016oyh, Boucaud:2017obn,Sternbeck:2017ntv}.~However, due to the breaking of rotational 
symmetry, the continuum tensor bases cannot be applied in discretised spacetime, at least not for general kinematics.~This has been explicitly demonstrated for the 
lattice gluon propagator in Landau gauge \cite{Leinweber:1998uu}.~There are a few reasons why this practice persists, despite the errors that it might induce on calculated 
vertex form factors.~One is that, for most vertex functions, the correct alternative to using a continuum basis is simply unknown, despite some clues from lattice perturbation
theory \cite{Rothe:1992nt}.~The other important justification is that some lattice studies are almost exclusively interested in the infrared region \cite{Athenodorou:2016oyh,
Boucaud:2017obn}, where discretisation effects are expected to be small and can arguably be ignored.~As an alternative to continuum bases, some authors have used tree-level
tensor elements from lattice perturbation theory \cite{Cucchieri:2004sq, Cucchieri:2006tf, Ilgenfritz:2006he, Maas:2007uv,Cucchieri:2008qm}, which however do not provide a 
complete representation for most vertex functions. 

In this paper, we attempt to put these matters on a firmer footing, in terms of discretisation error estimates.~We present a simple method, based on vertex
reconstruction, that enables one to quantify how (un)well some basis describes a given correlation function.~We apply the method to the lattice Landau gauge gluon
propagator and three-gluon vertex, and demonstrate that, for general kinematics, these functions are described relatively poorly by the continuum tensor bases.~However, we
also show numerically that there exist special kinematic configurations for which these functions can be described, with virtually no errors, in terms of continuum basis
elements.~We demonstrate analytically why this last statement holds, and argue that it is also applicable, with some caveats, to other QCD correlators.~The possibility to
describe the tensor structure of a given lattice correlator with virtually no discretisation artifacts is of particular interest for lattice studies of the QCD running 
coupling \cite{Alles:1996ka, Boucaud:1998bq, Boucaud:2013jwa, Bloch:2003sk, Blossier:2012ef, Blossier:2013ioa, Boucaud:2013bga}, where elimination of uncertainties in the
ultraviolet energy region is of paramount importance.   

The paper is organised as follows.~In Section \ref{sec:setup} we provide details of our lattice setup.~In Section \ref{sec:reconstruct} we describe the reconstruction
procedure, and test the method by using it on the Landau gauge lattice gluon propagator.~In Section \ref{sec:three_glue} we employ the reconstruction approach to probe the
tensor elements of the Landau gauge three-gluon vertex, and comment on our findings.~Some further discussions, indirectly related to the results presented here, as well as
conclusions are provided in Section \ref{sec:conclude}.~The important technical details have been relegated to two appendices, while the third Appendix contains some of our 
results for vertex dressing functions.                 

\section{Numerical setup}\label{sec:setup}

\subsection{Generation of configurations}\label{sec:generation}

In this work we will consider a lattice $SU(2)$ gauge theory in two and three dimensions, with periodic boundary conditions and an equal number of points $N$
in all directions.~The gauge field configurations used in our simulations have been generated with the standard gauge action of Wilson \cite{Wilson:1974sk}, as well 
as with an $\mathcal{O}(a^2)$ tree-level improved theory \cite{Symanzik:1983dc,Symanzik:1983gh, Weisz:1982zw, Weisz:1983bn, Luscher:1984xn}.~Denoting the Wilson and
improved gauge actions as $S_W$ and $S_I$, respectively, one has
\begin{align}\label{eqn:gauge_actions}
& S_W = \frac{\beta}{N_c} \sum_\text{plaq} Re  \left[ \, \text{Tr} \left(\mathbb{1} - U_\text{plaq} \right)\right] \, , \nonumber \\[0.1cm]
& S_I = \frac{5\,\beta}{3\,N_c} \sum_\text{plaq} Re  \left[ \, \text{Tr} \left(\mathbb{1} - U_\text{plaq} \right)\right] - 
        \frac{\beta}{12N_c} \sum_\text{rect} Re  \left[ \, \text{Tr} \left(\mathbb{1} - U_\text{rect} \right)\right] \, ,
\end{align}   

\noindent
where $N_c = 2$, $U_\text{plaq}$ is a Wilson plaquette operator, and $U_\text{rect}$ stands for $1 \times 2$ and $2 \times 1$ rectangle operators.
More explicitly, we have
\begin{align}\label{eqn:plaq_rect}
U_\text{plaq}(x) \, = & \,\,\, U_\mu(x)\,U_\nu(x+\hat{\mu})\,U^\dagger_\mu(x+\hat{\nu})\,U^\dagger_\nu(x) \, , \nonumber \\ 
U_\text{rect}(x) \, = & \,\,\, U_\mu(x)\,U_\nu(x+\hat{\mu})\,U_\nu(x+\hat{\nu}+\hat{\mu})\,U_\mu^\dagger(x+2\hat{\nu})\,U_\nu^\dagger(x+\hat{\nu})\,U^\dagger_\nu(x) \,\, + \,\, 
  \nonumber \\        & \,\,\, U_\mu(x)\,U_\mu(x+\hat{\mu})\,U_\nu(x+2\hat{\mu})\,U_\mu^\dagger(x+\hat{\mu}+\hat{\nu})\,U_\mu^\dagger(x+\hat{\nu})\,U^\dagger_\nu(x) \,\, .
\end{align} 

In \eqref{eqn:plaq_rect}, all of the links $U_\sigma$ are elements of an $SU(2)$ gauge group.~They are parametrised as $U \equiv U_0\,\mathbb{1} + i\,\vec{U}\cdot
\vec{\sigma}$, where $\mathbb{1}$ is the identity matrix and $\vec{\sigma} \equiv (\sigma^1,\,\sigma^2,\,\sigma^3)$ are the Pauli matrices.~The coefficients $(U_0, \vec{U})
$ are real numbers, and one has $U_0^2 + \vec{U}^2 = 1$.~The gauge actions of \eqref{eqn:gauge_actions} formally become equivalent to the continuum Yang-Mills theory in
the limit $a \rightarrow 0$, if one defines the lattice coupling as $\beta \equiv 4/(a^2 \, g^2)$ (in 2D), or $\beta \equiv 4/(a \, g^2)$ (in 3D).~Here, $g$ is a bare
coupling constant.

For configuration updates, we used a multi-hit variant of the Metropolis algorithm, with 12 hits (update suggestions) per one staple evaluation.~Parameters
of the algorithm were tuned such that, on average, approximately half of all suggested updates was accepted.~Starting from a cold configuration, we performed 5000 update
steps for thermalisation, for all the volumes and $\beta$ values considered in this work.~Upon thermalisation, we kept all of the subsequent configurations for measurements,
9600 for each $(N,\beta)$ pair, and performed an integrated autocorrelation time analysis when calculating statistical uncertainties.~For an estimation of the integrated
autocorrelation time $\tau_\text{int}$, we used an automatic windowing procedure outlined in section 3.3 of \cite{Wolff:2003sm}, with parameter $S=2.5$\,.~For the quantities
studied in sections \ref{sec:reconstruct} and \ref{sec:three_glue} of this paper, the biggest obtained $\tau_\text{int}$ was slightly larger than 1 (recall that $\tau_\text{int}
= 0.5$ implies there are no autocorrelations).  

Comparisons of Wilson and $\mathcal{O}(a^2)$ improved setups were done at constant physics, i.\,e.~for each $\beta$ used in the Wilson approach, we tried to find a
corresponding value in the improved theory, such that the lattice spacings are roughly the same (in physical units) for the two cases.~To determine the spacing $a$ in
physical units, the measurements of the static quark-antiquark potential were used.~The scale was set via the string tension, with the value $\sqrt{\sigma} = 0.44$ GeV.~To
improve the signal quality for the potential, we used APE smearing  \cite{Albanese:1987ds}:~the associated parameter values are collected in Table \ref{tab:config_details}.~In
case of the Wilson gauge action, we also compared the dimensionless quantity $\sqrt{\sigma}\,a$ from our simulations with the analytic result of \cite{Dosch:1978jt} (for 2D
theory), as well as with a fit of equation (67) from \cite{Teper:1998te} (for 3D theory).~In all cases we obtained reasonable agreement of results, see Table \ref{tab:config_details}
for details. 
                     
\subsection{Gluon potential and gauge fixing}  

We use a standard linear definition for the lattice gluon potential $A_\mu$, which is an element of the $SU(2)$ Lie algebra:    
\begin{align}
A_\mu(x) \, \equiv \, \frac{1}{2} \left[ U_\mu(x) - U_\mu^\dagger(x) \right] \, = \, i \, \vec{U}_\mu(x) \cdot \vec{\sigma} \, .
\end{align}      

The colour components of $A_\mu(x)$ are obtained as  
\begin{align}\label{eqn:col_comp}
A_\mu^b(x) \, \equiv \,  \frac{1}{2i} \text{Tr} \left[A_\mu(x)\,\sigma^b\right] \, , \,\,\, b = 1\ldots 3 \, .
\end{align}     

The correlation functions in which we are interested are gauge-dependent, and to evaluate them we fix the thermalised configurations $\{ U \}$ to Landau gauge.~The details
on how this is done can be found in \cite{Wilson:1979wp,Davies:1987vs,Cucchieri:1995pn}.~More precisely, we use equation (3.3) of \cite{Davies:1987vs}, with an expansion
to leading order in $\alpha$, and subsequent reunitarisation.~The free parameter $\alpha$ can be tuned to improve convergence, and its optimal values are collected for
each set of considered gauge field configurations in Table \ref{tab:config_details}.~We are using this so-called Cornell method to fix the gauge due to the algorithm's
straightforward implementation in a parallel environment.~The iterative gauge-fixing process is stopped when the convergence criterion 
\begin{align}\label{eqn:conv_crit}
\frac{1}{V} \sum_x \sum_{b=1}^3 \left[ \nabla \cdot A (x) \right]^2_b \, \leq \, 10^{-14} \, , 
\end{align} 

\noindent
is satisfied.~In the above expression, $V \equiv a^d N^d$ is the lattice volume and $\nabla \cdot A_\mu^b (x)$ stands for the colour components of the lattice
divergence of $A_\mu$.~More precisely, one has 
\begin{align}\label{eqn:a_diverge}
\nabla \cdot A_\mu^b (x) \, \equiv \, \sum_{\mu = 1}^d \left[ A^b_\mu(x) - A^b_\mu(x - e_\mu) \right] \, .
\end{align} 

With \eqref{eqn:conv_crit} one approximates, in terms of lattice quantities, the continuum Landau gauge condition $\partial_\mu A_\mu(x)= 0$.~With the gauge-fixing criterion
thus specified, we can turn to the final ingredient needed for the evaluation of $n$-point gluon correlators in momentum space, which is the Fourier transform of $A^b_\mu(x)$.
It is defined as  
\begin{align}\label{eqn:fourier_glue}
\widetilde{A}_\mu^b(k) \, \equiv \, \sum_x & A^b_\mu(x) \exp \left[2\pi i (k\cdot x + k_\mu/2 ) \right] \, , \quad \text{with} \nonumber \\ 
k_\mu & \equiv \frac{2\pi n_\mu}{a N} \, , \quad n_\mu \in [0,N-1] \, .
\end{align}

In \eqref{eqn:fourier_glue}, the $k_\mu/2$ modification is applied in order to recover the continuum Landau gauge condition with $\mathcal{O} (a^2)$ corrections, instead of
$\mathcal{O}(a)$ ones \cite{Alles:1996ka}.~Namely, with the lattice divergence of \eqref{eqn:a_diverge}, and the Fourier transform $\widetilde{A}_\mu^b(k)$ as defined in 
\eqref{eqn:fourier_glue}, the lattice version of the momentum space Landau gauge condition takes the form
\begin{align}\label{eqn:landau_mom}
\sum_{\mu = 1}^d \, \hat{p}_\mu \widetilde{A}_\mu^b(p) \, = \, 0 \, ,
\end{align}


\noindent
where $\hat{p}_\mu = 2\sin(p_\mu/2)$.~The above relation is formally equivalent to $p_\mu \widetilde{A}_\mu(p) = 0$ up to order $\mathcal{O}(a^2)$.~In actual
simulations the number on the r.\,h.\,s.~of \eqref{eqn:landau_mom} will not be exactly 0, but will have some value on the order of $10^{-5}$ or $10^{-6}$, as dictated by
the gauge-fixing criterion \eqref{eqn:conv_crit}. 


\section{Vertex reconstruction and lattice gluon propagator}\label{sec:reconstruct}

We wish to test the applicability of describing the lattice correlators, primarily the three-gluon vertex, with continuum tensor bases.~A question arises as to how can this 
be done in practical terms, i.\,e.~how one can check if some basis is suitable for a description of a given vertex function.~One approach is presented in \cite{Leinweber:1998uu}, 
where it was applied to the Landau gauge lattice gluon propagator.~The technique employed there can be useful, but it only works for vertices with a single tensor element.~Here
we propose a method based on vertex reconstruction, which can (in principle) be used for arbitrary correlators. For the gluon propagator, our approach reduces to the same steps
used in \cite{Leinweber:1998uu}.

We denote a generic lattice correlation function with $\Gamma_\mu(p)$, where the superindex $\mu$ stands for any applicable Lorentz indices, and $p$ subsumes the independent
momentum variables.~We wish to test if $\Gamma_\mu(p)$ can be described with a basis $\tau^{\,j}_\mu(p)$, with index $j$ denoting individual tensor elements.~That is, we wish
to see if the relation 
\begin{align}\label{eqn:arbit_correl}
\Gamma_\mu(p) \, = \, \sum_j \, \mathcal{F}_j(p) \, \tau^{\,j}_\mu(p) \, ,
\end{align}
\noindent
holds, with $\mathcal{F}_j$ being a dressing/coefficient function (or form factor) of a tensor element $\tau^{\,j}_\mu$.~One way to do this is to attempt a vertex 
reconstruction.~Explicitly, one constructs the projectors for the basis $\tau_\mu$, and projects out the functions $\mathcal{F}$ from the lattice vertex $\Gamma_{\mu
}$.~One then reconstructs the correlator, via \eqref{eqn:arbit_correl}, from the dressings $\mathcal {F}$ and the $\tau_\mu$ basis.~Finally, one compares the reconstructed
and the original vertex.~If the relation \eqref{eqn:arbit_correl} is correct, then no information will be lost when computing the $\mathcal{F}$ functions.~Consequently, the
reconstructed vertex will be equal to the original one.~Any discrepancy between the reconstructed and original vertex points to an inadequacy of the basis $\tau_\mu$, and the
``size'' of the discrepancy is an indication on how unsuitable the basis is, for given kinematics.~Let us test these ideas on the gluon two-point function.~The continuum,
infinite-volume version of the Landau gauge gluon propagator is given by 
\begin{align}\label{eqn:gluon_cont_tensor}
D^{\,\text{cont}, \, ab}_{\mu\nu, \, p} = \left(\delta_{\mu\nu} - \frac{p_\mu p_\nu}{p^{\,2}}\right)\delta^{ab}\,D(p^2) \,,
\end{align}
\noindent
with colour indices $a,\,b$.~On the lattice, one can deduce the tensor structure of the Landau gauge propagator by combining the correlators definition with 
\eqref{eqn:landau_mom}.~The lattice gluon propagator is given by
\begin{align}\label{eqn:latt_gluon}
D_{\mu\nu}^{\,ab}(p) = \frac{1}{V} \left\langle \widetilde{A}^a_\mu(p) \, \widetilde{A}^b_\nu(-p)   \right\rangle \,,
\end{align}
   
\noindent
with $V$ the lattice volume and $\widetilde{A}(p)$ defined in \eqref{eqn:fourier_glue}.~From the constraint of \eqref{eqn:landau_mom} and the definition of 
\eqref{eqn:latt_gluon}, one can straightforwardly show that the lattice gluon two-point function in Landau gauge has the form (up to corrections dictated by
numerical gauge-fixing):
\begin{align}\label{eqn:gluon_latt_tensor}
D^{\,ab}_{\mu\nu}(p) = \left(\delta_{\mu\nu} - \frac{\hat{p}_\mu \, \hat{p}_\nu}{\hat{p}^{\,2}}\right)\delta^{ab}\,D(p^2) \,.
\end{align} 
 
The structure of \eqref{eqn:gluon_latt_tensor} will remain the same regardless of the employed lattice action, as long as the same gauge-fixing algorithm is used
for all simulations.~We assume the propagator to be diagonal in colour space, as shown above, and will henceforth consider the colour-averaged quantities $D_{\mu\nu}
\equiv \frac{1}{3} \sum_a D^{\,aa}_{\mu\nu}$.~This leaves only the tensorial part.~For both the representation of \eqref{eqn:gluon_cont_tensor}, and the one of
\eqref{eqn:gluon_latt_tensor}, the form factor $D(p)$ can be projected out with a simple $D$-dimensional Kronecker tensor $\delta_{\mu\nu}$.~In other words,
one has
\begin{align}\label{eqn:project_glue}
D(p) \, = \, \frac{1}{\mathcal{N}} \, \delta_{\mu\nu} \, D_{\mu\nu}(p) \, , 
\end{align}   

\noindent
with implied summation over repeated indices.~For $p=0$, the normalisation factor $\mathcal{N}$ equals $D$ (the number of dimensions), otherwise it is $D-1$ 
\cite{Alles:1996ka}.~Since the projector of \eqref{eqn:project_glue} is momentum-independent, the discussion of tensor structure is actually superfluous for the Landau
gauge gluon propagator.~Put differently, a detailed consideration of the propagators tensor representation has no bearing on the way that one calculates the form factor
$D(p)$.~Nevertheless, taking a closer look at this two-point function is useful for demonstrating the basic ideas of our method.

\newpage

~For the reconstruction part of our approach, we take the propagator dressing of \eqref{eqn:project_glue}, and obtain the reconstructed correlator by plugging in $D(p)$ into
either of equations \eqref{eqn:gluon_cont_tensor} or \eqref{eqn:gluon_latt_tensor}.~The end result is then compared to the original propagator, i.\,e.~$D^\text{\,calc}_{\mu\nu}
\sim \widetilde{A}_\mu \widetilde{A}_\nu$.~Since we do not wish to compare the two-point functions for each individual value of indices $\mu$ and $\nu$, we will consider the
index-averaged quantities, namely 
\begin{align}\label{eqn:prop_ratio}
\frac{D_{\left|\left\langle\mu\nu\right\rangle\right|}^\text{\,calc}}{D_{\left|\left\langle\mu\nu\right\rangle\right|}^\text{\,recon}} \,\, \equiv \,\,
\frac{\sum_\mu \sum_\nu | D_{\mu\nu}^\text{\,calc}|} { \sum_\mu \sum_\nu |D_{\mu\nu}^\text{\,recon}|} \, , 
\end{align} 

\noindent
with $|.|$ denoting a (complex number) absolute value.~When evaluating the ratios like the one above, we will always use the absolute value of propagators and
vertices.~There are multiple reasons for this, and here we mention two of them.~Firstly, for diagonal momenta (i.\,e.~$p_\mu = p_\nu$ for all $\mu, \nu$), performing an
index average for the reconstructed Landau gauge correlators would always yield zero, without the absolute value.~For the gluon propagator, this can be seen by taking an
ordinary (no absolute value) index average of the r.\,h.\,s.~of either of equations \eqref{eqn:gluon_cont_tensor} or \eqref{eqn:gluon_latt_tensor}, for diagonal momenta.~The
second reason is that the signal quality is generally better for absolute value of correlators than the correlators themselves.~We discuss the second point in more detail at
the end of section \ref{sec:vertex_2d}.~To confirm that the index-averaging procedure does not introduce a large bias for the results, we've also performed calculations where
propagators were compared component-wise (e.\,g.~$D^\text{\,calc}_{11}/D^\text{\,recon}_{11}$, etc.), and checked that such comparisons yield (on average) results similar to
the ratio of \eqref{eqn:prop_ratio}.~The biggest relative difference in results between the two methods was on the order of one percent.    

\begin{figure}[!t]
\begin{center}
\graph[width = 0.44\tew]{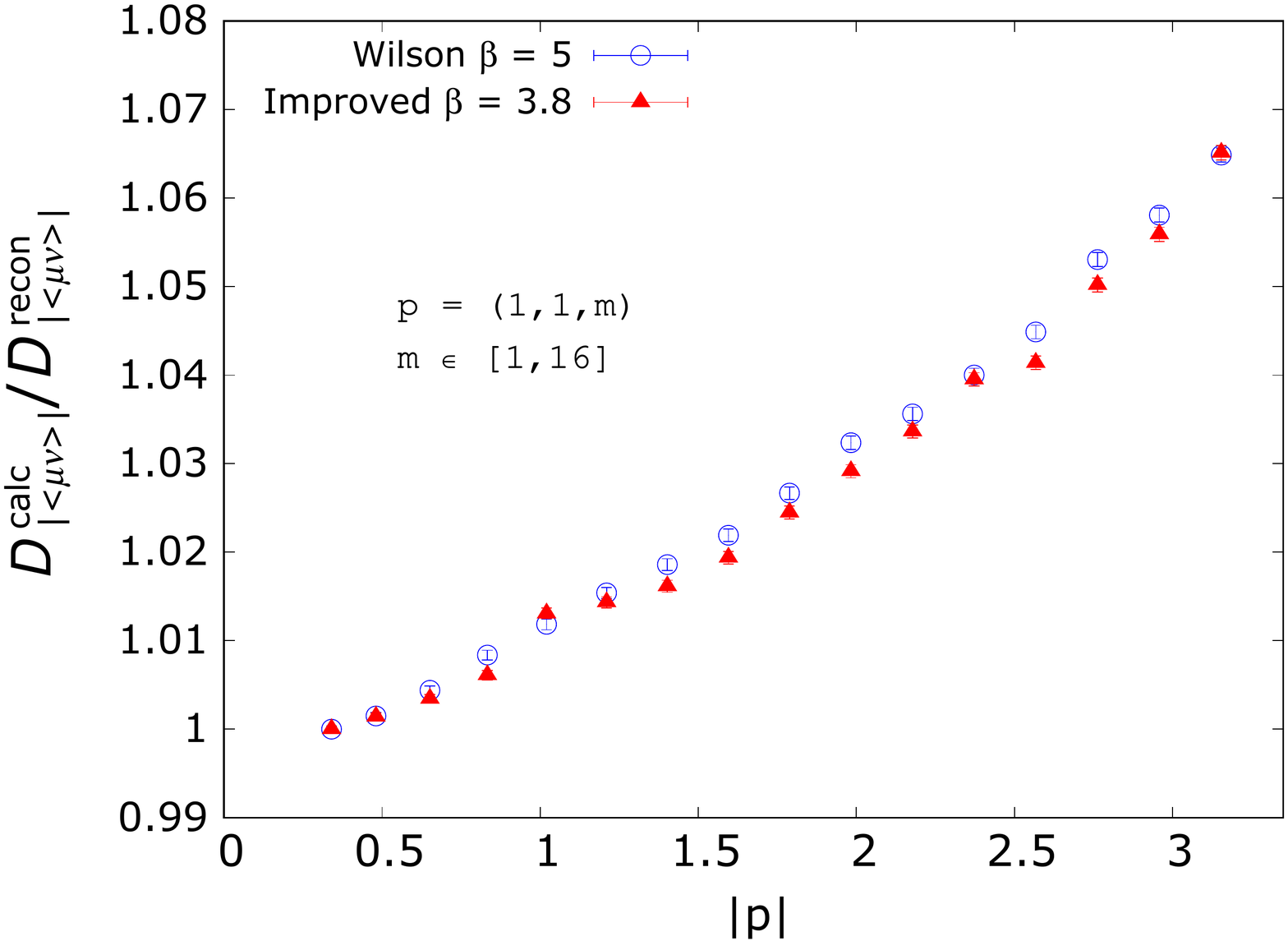}\graph[width = 0.44\tew]{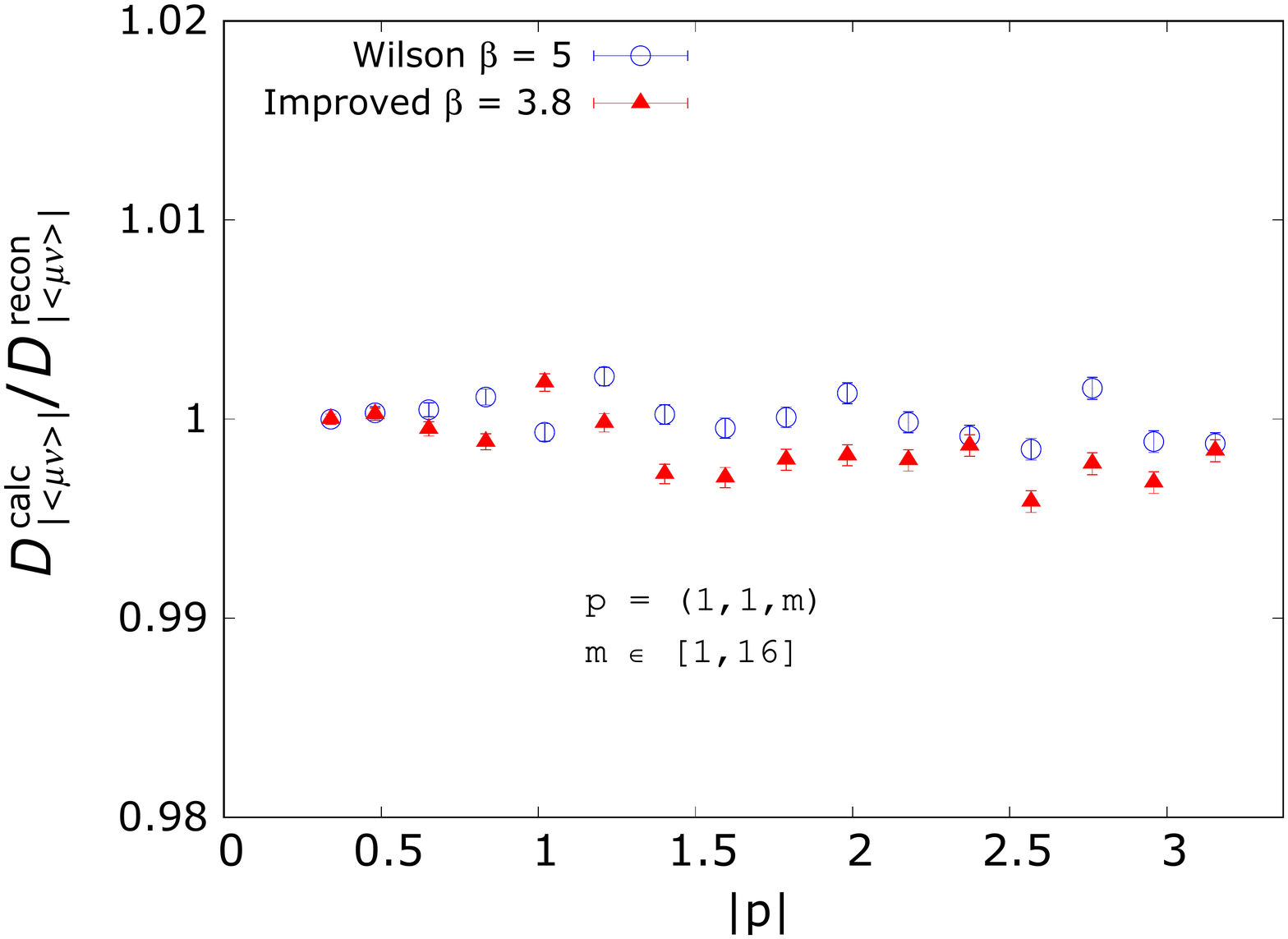}\\
\caption{Comparison of calculated and reconstructed gluon on a $32^3$ lattice, for near-axis momentum $p$ and with $|\,p\,| \equiv \sqrt{\smash[b] p^2}$.~Results
are in lattice units, with $p$ given in terms of components of vector $n_\mu$ of \eqref{eqn:fourier_glue}.~\textit{Left}:~Propagator reconstruction according to
\eqref{eqn:gluon_cont_tensor}.~\textit{Right}:~Reconstruction according to \eqref{eqn:gluon_latt_tensor}.~Shown are data for Wilson and $\mathcal{O}(a^2)$ improved 
actions.} 
\label{fig:gluon_side}
\end{center}
\end{figure}

\begin{figure}[!t]
\begin{center}
\graph[width = 0.44\tew]{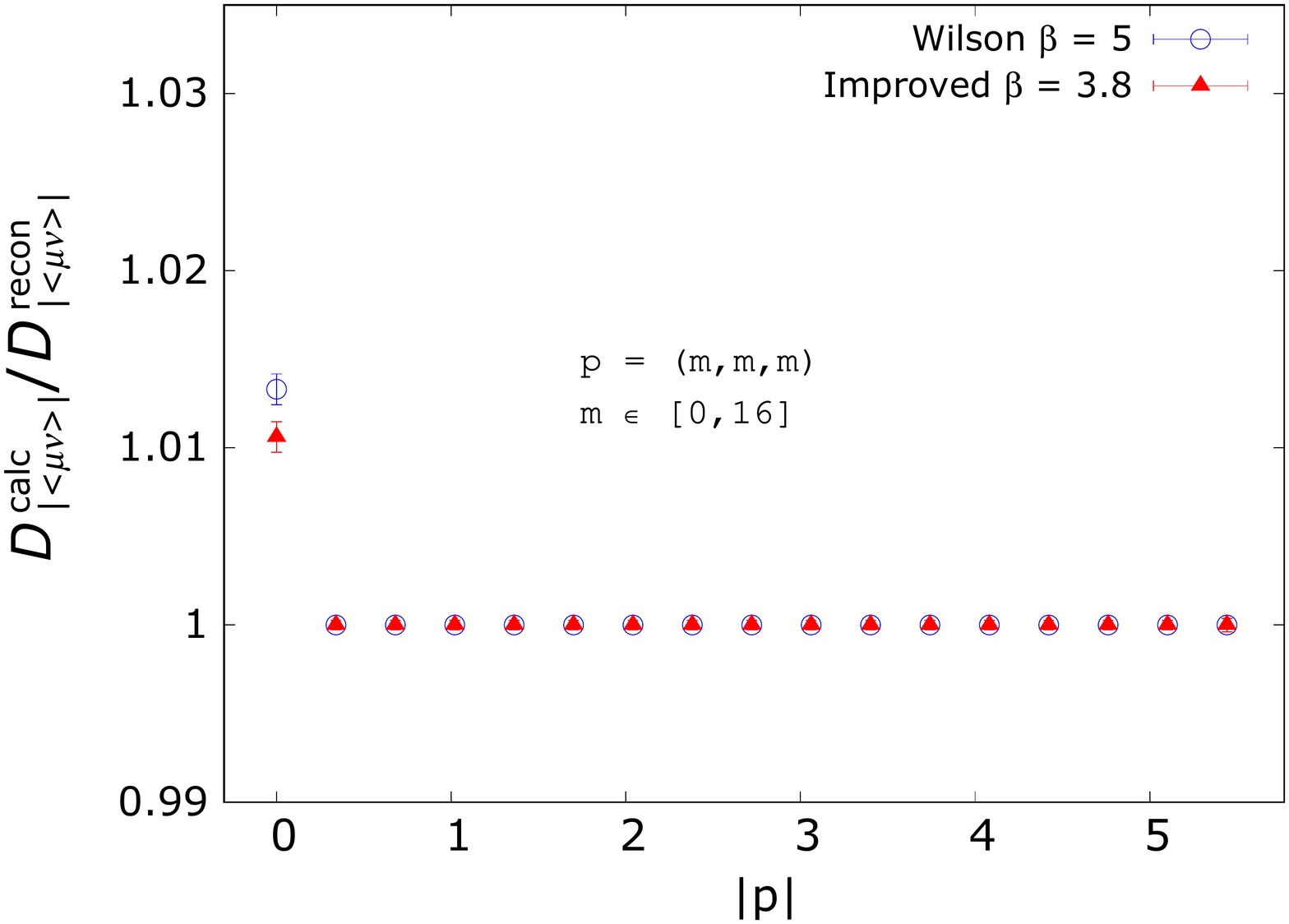}\\
\caption{Comparison of calculated and reconstructed gluon on a $32^3$ lattice, for diagonal momentum $p$ and with reconstruction according to \eqref{eqn:gluon_cont_tensor}.~Results 
are in lattice units, with $p$ given in terms of components of vector $n_\mu$ of \eqref{eqn:fourier_glue}.~$|\,p\,|$ stands for $\sqrt{\smash[b] p^2}$.} 
\label{fig:gluon_diag}
\end{center}
\end{figure}

A remark is in order regarding our notation.~In equation \eqref{eqn:prop_ratio}, $D^\text{\,calc}_{\mu\nu}$ does \textbf{not} stand for a Monte Carlo average, akin to the one
of \eqref{eqn:latt_gluon}.~It instead denotes a product of vector potentials, considered for each gauge field configuration separately.~The same goes for the reconstructed
gluon propagator and the whole ratio in \eqref{eqn:prop_ratio}:~the ratios are evaluated on the level of individual configurations, and in the end these results are averaged to
get the final estimate, together with the associated uncertainty.~For better statistics, we also perform averages over permutations of momentum components, of which there are 2
in two dimensions, and 6 in three dimensions\footnote{As an example of permutations in 3D, one may look at a momentum $p$ with components $p = (a,b,c)$.~To each result for $p$ we
add results for permuted versions, i.\,e.~for momenta $p' = (a,c,b)$, $p''= (b,a,c)$, and three others, and make an average of this sum.}.~Due to hypercubic symmetry (a symmetry
under permutations and reflections of coordinates), the dressing function $D(p)$ of \eqref{eqn:project_glue} should remain unchanged when components of $p$ are interchanged,
thus justifying the aforementioned permutation average.~The final results for momenta near the lattice axis in a 3D theory are given in Figure \ref{fig:gluon_side}.~In 
Fig.~\ref{fig:gluon_side} we do not consider the momenta exactly along the axis, in order to avoid finite volume effects, see the first data point of Fig.~\ref{fig:gluon_diag}. 

The plots in Fig.~\ref{fig:gluon_side} show the behaviour that one would expect, based on our previous discussions.~The data indicate that the reconstruction method
works for the gluon propagator.~To get more valuable insight, one can use the continuum tensor of \eqref{eqn:gluon_cont_tensor} for reconstruction, but for diagonal
momenta.~The result is given in Fig.~\ref{fig:gluon_diag}, and it suggests that along the lattice diagonal, one can describe the lattice gluon with a continuum tensor
structure.~The explanation for this is straightforward.~For diagonal kinematics, the non-trivial part of the transverse projector $T_{\mu\nu}^p = \delta_{\mu\nu} - p_\mu
p_\nu/p^2$ [\,bracketed object on the r.\,h.\,s.~of equations \eqref{eqn:gluon_cont_tensor} and \eqref{eqn:gluon_latt_tensor}\,] is momentum-independent.~To be more explicit,
with $p^\mu = p^\nu$ (for all $\mu,\nu$), one gets\footnote{We are grateful to Attilio Cucchieri for pointing this out to us.} 
\begin{align}\label{eqn:gluon_discrete}
\frac{p_\mu p_\nu}{p^2} \, = \, \frac{\hat{p}_{\mu} \hat{p}_{\nu}}{\hat{p}^2} \, = \, \frac{1}{D} \, , \quad \text{for all} \,\,\, \mu, \nu \, ,
\end{align} 

\noindent
where $D$ is the number of dimensions.~The above equation might be somewhat confusing, due to a ``loss of indices'' on the right-hand side.~To clarify things, let us look at an explicit 
example in two dimensions, with diagonal momentum $p = (p_1, p_2) = (m,m)$, and its sine-transformed version $\hat{p} = (\hat{p}_1, \hat{p}_2) = (\hat{m}, \hat{m})$, with $\hat{m} = 
2 \sin(m/2)$.~For such kinematics, it holds that 
\begin{align}\label{eqn:exp_example}
\frac{p_1^2}{p^2} \, = \, \frac{p_2^2}{p^2} \, = \, \frac{p_1 \, p_2}{p^2} \, = \, \frac{m^2}{D\,m^2}  \, = \, \frac{\hat{p}_1^2}{\hat{p}^2} \, = \, \frac{\hat{p}_2^2}{\hat{p}^2} \,
= \, \frac{\hat{p}_1 \, \hat{p}_2}{\hat{p}^2} \, = \, \frac{\hat{m}^2}{D\,\hat{m}^2} \, = \,  \frac{1}{D} \, .
\end{align}  

Relations \er{eqn:exp_example} are equivalent to \er{eqn:gluon_discrete}, if one explicitly writes out \er{eqn:gluon_discrete} for all possible values of indices $\mu$ and $\nu
$.~Generalisation of the above statements to higher dimensions is straightforward.~It should be fairly obvious that the result \er{eqn:gluon_discrete} is discretisation-independent,
meaning that the exact form of $\hat{p}_\mu$, and the details of the lattice formulation, are unimportant.~As we show later, for the three-gluon vertex there are several kinematic
configurations for which the same observation holds.~Regarding the results in Fig.~\ref{fig:gluon_diag}, we additionally point out that the discrepancy at $p=0$ is most likely due to
finite-volume artifacts, see section II of \cite{Leinweber:1998uu} for a more detailed discussion.~Some other effects that might also contribute at zero momentum are discussed in 
Section \ref{sec:conclude}.

Before moving further, we wish to address the absence, in Figures \ref{fig:gluon_side} and \ref{fig:gluon_diag}, of any appreciable differences between data for the Wilson
action gluon and the $\mathcal{O}(a^2)$ improved one.~Two main factors contribute to this result.~First, in our plots we consider the ratios of propagators, where the correlator
dressing function $D(p^2)$ (which is different for the two kinds of lattice gauge action) should drop out.~Second, the gauge field configurations coming from both $S_W$ and
$S_I$ actions, have been numerically subjected to the constraint of \eqref{eqn:landau_mom}, which ensures that the two kinds of propagator have an identical tensor structure
given in \eqref{eqn:gluon_latt_tensor}.


\section{The three-gluon vertex}\label{sec:three_glue}

\subsection{Colour and tensor structure in the continuum}\label{sec: colour_tensor}

The lattice three-gluon vertex is defined as 
\begin{align}\label{eqn:g3_lattice}
\Gamma^{\,abc}_{\mu\nu\rho}(p, q, r) \, = \, \frac{1}{V} \left \langle \widetilde{A}^{a}_\mu(p) \, \widetilde{A}^{b}_\nu(q) \, \widetilde{A}^{c}_\rho(r)
\right\rangle  ,
\end{align} 

\noindent
where $r = - (p + q)$.~Actually, the above quantity is not the one-particle-irreducible (1PI) vertex, but simply a gluon three-point function:~to obtain the true 
1PI vertex, one needs to amputate the gluon legs, see e.\,g.~\cite{Parrinello:1994wd}.~For most of our results, this distinction is unimportant, as we will be looking at
vertex ratios where the gluon propagators coming from the amputation would anyway drop out.~Before presenting our results for the function of \eqref{eqn:g3_lattice}, we 
need briefly to discuss its colour and tensor decomposition in the continuum.~Let us start with the colour part.~In general, the continuum three-gluon vertex has the form
(we temporarily suppress the momentum dependencies):
\begin{align}\label{eqn:colour_basis}
\Gamma_{\mu\nu\rho}^{\,abc} \, = \, f^{\,abc} \, \Gamma^\text{\,a}_{\mu\nu\rho} + d^{\,abc} \, \Gamma^\text{\,s}_{\mu\nu\rho} \, ,
\end{align} 

\noindent
with $f^{\,abc}$ and $d^{\,abc}$ the antisymmetric and symmetric structure constants, respectively.~$\Gamma^\text{\,a/s}_{\mu\nu\rho}$ are the corresponding
tensor elements.~Orthogonality of the colour constants (i.\,e.~$f^{\,abc}d^{\,abc} = 0$) can be used to project out the desired tensor piece.~When doing vertex
reconstruction, the colour symmetric and antisymmetric parts can be analysed independently of each other.~For the $SU(2)$ group which we are considering, these
matters are simpler since $d^{\,abc} = 0$ .~It is also possible that for other gauge groups, like $SU(3)$, the symmetric contributions to the vertex are negligibly 
small or even completely vanishing.~Results that might point to this conclusion can be found in \cite{Boucaud:1998bq, Huber:2017txg, Davydychev:1996pb, Davydychev:1997vh,
Binger:2006sj, Smolyakov:1980wq, Blum:2015lsa}.~In either case, we extract the tensor part of the correlator with a contraction $\Gamma_{\mu\nu\rho} = (f^{\,abc}/6) \cdot 
\Gamma^{\,abc}_{\mu\nu \rho}$, where $\Gamma^{\,abc}_{\mu\nu\rho}$ is given in \eqref{eqn:g3_lattice}.~The prefactor of (1/6) in the colour projection takes care of 
normalisation:~it can be deduced by applying the identity $f^{\,acd}f^{\,bcd} = N\, \delta^{ab}$, valid for arbitrary $SU(N)$ groups, to the specific case of $SU(2)$ 
gauge transformations.       

This brings us to the tensor part.~For covariant gauges and a number of dimensions greater than 2, the three-gluon vertex can be decomposed into 14 linearly independent
tensor elements.~For Landau gauge with more than 2 spacetime dimensions, the number of dynamically relevant tensor structures is reduced to 4, due to transversality 
conditions.~In our numerics we mostly employ the transverse orthonormal (ON) basis, used for the first time in \cite{Eichmann:2014xya}.~That paper gives a full account 
on how the basis is constructed, but we repeat the main steps in our Appendix \ref{sec:ortho_basis} as well.~In the same Appendix we prove, using the ON basis, that in
two dimensions a single tensor element is adequate to describe the Landau gauge three-gluon vertex.~Summing up, for our study in three dimensions we use the decomposition 
\begin{align}\label{eqn:cont_tensor}
\Gamma_{\mu\nu\sigma}(p,q,r) \, = \, \sum_{j=1}^4 \, \mathcal{B}_{\,j}(p,q,r) \, \rho^{\,j}_{\mu\nu\sigma}(p,q,r) \, ,
\end{align} 
\noindent
with elements $\rho^{\,j}_{\mu\nu\sigma}$ given in equation (A11) of \cite{Eichmann:2014xya}, as well as in equation \eqref{eqn:on_basis} of our Appendix \ref{sec:ortho_basis}.~In
two dimensions, only the tensor $\rho^{\,2}_{\mu\nu\sigma}$ is needed to represent the three-gluon coupling, as all the other ones vanish.~Since the basis $\rho^{\,j}_{\mu\nu\sigma}$
is orthonormal, it is straightforward to get the corresponding form factors from the calculated vertex.~Namely, one has 
\begin{align}\label{eqn:project_on}
\mathcal{B}_{\,j}(p,q,r) \, = \, \rho^{\,j}_{\mu\nu\sigma}(p,q,r) \cdot \Gamma_{\mu\nu\sigma}(p,q,r) \, , \qquad j = 1\ldots 4 \, . 
\end{align}

Henceforth, we employ the Einstein summation convention, unless stated otherwise.~The ON basis is useful for numerics and vertex reconstruction, but it is not very
``friendly'' for certain analytic manipulations.~We are mainly referring to our intent to demonstrate, for the three-gluon correlator, some results akin to equation
\eqref{eqn:gluon_discrete} for the gluon propagator.~Such relations can be proved with the ON basis as well, but for calculations of this type we prefer to use another
tensor decomposition for the vertex, where some arguments become more transparent.~We refer to the said decomposition as the ``Simple'' one, and show the construction of
corresponding elements in Appendix \ref{sec:simple_basis}.~The connection between the ON and Simple basis is also provided there.  

Finally, before moving on to the results for the three-gluon vertex, we wish to emphasise an important part of our numerical procedure.~Namely, apart from the lattice 
implementation of the Landau gauge condition \er{eqn:landau_mom}, we additionally act on the product \er{eqn:g3_lattice} explicitly with transverse projectors, in a
somewhat continuum fashion: 
\begin{align}\label{eqn:add_trans}
\Gamma^{\,abc, \, \text{tr}}_{\mu\nu\rho}(p, q, r) \, = \, T_{\alpha\mu}^{\,p,\,l} \, T_{\beta\nu}^{\,q,\,l} \, T_{\gamma\rho}^{\,r,\,l} \cdot 
\Gamma^{\,abc}_{\alpha\beta\gamma}(p, q, r) \, .
\end{align}

In the above relation, $T_{\alpha\mu}^{\,p,\,l}$ is a projection operator with lattice-adjusted momentum (note the superscript `$l$'), i.\,e.~$T_{\alpha\mu}^{\,p,\,l} =
\delta_{\mu\alpha} - \, \hat{p}_\alpha\hat{p}_\mu/\hat{p}^2$, where $\hat{p} = 2\sin(p/2)$.~Additionally, $\Gamma^{\,abc}_{\alpha\beta\gamma}$ stands for the product 
\er{eqn:g3_lattice}, with renamed Lorentz indices, and $\Gamma^{\,abc, \, \text{tr}}_{\mu\nu\rho}$ is the vertex that we will be working with from now on.~We perform
the above operation because the lattice transversality condition \er{eqn:conv_crit} may not be quantitatively good enough, for certain kinematics, when it comes to our
vertex reconstruction procedure.~We will clarify this last point at the end of the following section.  
     
\subsection{Vertex results in two dimensions}\label{sec:vertex_2d}

The setup of our calculations for the three-gluon coupling is an extension of the procedure we outlined for the gluon propagator.~The lattice vertex is calculated 
as a product $\Gamma^{\,abc}_{\mu\nu\rho} \sim \widetilde{A}_\mu^a\,\widetilde{A}_\nu^b\,\widetilde{A}_\rho^c$, and its colour dependence is taken care of with the 
$f^{\,abc}$ projection.~We attempt to reconstruct the remaining tensor piece with appropriate tensor bases, and form the ratios of index-averaged quantities.~The
index average of (say) a calculated vertex is defined as 
\begin{align}\label{eqn: gamma_aver}
\Gamma_{\left|\left\langle\mu\nu\rho\right\rangle\right|}^\text{\,calc} \, = \,  \sum_{\mu\nu\rho}| \Gamma_{\mu\nu\rho}^\text{\,calc}| \, ,
\end{align}

\noindent
where $|.|$ again denotes a complex number absolute value.~As in the case of the gluon propagator, the values for vertex ratios are obtained for each gauge field configuration
separately, and these results are averaged over to obtain the final answer and the corresponding error estimate.~As noted in the previous section, in a two-dimensional theory
only the tensor element $\rho^{\,2}_{\mu\nu\sigma}$ of \eqref{eqn:on_basis} is required for a reconstruction in the continuum.    
 
Owing to momentum conservation, $r = - (p + q)$, only two out of three momenta that enter the three-gluon vertex are independent.~In our simulations we take these vectors
to be $p$ and $q$.~For improved statistics we perform permutation averages, wherein to each result for momenta $(p,q)$ we add results where components of $p$ and $q$ have 
been permuted in various ways.~We do not consider such permutations for each momentum $p$ and $q$ separately, but instead perform the same transformation on both vectors.~Thus,
as in the case of the gluon propagator, we average over a total of 2 permutations in 2D, and 6 permutations in 3D.   

One final notion we need to introduce before discussing the results is that of the ``sine improvement''.~From Landau gauge condition \eqref{eqn:landau_mom} and the
definition of lattice three-gluon vertex \eqref{eqn:g3_lattice}, it is clear that this correlator should satisfy [\,see also \er{eqn:add_trans}\,]:
\begin{align}\label{eqn:3g_landau_latt}
\hat{p}_{\mu} \, \hat{q}_{\nu} \, \hat{r}_{\rho}\,\Gamma_{\mu\nu\rho}(p,q,r) \, = \, 0 \, .
\end{align}

The continuum Landau gauge vertex obeys the same relation as above, but with $(\hat{p},\,\hat{q},\,\hat{r})$ replaced with $(p,q,r)$.~The analogy suggests that, to
describe the tensor structure of the lattice correlator, one needs to use modified momenta, like $p \rightarrow \hat{p} = 2 \sin\,(p/2)$, when constructing the
vertex tensor elements.~However, for general kinematics, the sine modification cannot be carried out for all three momenta at once, since it would spoil the momentum
conservation condition $r = -(p+q)$ [\,since in general $\sin(x+y) \neq \sin(x) + \sin(y)$].~We still want to test if the sine correction can help with the reduction of
errors.~Aside from a normal reconstruction with independent momenta $(p,q)$, we also consider a sine-modified method, where vectors $(\hat{p},\hat{q})$ are used for the
tensor elements.~In our plots, we refer to the second procedure as ``sine''.~We will only display the sine results for Wilson gauge action, to prevent the graphs from
getting too cluttered.~An approach similar to our sine correction was already used for the lattice measurements of the three-gluon running coupling \cite{Boucaud:1998bq}.     
\begin{figure}[!t]
\begin{center}
\graph[width = 0.40\tew]{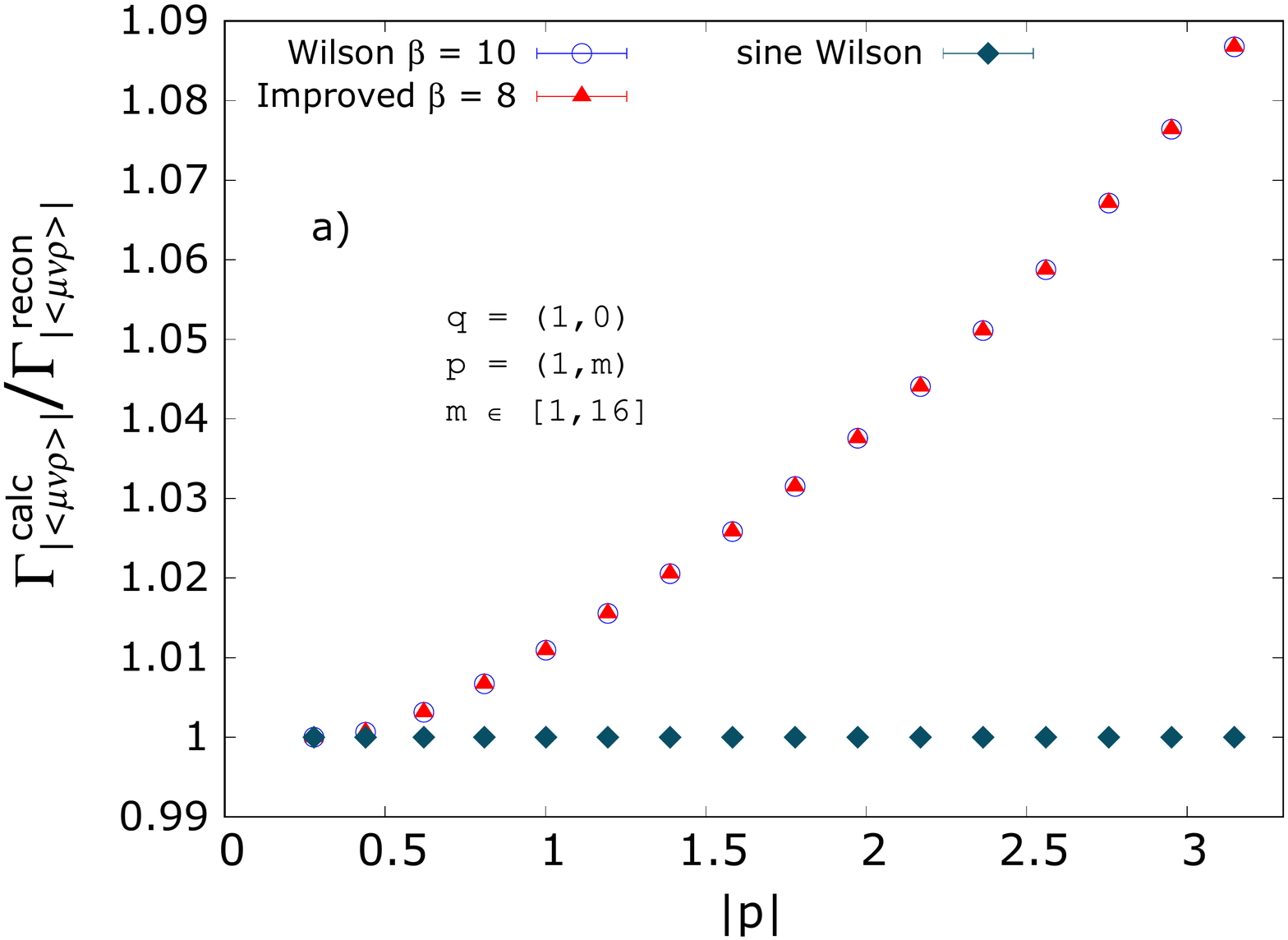}\graph[width = 0.40\tew]{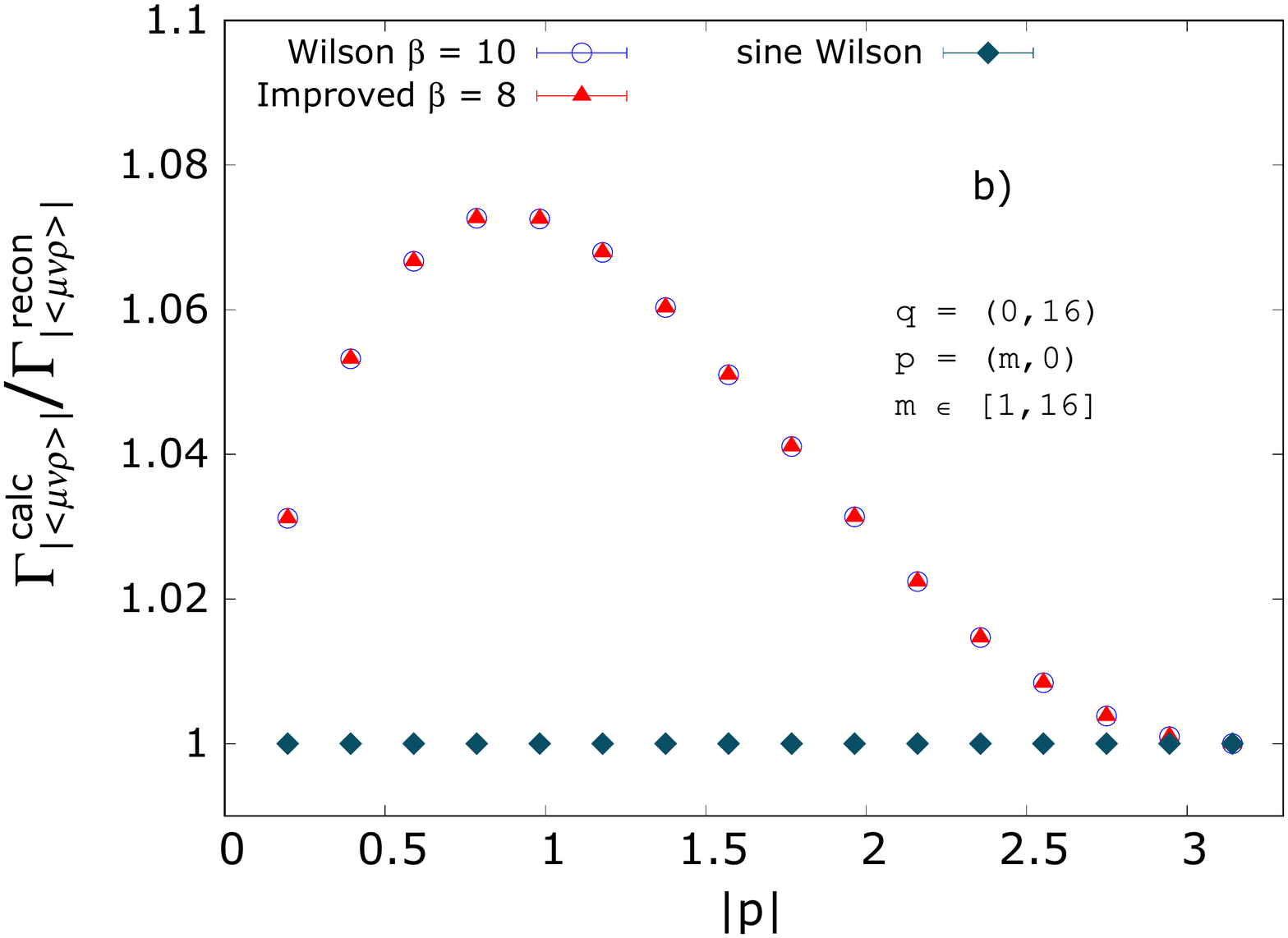}\\
\graph[width = 0.40\tew]{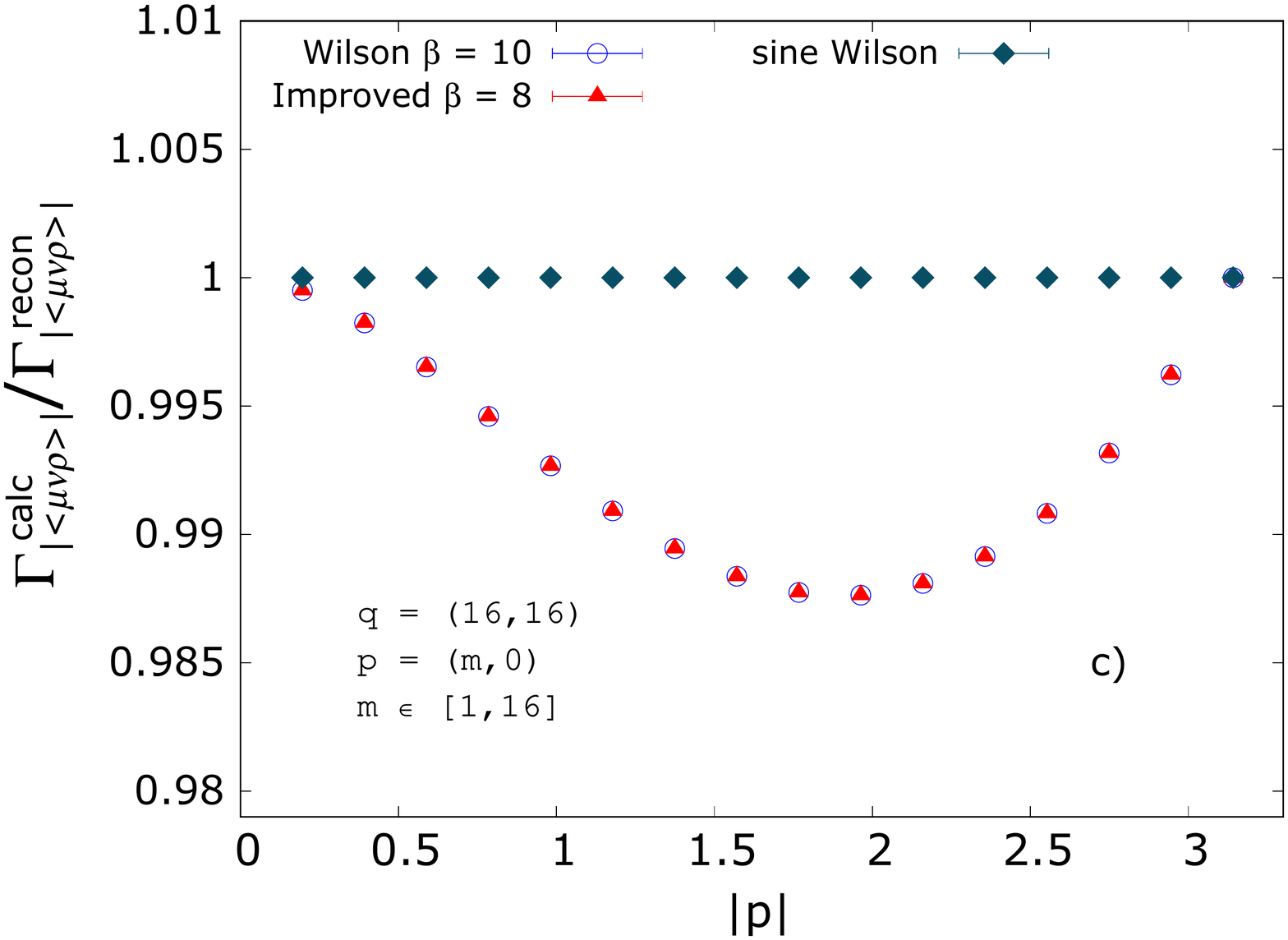}\graph[width = 0.40\tew]{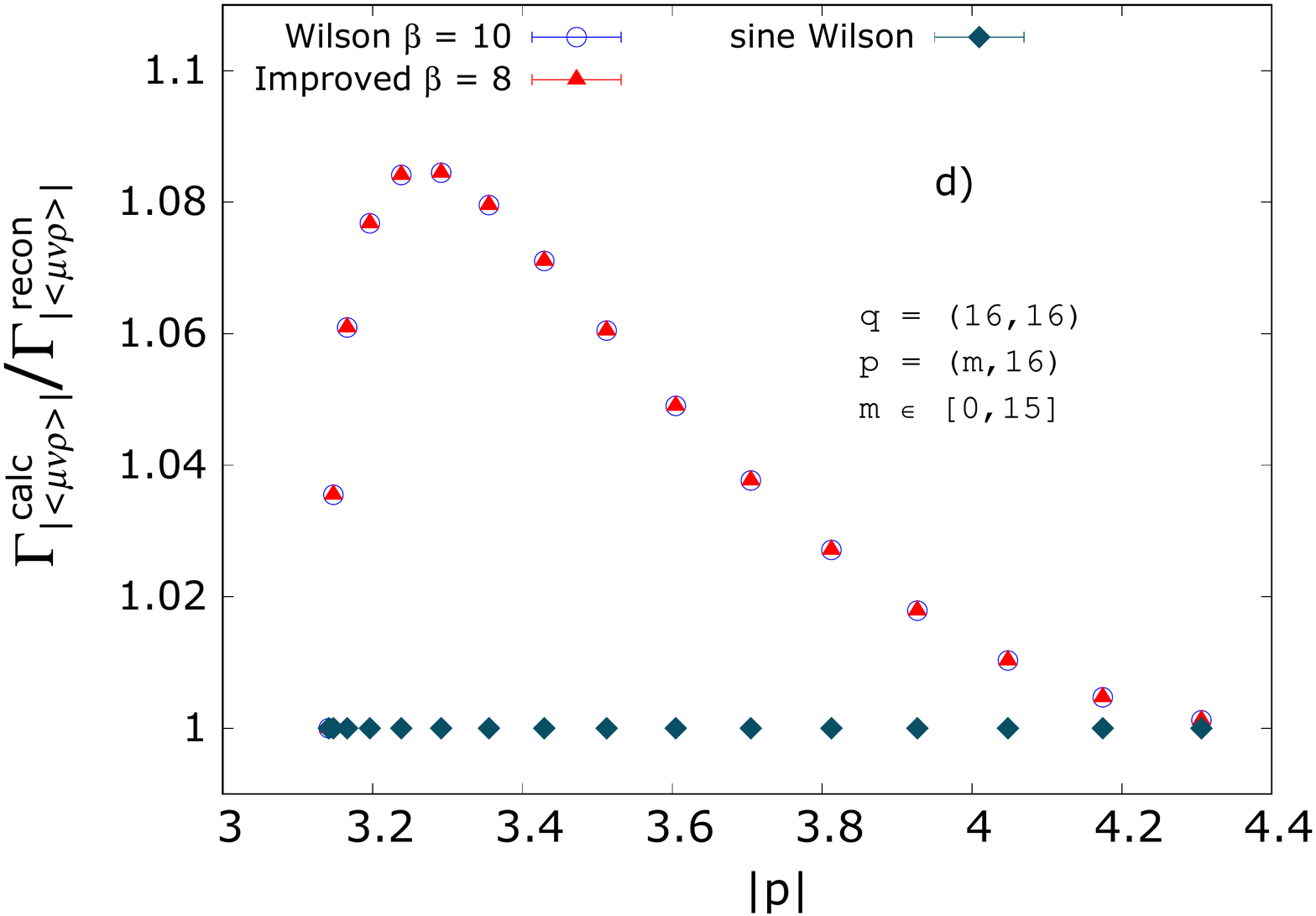}\\\caption{Ratios of calculated and reconstructed
vertices on a 32$^2$ lattice, as functions of $|\,p\,| = \sqrt{\smash[b]p^2}$.~Reconstruction was done with the ON basis tensor $\rho^{\,2}_{\mu\nu\sigma}$
of \eqref{eqn:on_basis}.~``Sine'' data are for a reconstruction with modified momenta $(\hat{p},\,\hat{q})$, where e.\,g.~$\hat{p} = 2\sin(p/2)$.~Results are in lattice
units, with all momenta given in terms of components of vector $n_\mu$ of \eqref{eqn:fourier_glue}.~See text for further discussion.} 
\label{fig:configs_3g_2d}
\end{center}
\end{figure}

This brings us to the results.~In Fig.~\ref{fig:configs_3g_2d} we show the plots of our data for several kinematic configurations on a two-dimensional lattice.
In the first plot, the vector $q$ has relatively small components.~Consequently, the sine improvement can be applied to all momenta, while approximately keeping
the momentum conservation intact, i.\,e.~$\hat{r} \approx - (\hat{p} + \hat{q})$.~This is why the sine-adjusted reconstruction works well in the whole examined range
of $p$ values.~For other plots in the Figure, the interpretation of the data shown is not as straightforward, but it is also not exceptionally challenging.   

Firstly, one notes that for results in the graphs \ref{fig:configs_3g_2d}\,b) through d), the first and the last kinematic points show the smallest deviations between 
the reconstructed and calculated vertex, with all the points in between corresponding to greater discrepancies.~This is because, for the examined kinematic cuts, the 
first and final points exemplify what we shall refer to as generalised diagonal kinematics, where all the components of vertex momenta are equal to a single scale $s$, or
there is some combination of a single scale $s$ and vanishing components.~To clarify, let us take the example of the final kinematic point in plot \ref{fig:configs_3g_2d}\,c).
The corresponding vertex momenta are 
\begin{align}\label{eqn:demon}
p \, = \, (\pi,0) \, , \quad q \, = (\pi,\pi) \, , \quad r \, = - \, (0,\pi) \, .
\end{align}  

The first component of momentum $r$ is zero in \eqref{eqn:demon}, due to periodic boundary conditions.~One easily sees that, up to a sign, the components of all the momenta
in \eqref{eqn:demon} are equal to either zero, or to the same non-zero number $s$ (in this case $s = \pi$).~Additionally, one of the momenta is fully diagonal, meaning that all 
of its components are equal to each other [\,in the case of configuration \er{eqn:demon}, this is vector $q$\,].~We refer to kinematic choices akin to \eqref{eqn:demon} as 
generalised diagonal kinematics, and in Appendix \ref{sec: diagonal} we show that such momentum points are special in terms of vertex tensor representations.~More precisely, for
configurations like \eqref{eqn:demon}, one can use continuum bases to describe the tensor structure of a lattice three-gluon vertex, with virtually no errors coming from rotational
symmetry breaking.~These statements are corroborated by the data in plots \ref{fig:configs_3g_2d}\,b) through \ref{fig:configs_3g_2d}\,d), and their validity does not depend on
the use of the sine adjustment for momenta (i.\,e.~the use of the sine modification makes no difference for such kinematics).~Concerning the sine improvement itself, the results in
\ref{fig:configs_3g_2d}\,b)$-$\,d) indicate that it always mitigates the discretisation errors, and for the momentum points considered in those graphs it eliminates the errors
completely.~One should \textit{not} conclude from this that the sine function can entirely remove the discretisation artifacts, for arbitrary kinematics.~We will show some results 
to this effect in the next section.    

We wish to note one final thing concerning the graphs in Figure \ref{fig:configs_3g_2d}\,b) through \ref{fig:configs_3g_2d}\,d).~A careful consideration of the plots reveals that
some of the data points are missing, namely those corresponding to a situation where $p=q$.~The reason for leaving those points out is that for such kinematics, the reconstructed
three-gluon vertex in Landau gauge identically vanishes.~We demonstrate this fact analytically in Appendix \ref{sec:collinear}.     

As a final example of generalised diagonal kinematics in 2D, we look at the situation defined by 
\begin{align}\label{eqn:quasy_symm}
p \, = \, (s,0) \, , \quad q \, = (0,s) \, , \quad r \, = -\,(s,s) \, ,
\end{align}  
\begin{figure}[!t]
\begin{center}
\graph[width = 0.43\tew]{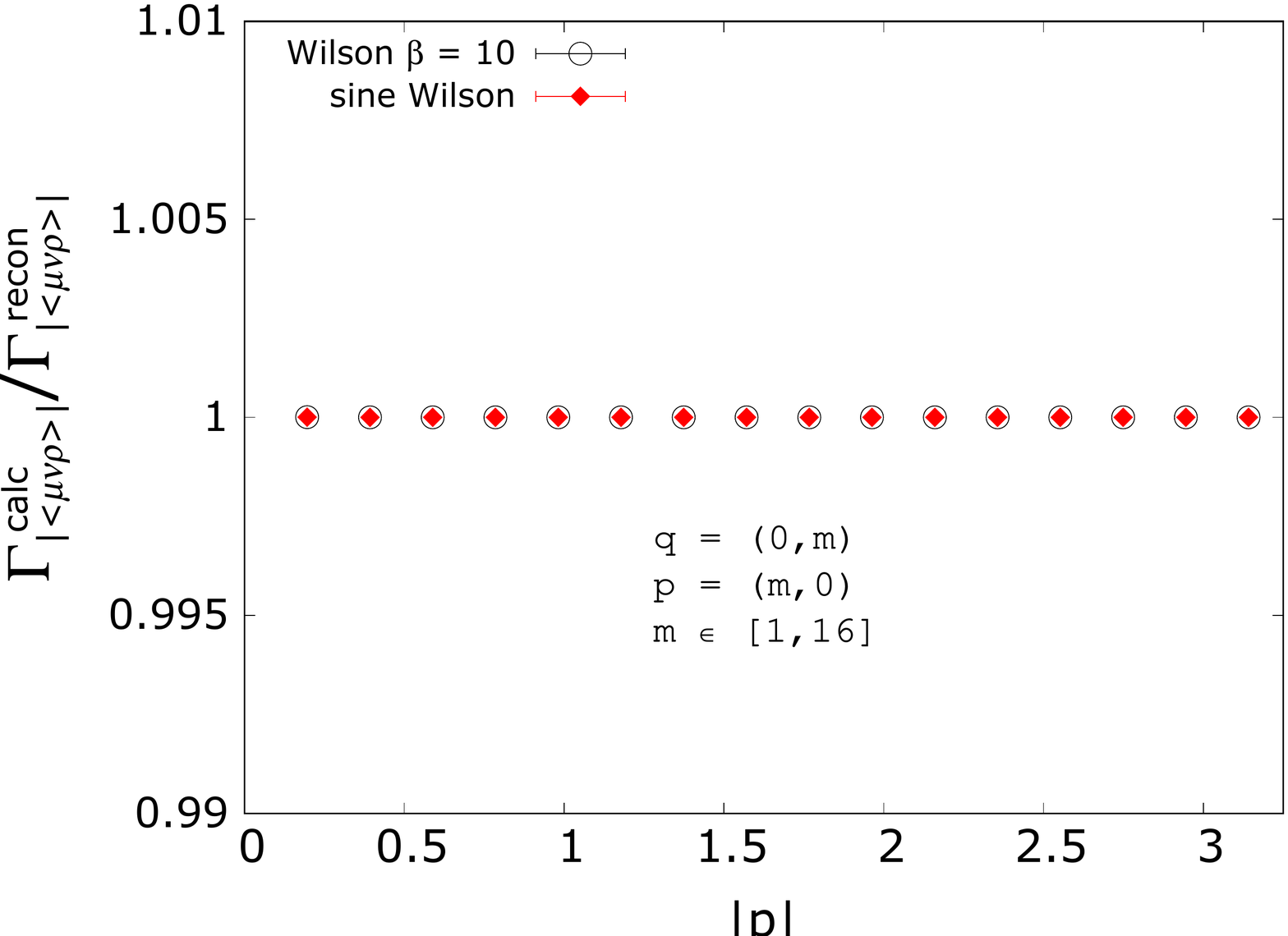}\\
\caption{Results of vertex reconstruction on a $32^2$ lattice, for a quasi-symmetric momentum configuration $(p^2,\,q ^2,\,r^2)$ = $(s^2,\,s^2,\,2s^2)$.~Reconstruction was
done with an ON basis tensor $\rho^{\,2}_{\mu\nu\sigma}$ of \eqref{eqn:on_basis}.~Results are in lattice units, with all momenta given in terms of components of vector
$n_\mu$ of \eqref{eqn:fourier_glue}.~$|\,p\,|$ stands for $\sqrt{\smash[b]p^2}$, and ``sine'' data refers to reconstruction with momenta $(\hat{p},\,\hat{q})$, where
e.\,g.~$\hat{p} = 2\sin(p/2)$.~See text for further discussion.} 
\label{fig:symmetric_2d}
\end{center}
\end{figure}
\noindent
\hspace{-0.13cm}where $s \equiv 2\pi \, n/(aN)$, with integer $n \in [1,N-1]$.~We refer to the above configuration as quasi-symmetric, since one has $(p^2,\,q^2,\,r^2)$ = $(s^2,
\,s^2,\,2s^2)$.~A fully symmetric case is not accessible on a lattice, in less than three dimensions.~Despite the lack of a full symmetry in terms of momentum invariants, the 
quasi-symmetric kinematic partitioning is interesting in its own right.~According to the analysis of Appendix \ref{sec: diagonal}, for kinematic situations like
\eqref{eqn:quasy_symm} it should be possible to describe the lattice three-gluon vertex exactly with continuum tensors alone, regardless of a particular value of $s$.~Also,
as mentioned before, the use of the sine adjustment should make no difference for these kinds of momentum configurations.~All these conclusions are confirmed by our results
in Figure \ref{fig:symmetric_2d}, which display an almost perfect agreement between the reconstructed and calculated vertex for all considered values of $s$. 
\begin{figure}[!b]
\begin{center}
\graph[width = 0.40\tew]{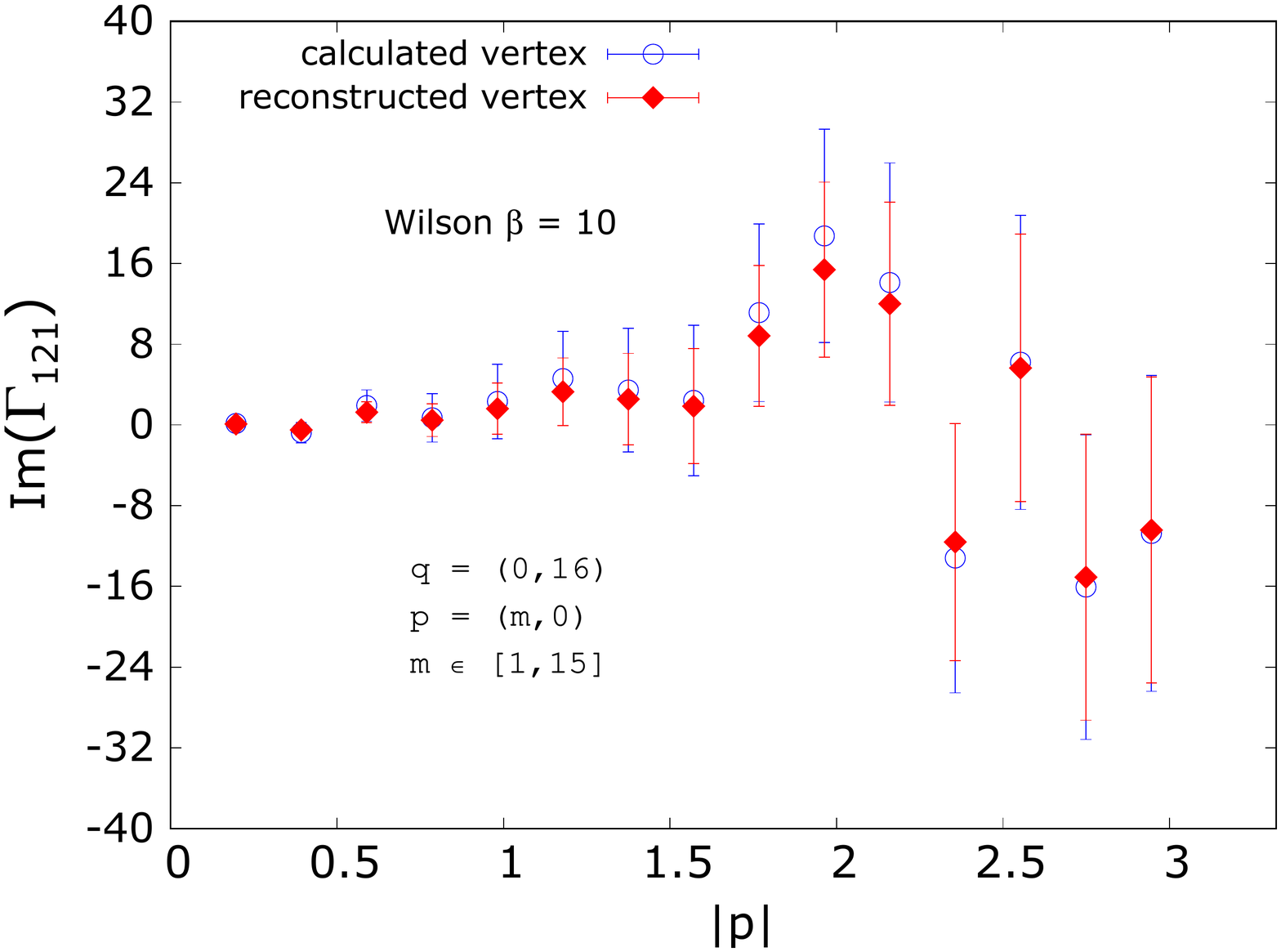}\graph[width = 0.40\tew]{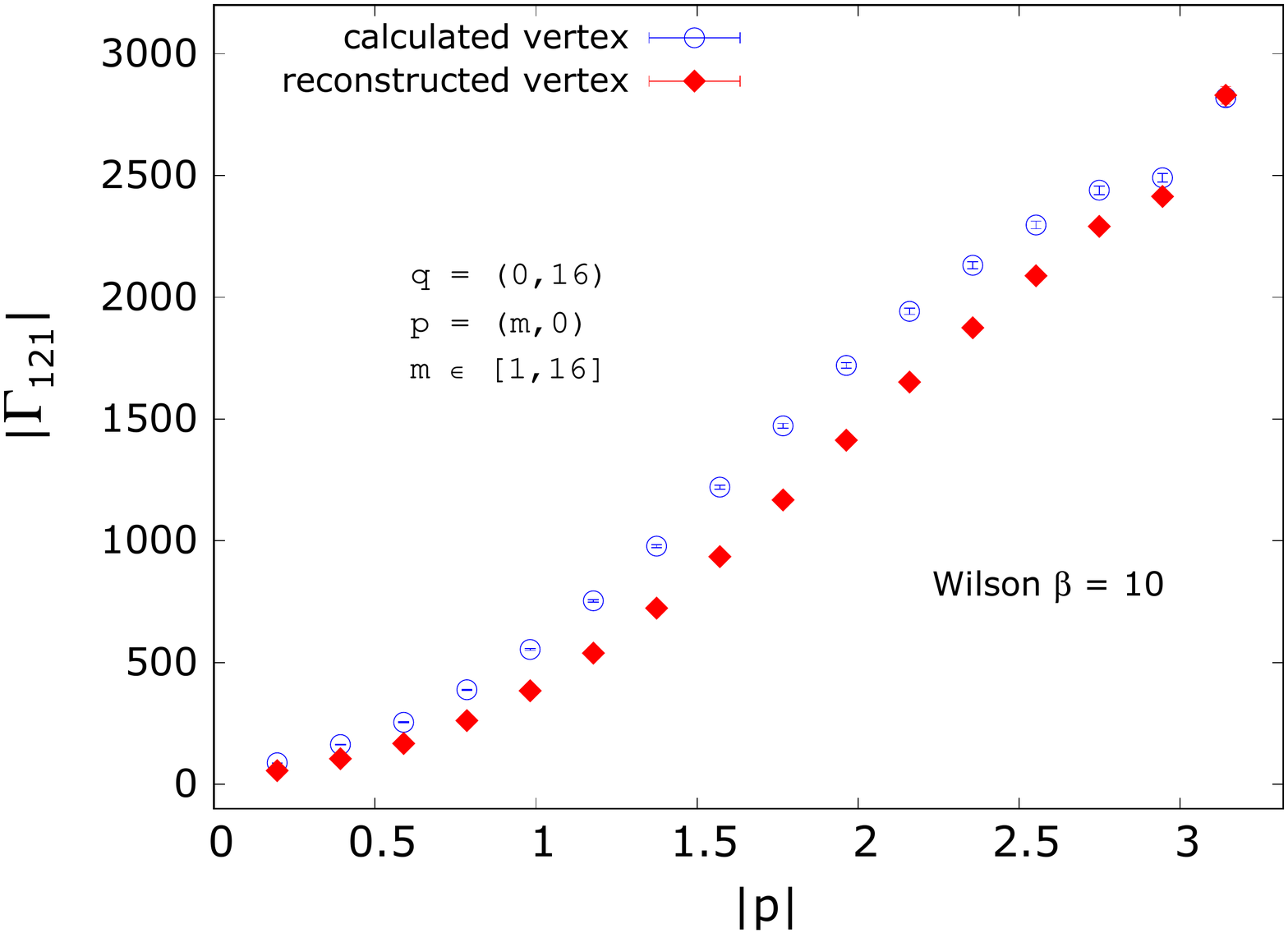}\\
\caption{\textit{Left:}~Monte Carlo (MC) average of an imaginary part of vertex component $\Gamma_{121}$, on a $32^2$ lattice.~\textit{Right:}~MC average of an 
absolute value of $\Gamma_{121}$.~Results are in lattice units, with momenta given in terms of vector $n_\mu$ of \eqref{eqn:fourier_glue}.} 
\label{fig:modulo_2d}
\end{center}
\end{figure} 
 
We would like to conclude this section by adressing two important issues.~The first is the apparent absence of error bars in most of our plots.~The uncertainties are present in
all the graphs, but are in most cases too small to be noticed on the overall plot scales.~This comes from the use of absolute values in calculations of quantities like \er{eqn:
gamma_aver}:~the absolute value makes all the contributions to Monte Carlo averages strictly non-negative, leading to very good signal quality.~We demonstrate this explicitly in
Figure \ref{fig:modulo_2d}, wherein we compare the data for the vertex component $\Gamma_{121}$, obtained in computations both with and without the $|.|$ modification.~The signal
is far more noisy in the case without the absolute value [\,left panel of Fig.\,\ref{fig:modulo_2d}\,], but both sets of data points lead to the same conclusions regarding the 
deviations between the calculated and reconstructed vertex component(s).~Thus, even though the absolute value average has a big influence on the final results of our simulations,
it does not distort the overall analysis, in terms of identification of special kinematic points on the lattice. 
     
\begin{figure}[!t]
\begin{center}
\graph[width = 0.42\tew]{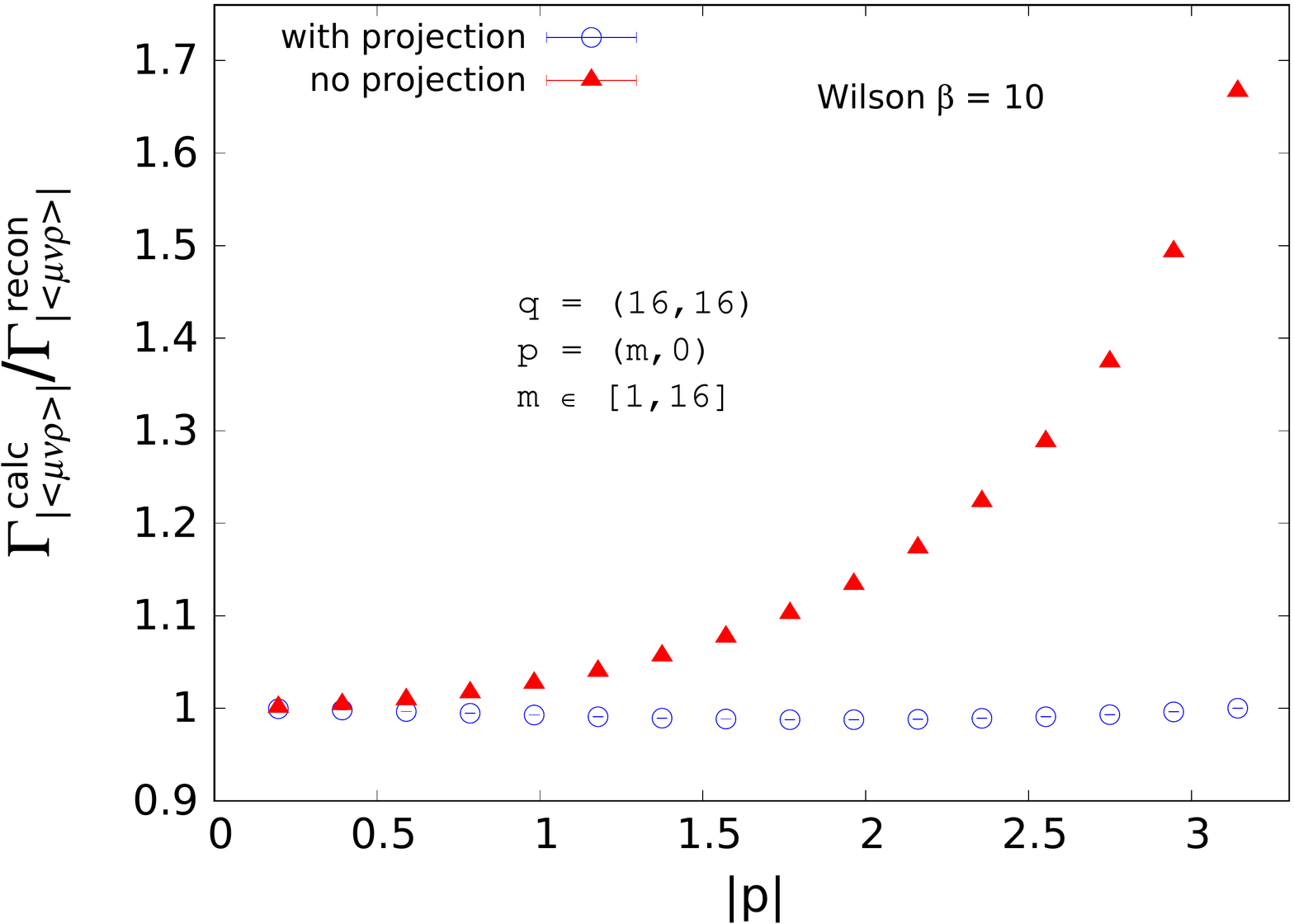}\\
\caption{Test of the influence of transverse projection \er{eqn:add_trans} on vertex reconstruction results, for particular kinematics on a 32$^2$ lattice.~For reconstruction,
we use the ON basis tensor $\rho^{\,2}_{\mu\nu\sigma}$ of \eqref{eqn:on_basis}.~Results are in lattice units, with all momenta given in terms of components of vector
$n_\mu$ of \eqref{eqn:fourier_glue}} 
\label{fig:transver_2d}
\end{center}
\end{figure}

The second issue which we would like to comment on here concerns the continuum-like transverse projection of \er{eqn:add_trans}.~We use this additional operation because the lattice
gauge-fixing condition may not be quantitatively good enough, for certain kinematics.~To clarify, with the convergence criterion \er{eqn:conv_crit} alone, the number `0' on the 
r.\,h.\,s.~of \er{eqn:landau_mom} is, in actual simulations, a small quantity on the order of $10^{-6} \,\, \text{to} \,\, 10^{-5}$.~For most purposes, this is certainly 
``transverse enough'', but in some cases it is desirable to perform also the projection \er{eqn:add_trans}, which makes the gluon field satisfy the transversality condition 
\er{eqn:landau_mom} within numerical precision, i.\,e.~the number `0' on the r.\,h.\,s.~of \er{eqn:landau_mom} becomes a quantity on the order of  $10^{-16}$.~To illustrate the impact
this may have, in Figure \ref{fig:transver_2d} we compare the reconstruction results with and without the additional projection \er{eqn:add_trans}, for a particular kinematic choice.~According to arguments of Appendix \ref{sec: diagonal}, the first and the last momentum points in Figure \ref{fig:transver_2d} correspond to special
kinematic configurations, and the vertex ratio should be close to unity in both cases.~However, the expected behaviour is seen only when the projection \er{eqn:add_trans} is applied,
whereas in its absence the ratio goes to values far from unity, for one of the momentum points.~This means that the reconstruction procedure itself is very sensitive to the numerical
accuracy at which the gluon transversality criterion is fulfilled.~Fortunately, the projection \er{eqn:add_trans} is cheap to implement numerically, and it should not introduce any
conceptual issues as it merely makes the condition \er{eqn:landau_mom} hold with better accuracy.  

\subsection{Vertex results in three dimensions}\label{sec: 3d_results}
%
\begin{figure}[!t]
\begin{center}
\graph[width = 0.40\tew]{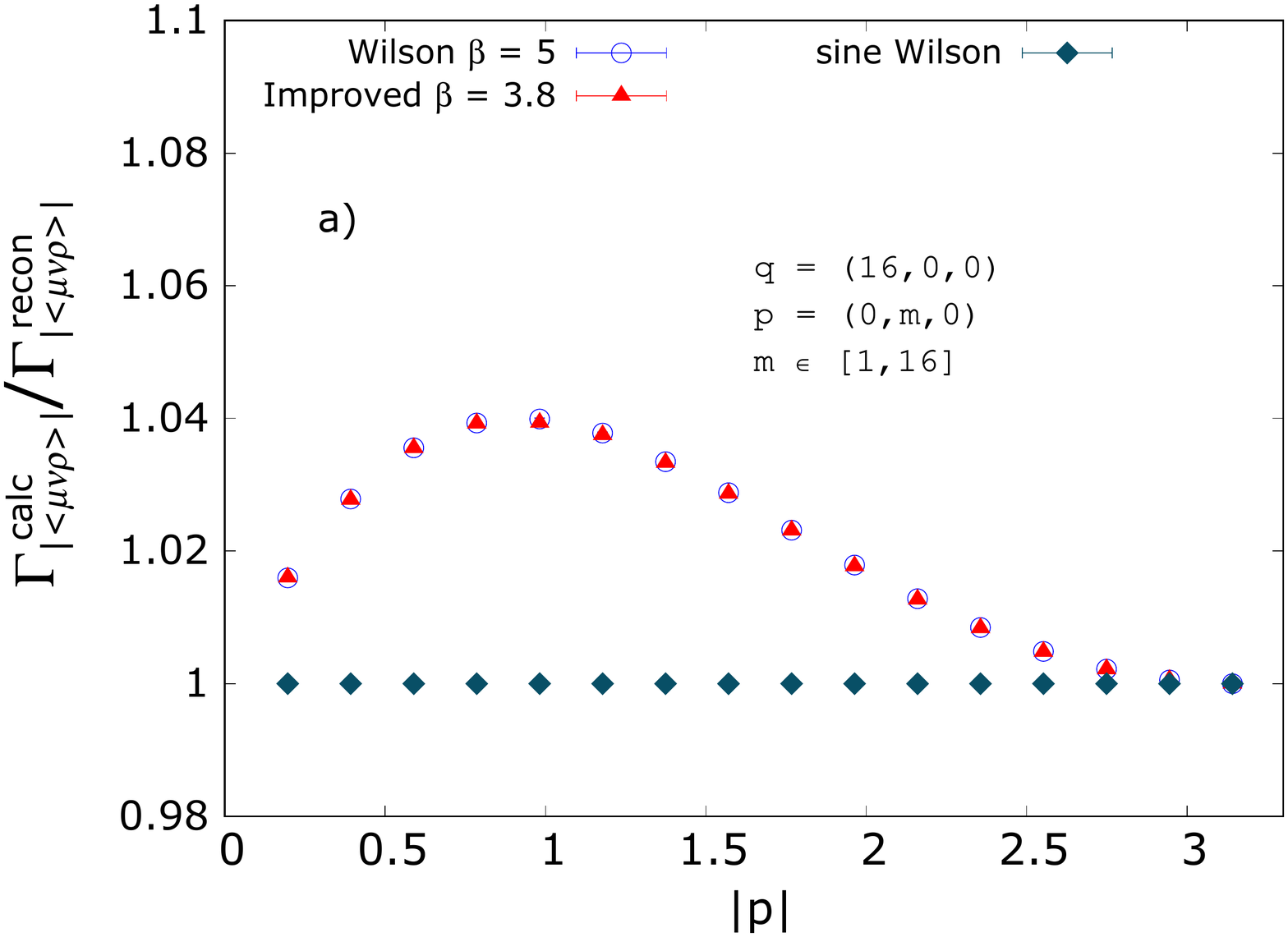}\graph[width = 0.40\tew]{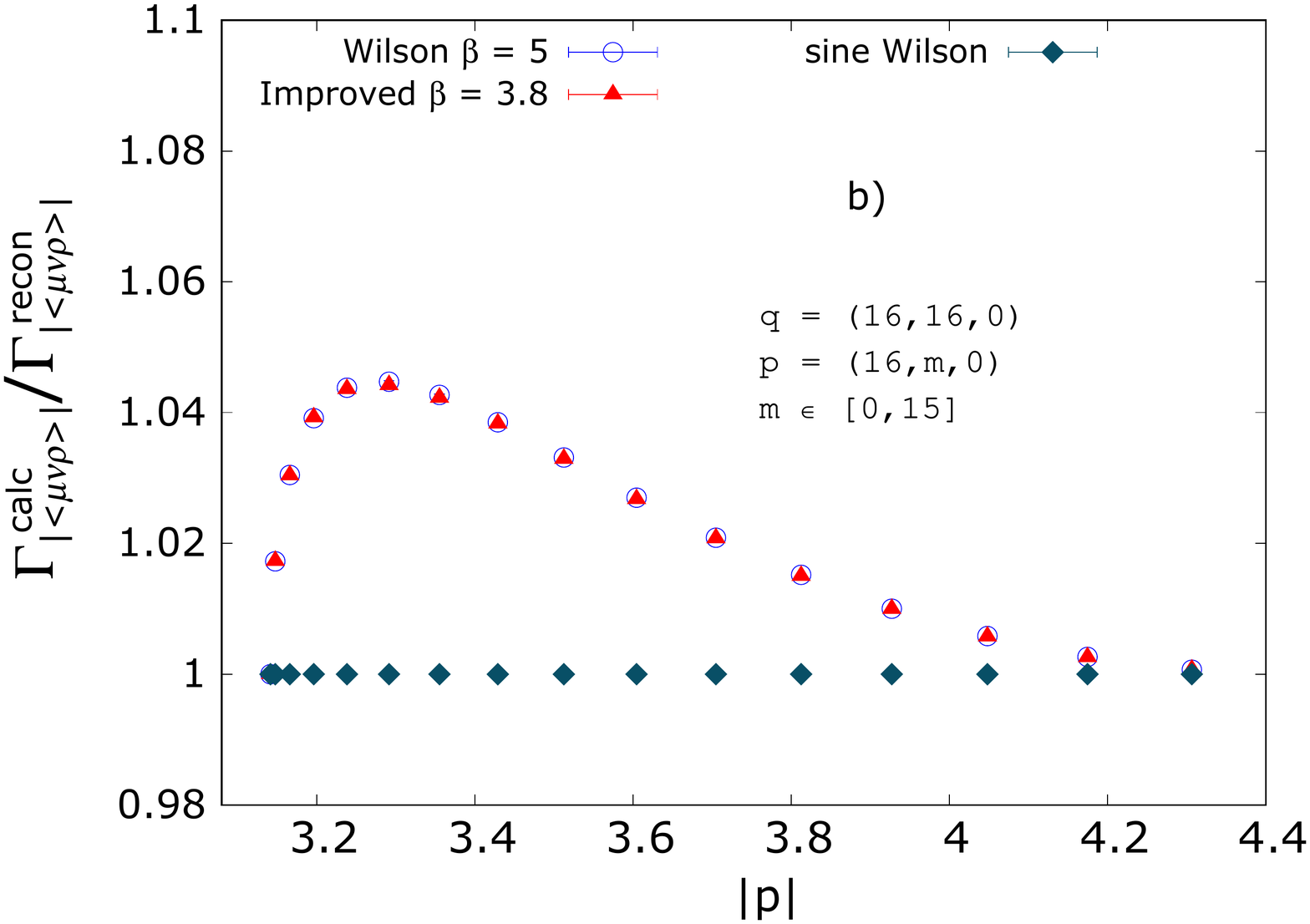}
\graph[width = 0.40\tew]{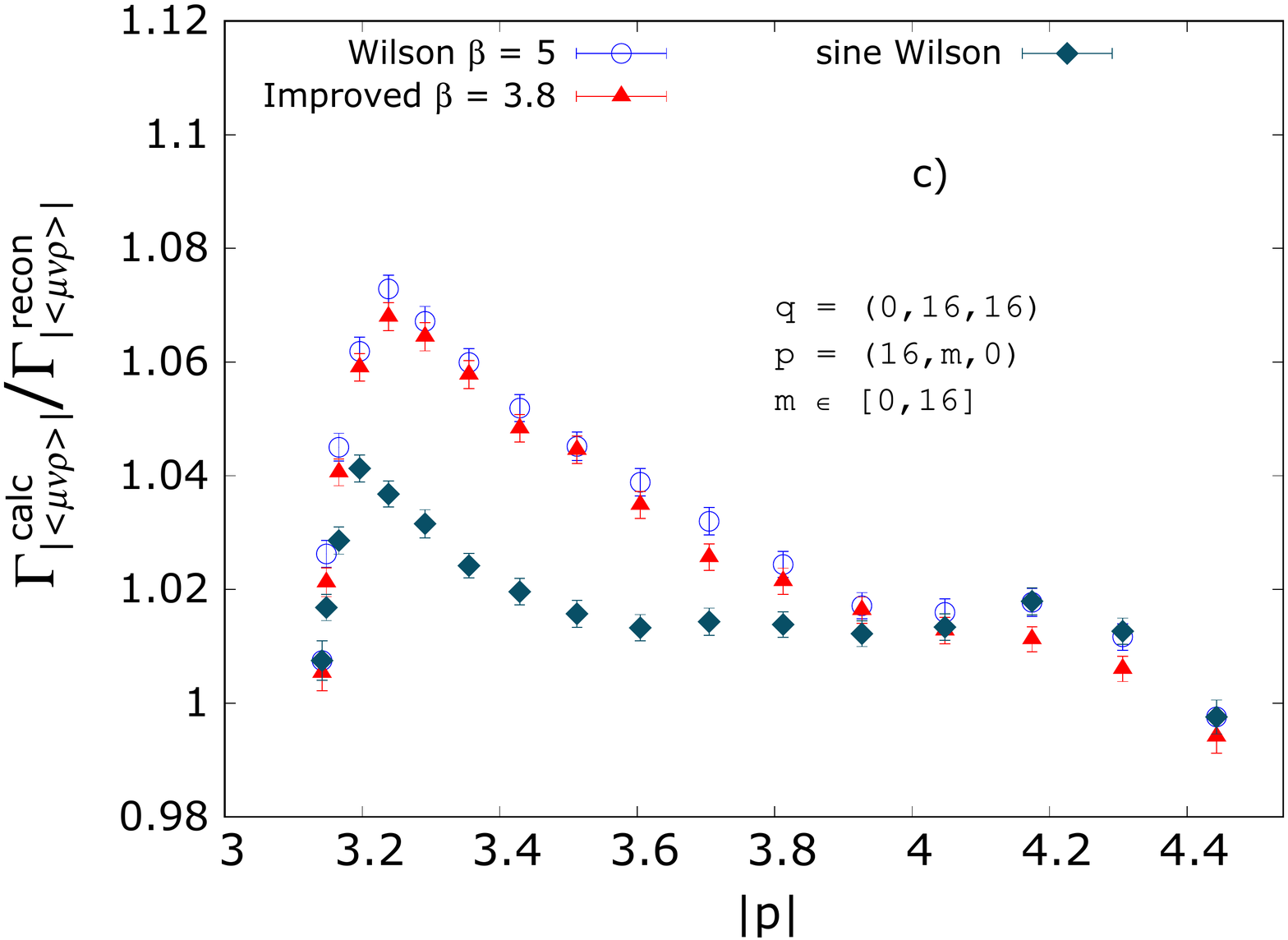}\graph[width = 0.40\tew]{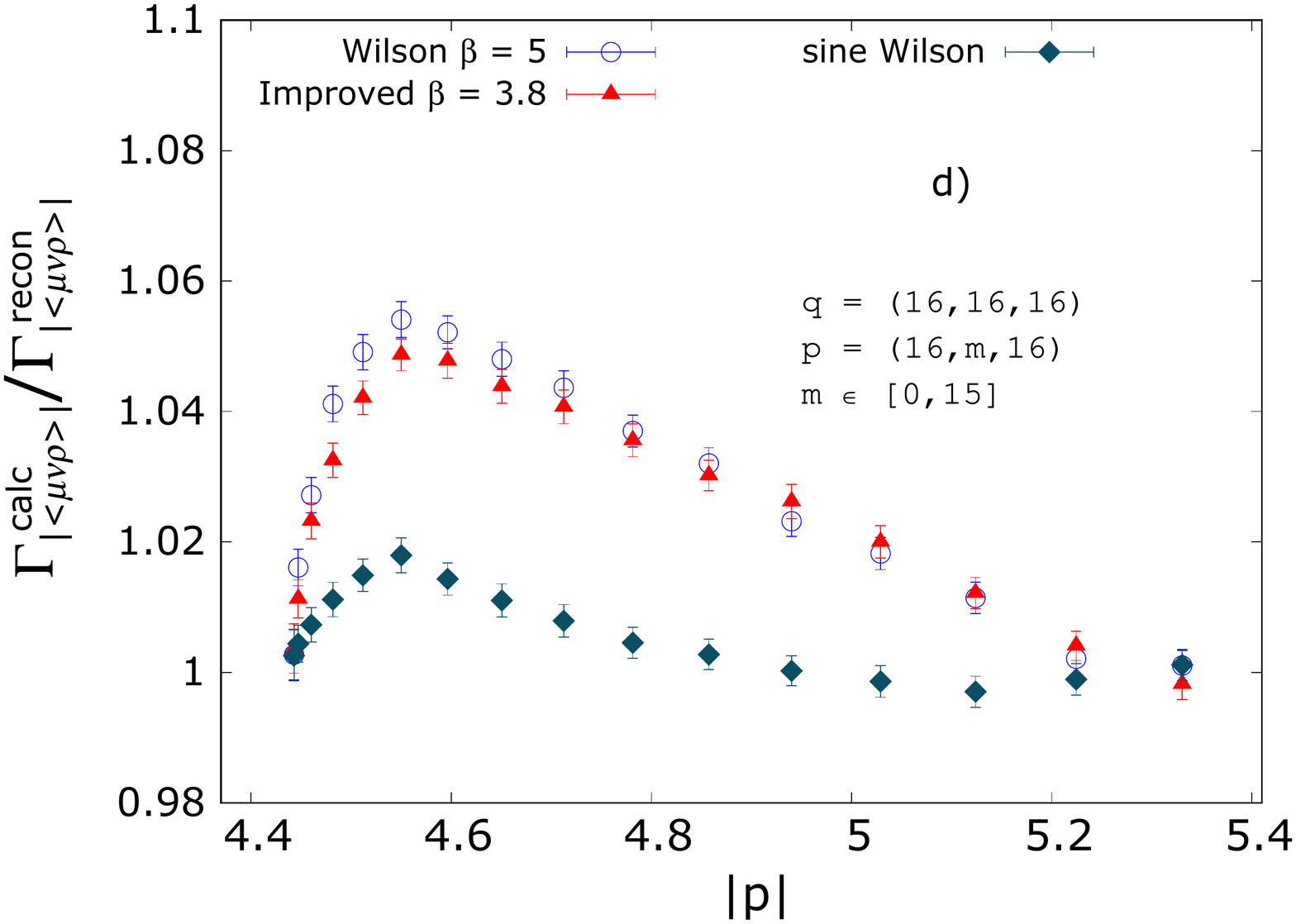}\\
\caption{Vertex ratios for certain kinematics a 32$^3$ lattice, as functions of $|\, p\,| = \sqrt{\smash[b]p^2}$.~Reconstruction was done with ON tensor elements
of \eqref{eqn:on_basis}.~Results are in lattice units, with momenta given in terms of vector $n_\mu$ of \eqref{eqn:fourier_glue}.~``Sine'' data refers to reconstruction with 
momenta $(\hat{p},\,\hat{q})$, where e.\,g.~$\hat{p} = 2\sin(p/2)$.~See text for further discussion.} 
\label{fig: configs_3d}
\end{center}
\end{figure}  
 
In Figure \ref{fig: configs_3d} we present the results of our simulations for certain three-dimensional kinematic configurations.~The data in all the graphs essentially confirm
our main conclusions regarding the special kinematic configurations on the lattice, as presented in Appendix \ref{sec: diagonal}.~Even so, there seem to be mild differences in 
signal quality between the first two and last two plots in the Figure.~For instance, the points in Fig.~\ref{fig: configs_3d}\,a) and b) feature an almost perfect agreement between 
the standard Wilson and improved gauge actions (without sine adjustment), which is absent in graphs c) and d).~Also, apart from a clearly pronounced maximum in vertex ratio results, 
the data in the upper panel of the Figure features no additional ``bumps'', which is not true for the plots in the lower panel, see e.\,g.~lower right side of Fig.~\ref{fig:
configs_3d}\,c). 

We are not entirely sure where the aforementioned minor differences in signal quality come from, but we think that they have to do with the examined kinematics, and the number of  
relevant tensor elements in vertex reconstruction.~As discussed in more detail at the end of section \ref{sec:ortho_basis}, the momenta considered in the first two plots of Figure
\ref{fig: configs_3d} are such that the ON basis element $\rho^{\,2}_{\mu\nu\sigma}$ (the only non-vanishing one in two dimensions) dominates over all the other ON tensor structures
of equation \eqref{eqn:on_basis}.~On the other hand, kinematics in the lower two graphs of the Figure do not lead to a single dominating basis element, meaning that all four tensors 
have equal importance in the reconstruction process.~This last fact can arguably result in a slight increase in fluctuations in the signal, compared to a situation with one significant 
tensor, since calculations feature a greater number of relevant ``moving parts'' which all contribute to the final outcome.~Whatever the reason for a mildly more noisy signal in graphs
\ref{fig: configs_3d}\,c) and d), these results still agree with our analysis of \ref{sec: diagonal}, within statistical uncertainties, and we will thus not comment on them further.    

Besides considerations of the lattice three-gluon vertex of Monte Carlo simulations, it would be good to have an alternative way to check some of the arguments made in \ref{sec: diagonal},
preferably without any statistical noise whatsoever.~One way to do this would be to apply the vertex reconstruction procedure to the tree-level three-gluon vertex from lattice perturbation
theory.~Being defined on a discretised spacetime, this object should suffer from the same rotational symmetry breaking effects as the three-gluon vertex of Monte Carlo calculations.~But
unlike the Monte Carlo correlator, the perturbative lattice vertex is inherently noise-free, and by subjecting it to vertex reconstruction one can solidify some of the claims concerning 
special kinematics on the lattice.~In Landau gauge, the perturbative lattice three-gluon vertex is \cite{Rothe:1992nt} 
\begin{align}\label{eqn: lattice_vertex}
&\Gamma^{\,\text{latt,\,Landau}}_{\mu\nu\rho}(p,q,r) \, = \, T_{\alpha\mu}^{\,p,\,l} \, T_{\beta\nu}^{\,q,\,l} \, T_{\gamma\rho}^{\,r,\,l} \cdot 
\Gamma^{\,\text{latt}}_{\alpha\beta\gamma}(p,q,r), \,\, \text{where} \nonumber \\[0.13cm]
& \Gamma^{\,\text{latt}}_{\alpha\beta\gamma}(p, q, r) \, = \, \delta_{\alpha\beta}\sin\left(\frac{p_\gamma - q_\gamma}{2}\right)\cos\left(\frac{r_\alpha}{2}\right)
\, + \, \delta_{\beta\gamma}\sin\left(\frac{q_\alpha - r_\alpha}{2}\right)\cos\left(\frac{p_\beta}{2}\right) \nonumber \\[0.1cm]
& \qquad \qquad \quad \,\,\, + \, \delta_{\gamma\alpha}\sin\left(\frac{r_\beta - p_\beta}{2}\right)\cos\left(\frac{q_\gamma}{2}\right) \,\, .
\end{align}  

\begin{figure}[!t]
\begin{center}
\graph[width = 0.39\tew]{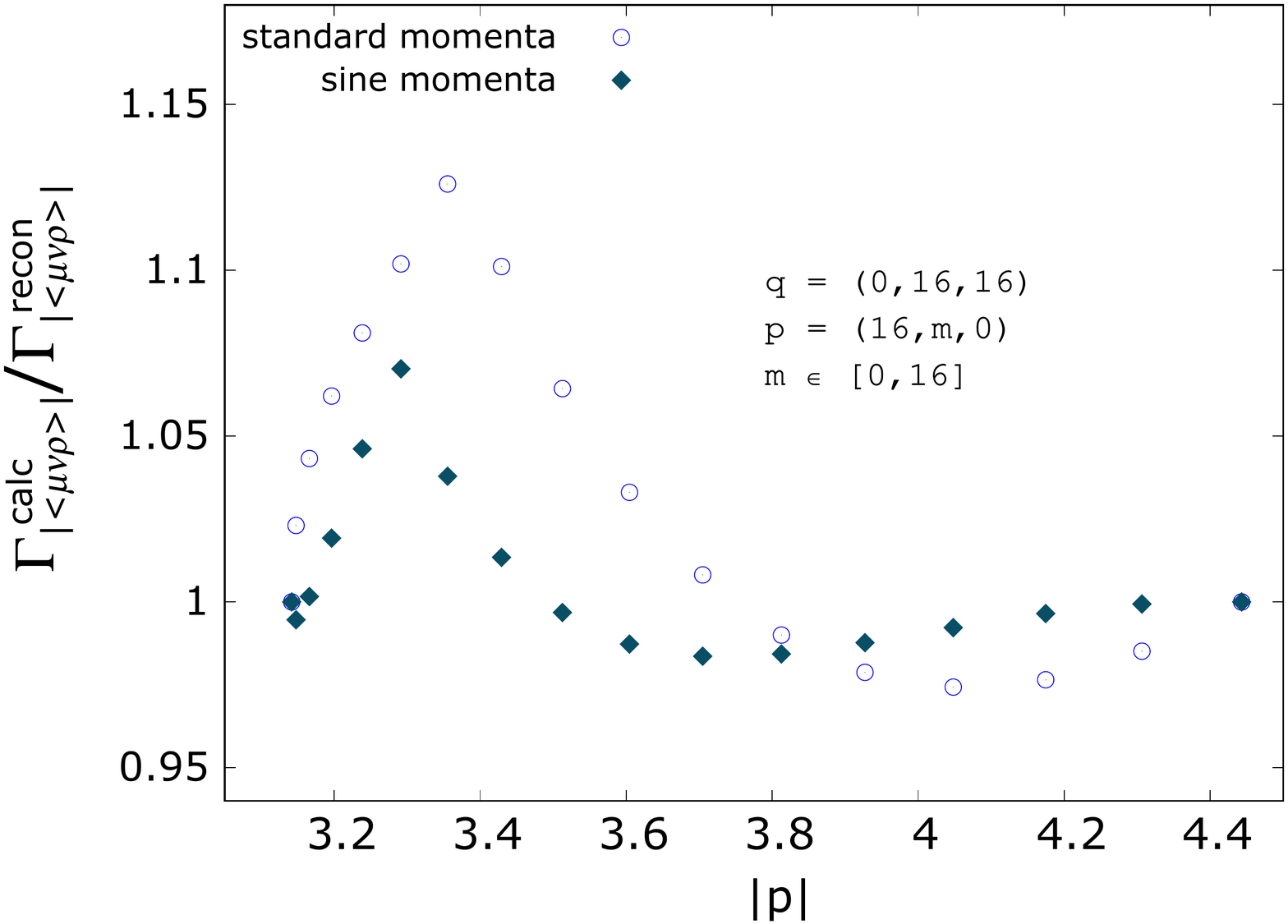}\graph[width = 0.39\tew]{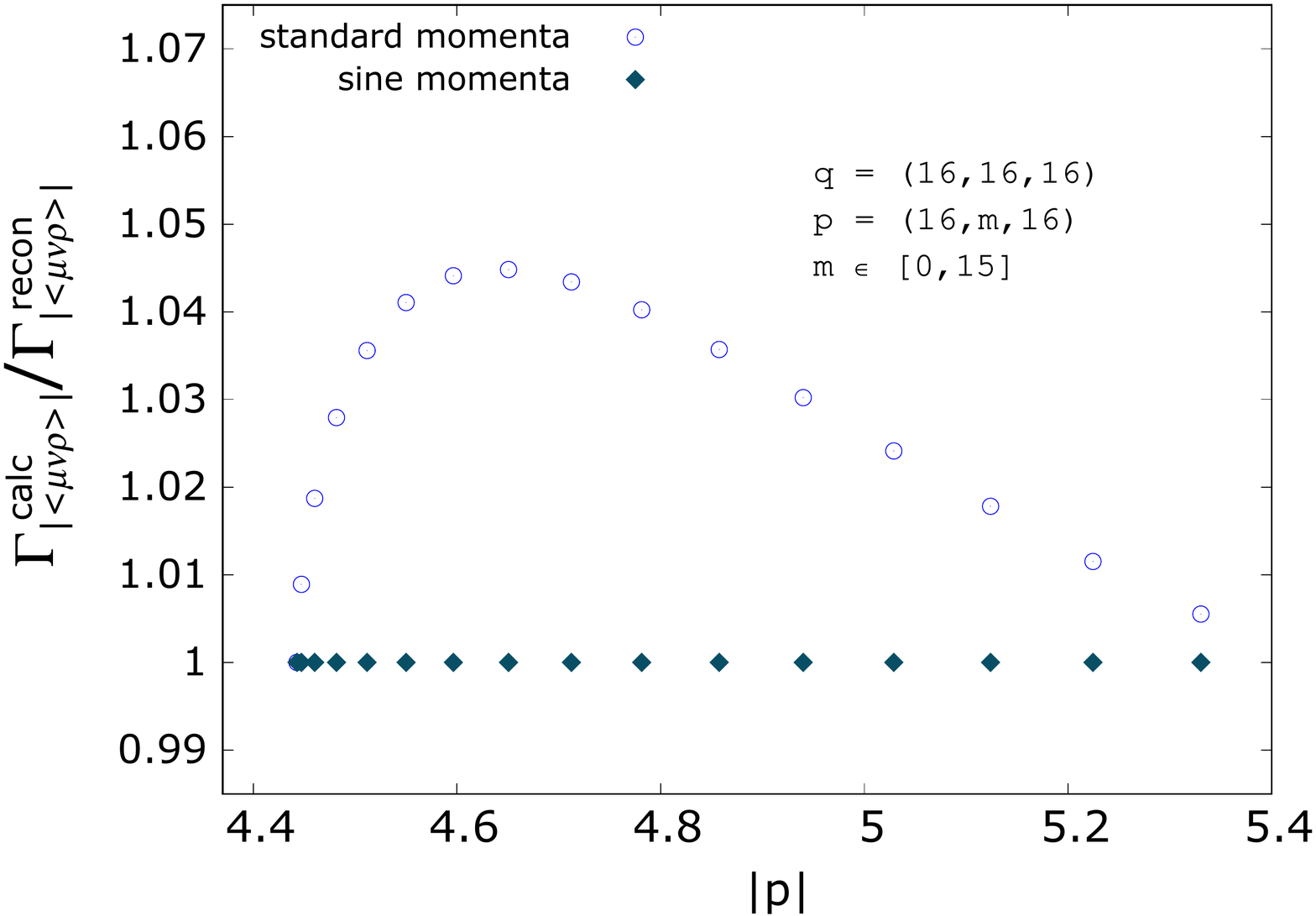}\\
\caption{Reconstruction results for the perturbative lattice vertex of \eqref{eqn: lattice_vertex}, as functions  of $|\, p\,| = \sqrt{\smash[b]p^2}$.~Reconstruction was done
with ON tensor elements of \eqref{eqn:on_basis}.~Results are in lattice units, with momenta given in terms of vector $n_\mu$ of \eqref{eqn:fourier_glue}.~``Sine'' data refers
to reconstruction with momenta $(\hat{p},\,\hat{q})$, where e.\,g.~$\hat{p} = 2\sin(p/2)$.} 
\label{fig:perturb_configs}
\end{center}
\end{figure}

In the above expression, we have ignored all of the multiplicative factors like the colour constants and similar, as they do not affect the forthcoming reconstruction results.~With $T_{
\alpha\mu}^{\,p,\,l}$ we denote the lattice-adjusted transverse projectors, introduced in \er{eqn:add_trans}.~In Figure \ref{fig:perturb_configs} we give the reconstruction results for 
the perturbative lattice vertex, for the same kinematics as examined in the lower panel of Figure \ref{fig: configs_3d}.~Concerning the special status of certain lattice kinematic 
configurations, the data of both Figures \ref{fig: configs_3d} and \ref{fig:perturb_configs} agree with the general arguments of Appendix \ref{sec: diagonal}.~From results in these
Figures one can also see that in three dimensions the sine modification does not always eliminate the discretisation errors completely.~Thus, one cannot rely solely on this adjustment,
when attempting to eradicate the errors due to rotational symmetry breaking in vertex tensor elements.      

\newpage

This brings us to the final two examples of special kinematics in this paper.~Both can be seen as a kind of a three-dimensional extension of the two-dimensional 
case given in \eqref{eqn:quasy_symm}.~These configurations are  
\begin{align}\label{eqn:3d_symm}
 i) \quad &p \, = \, (s, \, 0, \,s) \, , \quad q \, = (0, \, s, \, 0) \, , \quad  r \, = \, -(s,\,s,\,s) \, , \nonumber \\ 
 ii) \quad &p \, = \, (-s, \, 0, \,s) \, , \quad q \, = (s, \, -s, \, 0) \, , \quad  r \, = \, (0,\,s,\,-s) \, ,
\end{align}  

\hspace{-0.34cm}where $s \equiv 2\pi \, n/(aN)$, and integer $n$ takes on values $n \in [1,N-1]$.~The lower combination in \eqref{eqn:3d_symm} corresponds to a symmetric situation,
with momentum invariants $(p^2,\,q ^2,\,r^2)$ = $(2s^2,\,2s^2,\,2s^2)$.~The symmetric configuration is often considered in lattice and continuum studies of the three-gluon correlator,
see e.\,g.~\cite{Boucaud:1998bq,Boucaud:2013jwa,Athenodorou:2016oyh,Cucchieri:2006tf,Eichmann:2014xya,Blum:2014gna}.
\begin{figure}[!h]
\begin{center}
\graph[width = 0.39\tew]{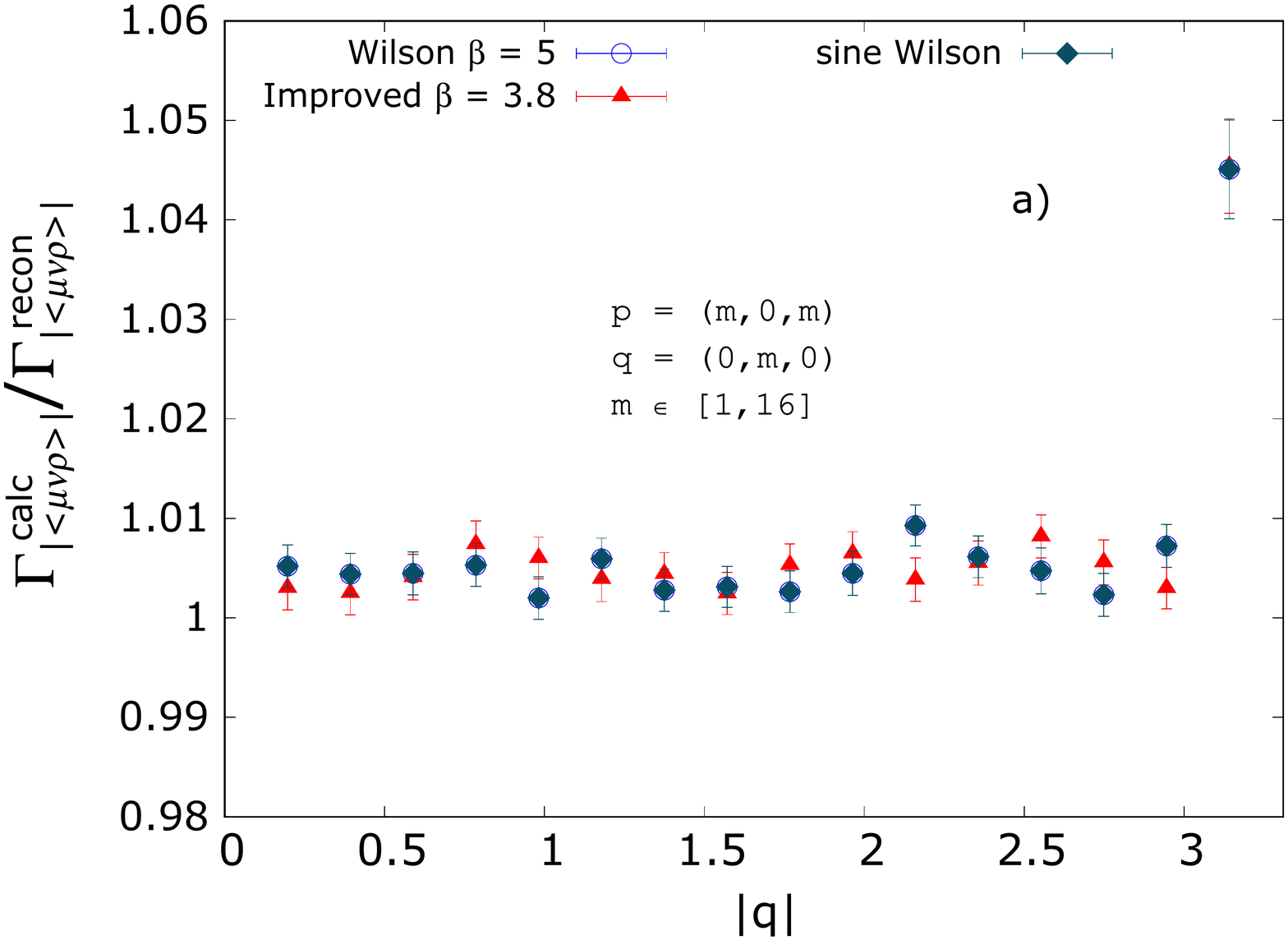}\graph[width = 0.39\tew]{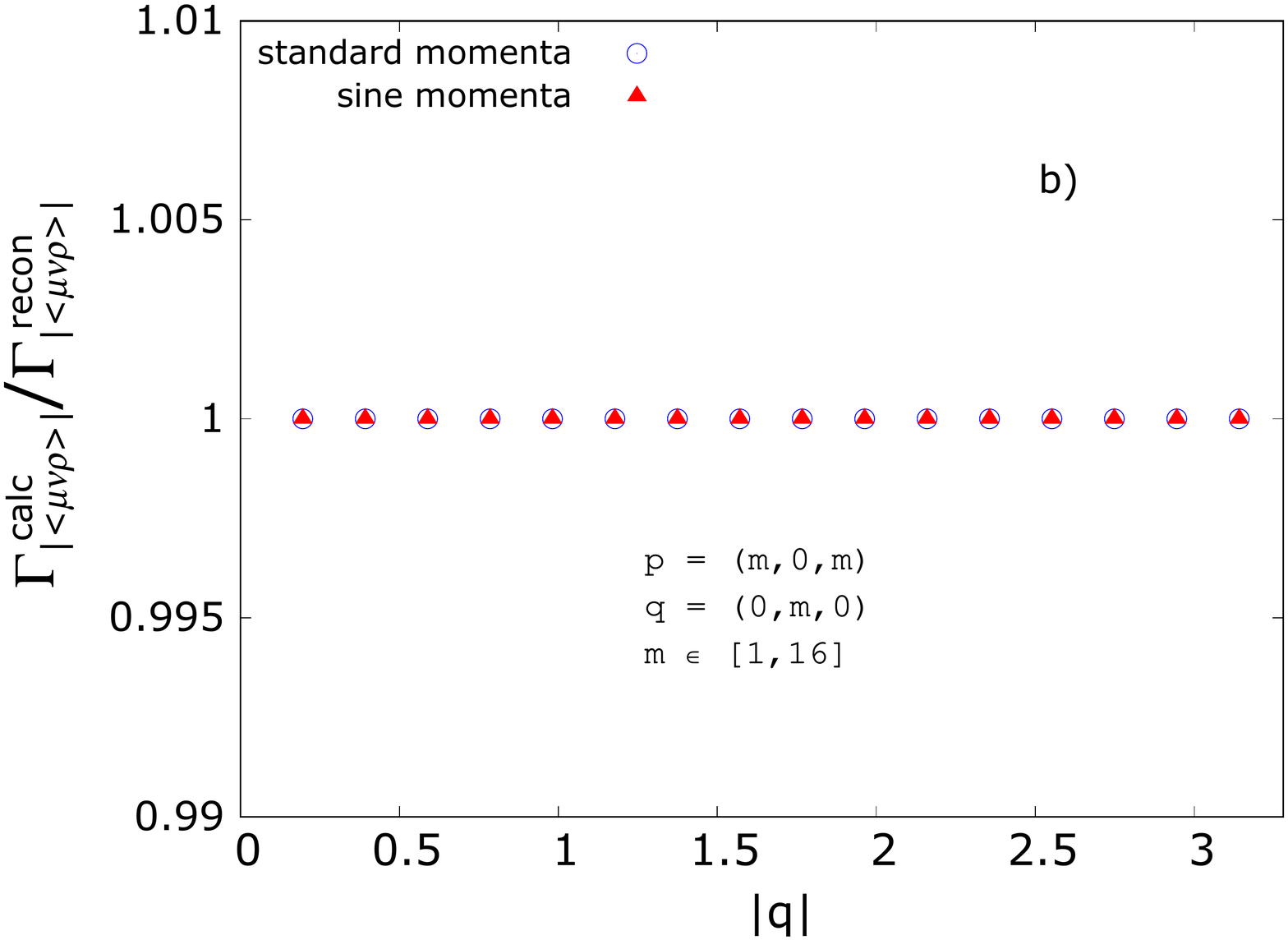}
\graph[width = 0.39\tew]{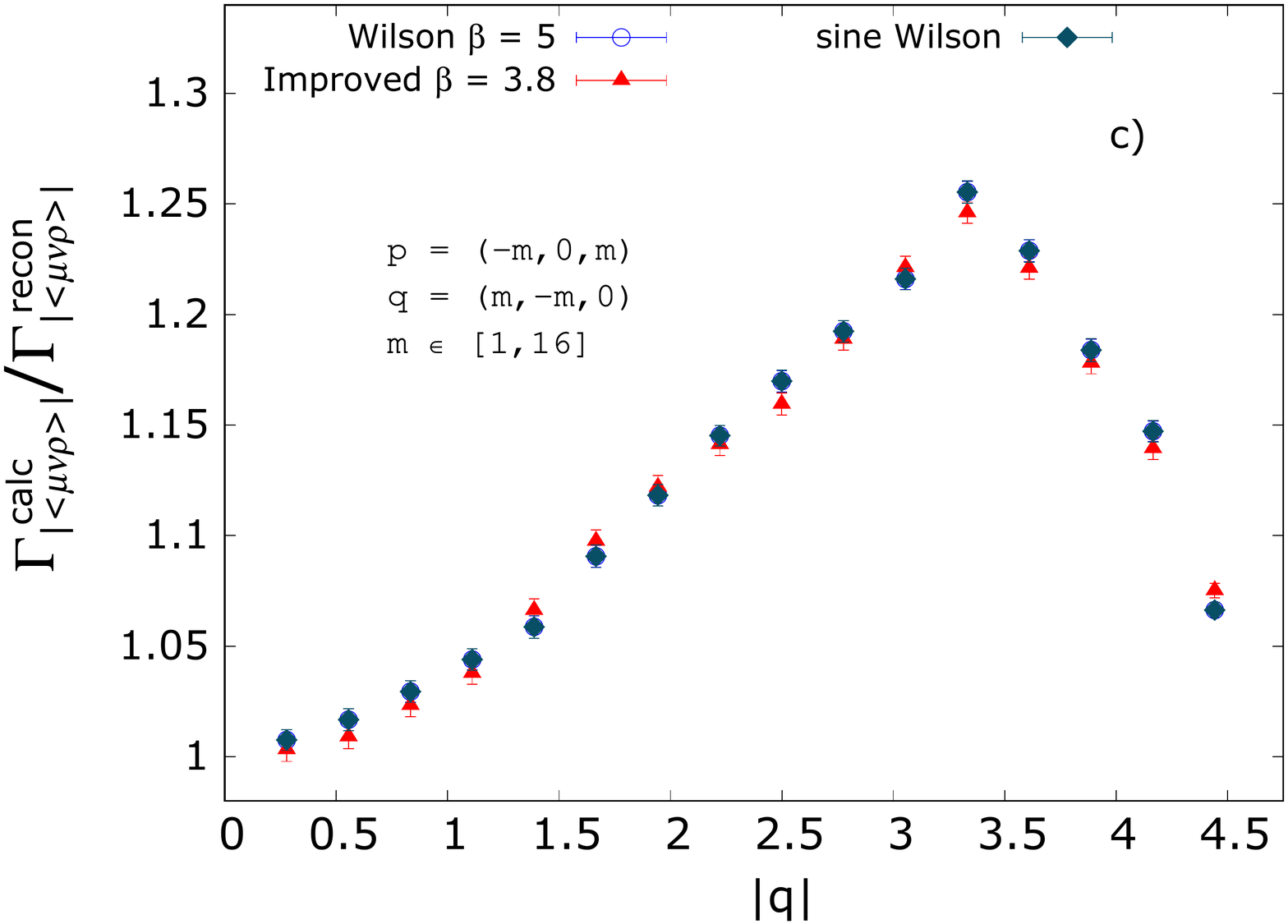}\graph[width = 0.39\tew]{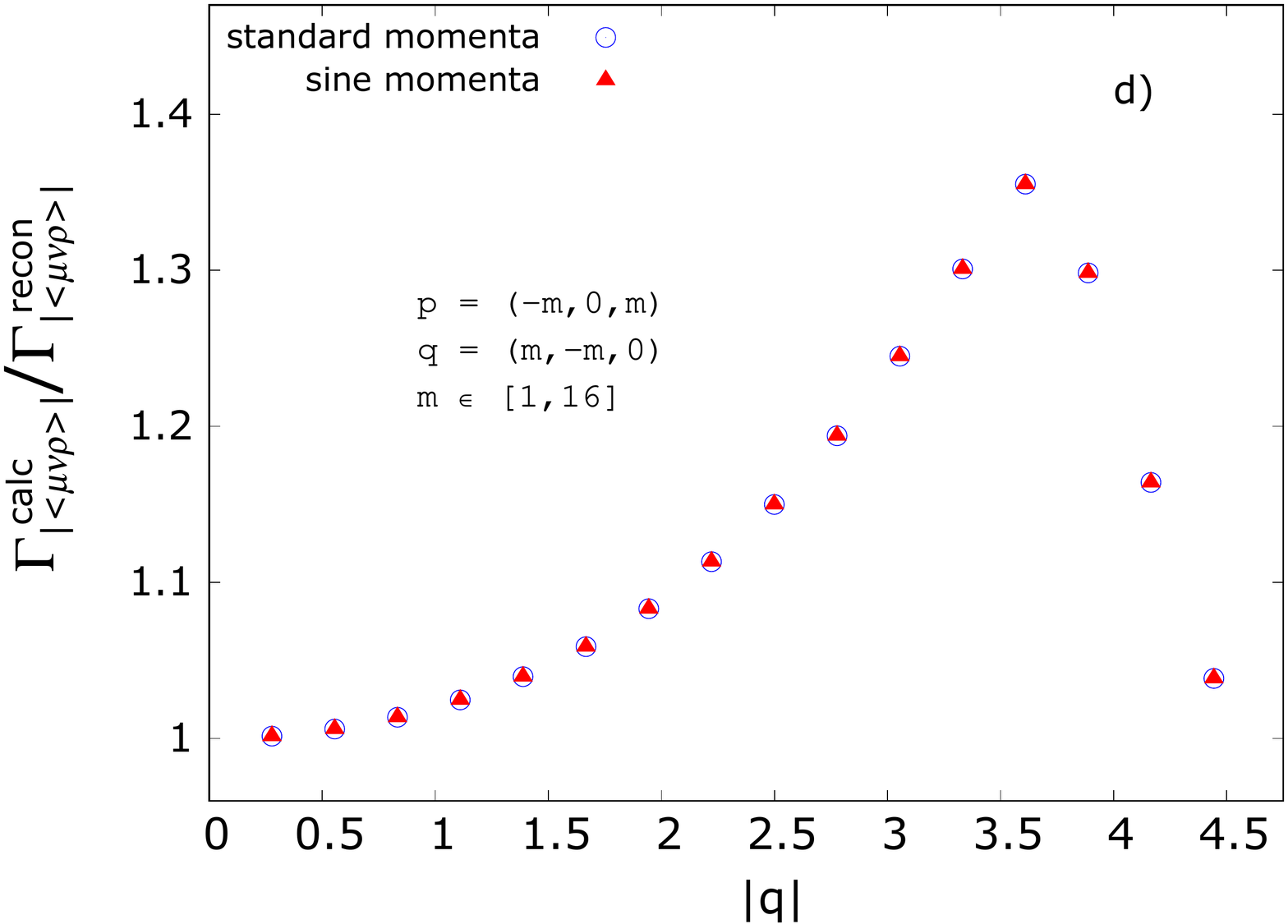}\\
\caption{Vertex ratios for kinematics of \eqref{eqn:3d_symm} on a 32$^3$ lattice, as functions of $|\, q\,| = \sqrt{\smash[b]q^2}$.~Reconstruction was done with ON tensor basis of
equation \eqref{eqn:on_basis}.~``Sine'' data refers to reconstruction with momenta $(\hat{p},\,\hat{q})$, where e.\,g.~$\hat{p} = 2\sin(p/2)$.~Plots a) and c) correspond to a vertex 
of Monte Carlo  calculations, b) and d) correspond to a perturbative lattice vertex \eqref{eqn: lattice_vertex}.} 
\label{fig:3d_symmetric}
\end{center}
\end{figure} 

~We show our reconstruction results for the kinematics of \eqref{eqn:3d_symm} in Figure \ref{fig:3d_symmetric}, for both the lattice Monte Carlo vertex, and the perturbative
one of \eqref{eqn: lattice_vertex}.~Data points in Figure \ref{fig:3d_symmetric} clearly indicate that the two kinematic cases in \eqref{eqn:3d_symm} are not equivalent, when 
it comes to vertex tensor representations.~For the upper configuration in \eqref{eqn:3d_symm} (the non-symmetric one), the continuum tensor elements seem to work rather well 
for all considered values of $s$.~This does not hold for a fully symmetric momentum partitioning, where differences between reconstructed and calculated vertex go up to around
30 to 40 percent, a clearly significant deviation.~Note that the symmetric case still has the property, shared with truly special kinematics, that the sine momentum adjustment
does not change the reconstruction results.~But this fact alone does not guarantee a continuum-like tensor description, as discussed in some detail in Appendix \ref{sec:
diagonal}.~We note that, for the proofs carried out in \ref{sec: diagonal}, the perturbative lattice vertex \er{eqn: lattice_vertex} plays a crucial role, since for that function
(in contrast to the Monte Carlo vertex) it is possible to show analytically why some kinematic configurations are special, i.\,e.~why the non-linear terms present in \er{eqn:
lattice_vertex} reduce to a sum of continuum tensor structures, for certain momentum points. 

Here we wish to briefly discuss the applicability of some of these ideas to other correlation functions of lattice QCD.~Most of the essential arguments presented in \ref{sec:
diagonal} should hold when working with vertex functions consisting purely of gluons (e.\,g.~a four-gluon correlator), or of ghost and gluon fields.~The situation becomes a bit
more subtle for vertices where quark and gluon degress of freedom are combined, since there is no \textit{a priori} reason that the tensor structures corresponding to these 
different kinds of fields should get modified in the same way, when going from continuum to discretised spacetimes.~Even so, for any given lattice action and any correlator of
interest (for instance, the quark-gluon vertex), it should be possible to carry out the same steps as in Appendix \ref{sec: diagonal}, which includes taking the corresponding
vertex from lattice perturbation theory, and checking if the said vertex reduces to a sum of continuum tensor elements, for certain choices of momenta.~It is our personal opinion
that the kinematics akin to those in the first line of \er{eqn:3d_symm} should ``work'' in this regard, for any lattice correlators, but we leave explicit demonstrations of this
for the future. 

\begin{figure}[!b]
\begin{center}
\graph[width = 0.40\tew]{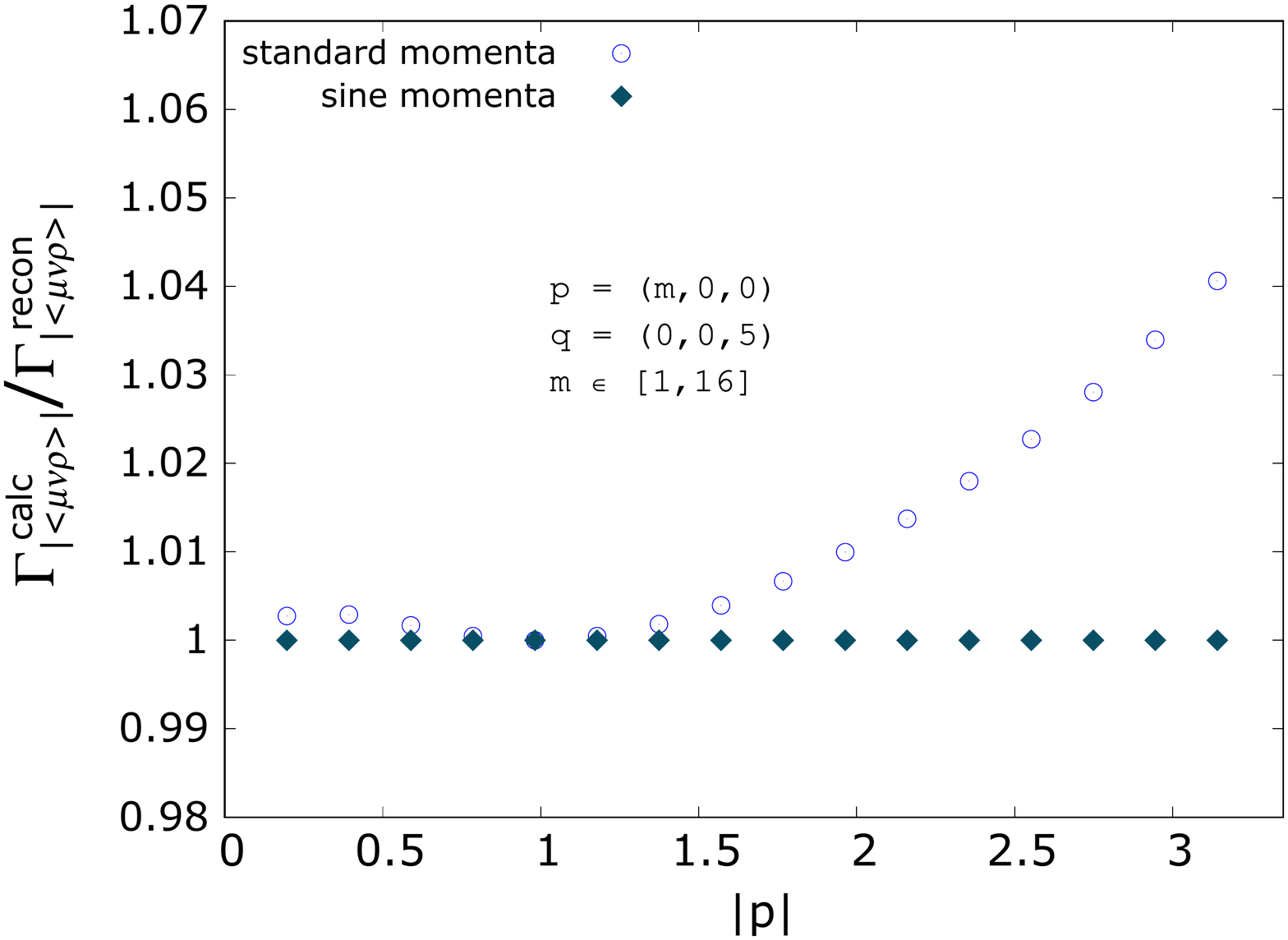}\graph[width = 0.40\tew]{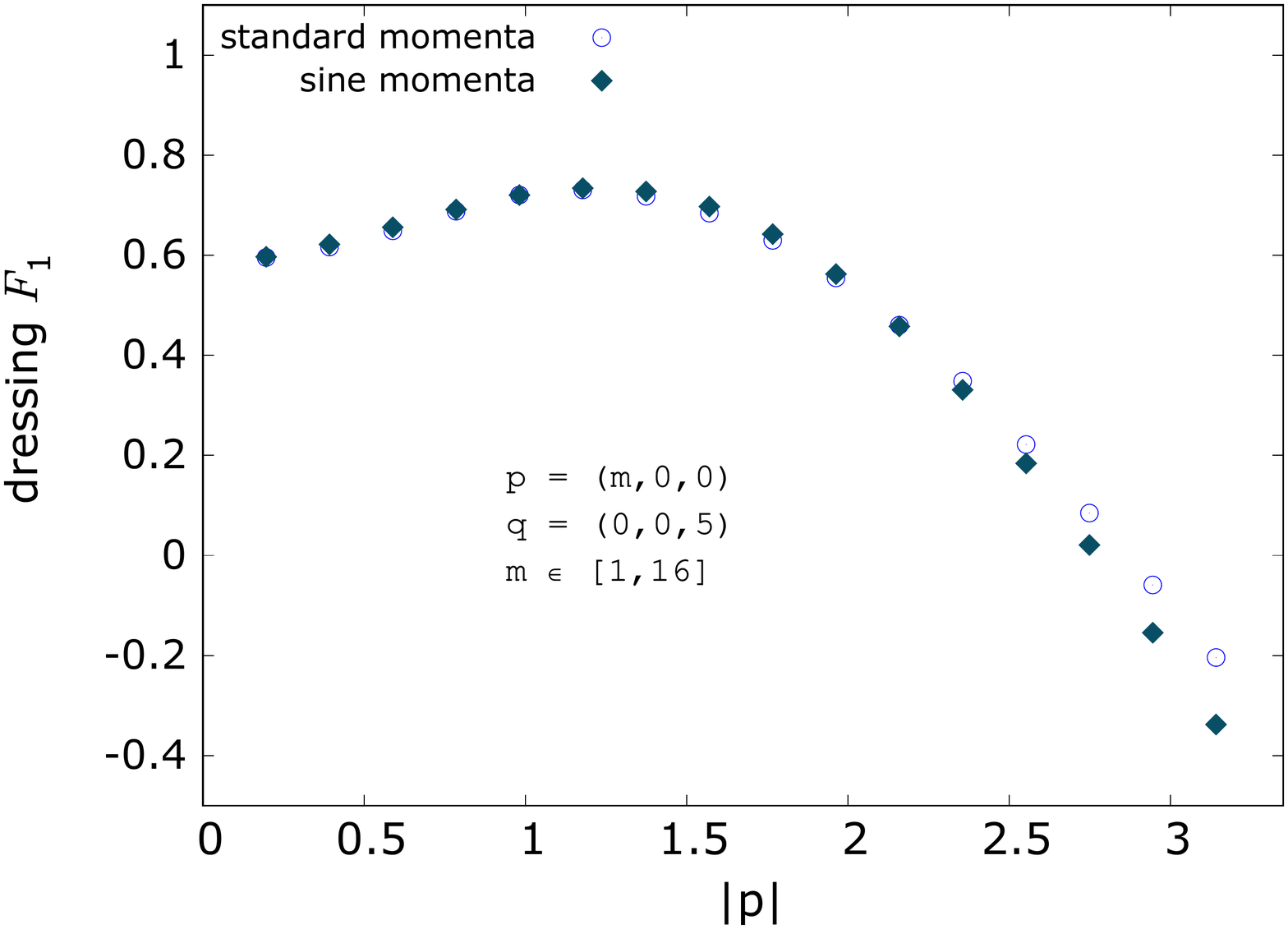}
\caption{\textit{Left}:~Reconstruction of perturbative vertex \er{eqn: lattice_vertex}, for particular kinematics on a 32$^3$ lattice.~\textit{Right}:~dressing function of 
the tree-level term $\tau_{\mu\nu\rho}^{\,1}$ of basis \er{eqn: bose_landau}, extracted from vertex \er{eqn: lattice_vertex}.~Reconstruction was done with ON tensor elements
\er{eqn:on_basis}, and function $F_1$ was calculated from ON form factors via rotation $R$ of \er{eqn: r_glory_one} and \er{eqn: r_glory_two}.~``Sine'' data refers to calculations
with momenta $(\hat{p},\,\hat{q})$, where e.\,g.~$\hat{p} = 2\sin(p/2)$.} 
\label{fig: 3d_dress}
\end{center}
\end{figure}

One final issue worth addressing here is the evaluation of vertex form factors.~In lattice Monte Carlo simulations, one would expect for some errors to arise when extracting the
three-gluon vertex dressing functions with continuum tensor elements, since the use of a continuum basis on a lattice incurs some loss of information.~Obviously, it would be good
to obtain at least some estimates for these errors, but this is generally a non-trivial task:~since there are no exact values for form factors on the lattice, one has no
benchmarks to employ when testing the sensitivity to unreliable tensor representations.~Fortunately, there are a few kinematic exceptions to this, and one of them was explored in 
Figure \ref{fig: configs_3d}\,a):~as can be seen from the corresponding data, for certain kinematics we have both the ``wrong'' and ``correct'' tensor descriptions, and we can test
the impact that using the wrong basis has on the values for vertex dressing.~In the aforementioned kinematic setup, correct and wrong tensors correspond, respectively, to continuum 
representations with and without the sine adjustment, see Figure \ref{fig: configs_3d}\,a).~Using these facts as a guide, we have calculated a particular dressing function of the
perturbative correlator \er{eqn: lattice_vertex}, for kinematics resembling those of \ref{fig: configs_3d}\,a), employing both wrong and correct tensor representations.~Comparison of
vertex form factors in the two cases is given in the right panel of Fig.~\ref{fig: 3d_dress}.~In the left panel of the Figure, we give the accompanying results of vertex reconstruction,
to serve as a reference point:~we wish to have a rough estimate of how discrepancies in tensor parametrisations translate to deviations in the calculated form factors.  

A comparison between the left and right panels in Figure \ref{fig: 3d_dress} leads to a somewhat unexpected conclusion, showing that the two kinds of deviations are not comparable
in size.~Namely, where the relative differences in vertex reconstruction peak at about 4 percent, the ones for correlator dressings peak at about 60 percent (i.\,e.~a value of 
roughly $-$0.20, compared to $\sim - 0.33$).~Of course, one should keep in mind that these are results for a particular kinematic configuration, and that they hold only for the
perturbative vertex of \er{eqn: lattice_vertex}:~we will discuss the case of the Monte Carlo vertex shortly.~Even so, the data of Figure \ref{fig: 3d_dress} enables us to make some
informed guesses on uncertainties for three-gluon correlator dressings, pertaining to Monte Carlo calculations.~A quick glance at our reconstruction results in this paper reveals 
that, in a majority of cases, the deviations seen in our vertex reconstructions peak at about 5 to 10 percent.~On the basis of results in \ref{fig: 3d_dress}, and being somewhat
conservative in our estimations, we would say that the corresponding errors for correlator dressings are no larger than 40 to 50 percent.~By extrapolation (keeping in mind that 
such extrapolations can be rather ``dangerous''), one could expect for uncertainties of similar size to be present in most of the lattice studies of the three-gluon interaction
kernel.~Now, an error of about 50 percent is obviously a significant number, but it needs to be put into context.~Most of the lattice investigations of vertex dressing functions 
e.\,g.~\cite{Cucchieri:2006tf, Maas:2007uv, Cucchieri:2008qm,Boucaud:2013jwa,Athenodorou:2016oyh} deal with statistical errors which are either comparable, or even significantly
larger than the expected uncertainty coming from the use of a continuum tensor basis.~Thus, even though the tensor-related issues have an appreciable quantitative impact, from a
practical perspective they can only become important once the signal for vertex form factors has undergone some serious improvements, compared to the current situation.~This is
also the reason why, in Figure \ref{fig: 3d_dress}, we chose to present and discuss the results for the perturbative lattice vertex, and not the one of Monte Carlo calculations.~The
signal quality for vertex dressings of the Monte Carlo vertex is such that no definitive conclusions can be drawn from them, since all the results (with and without the sine 
adjustment) practically agree with each other, within very large statistical error bars.~The said results are given and further elaborated on in Appendix \ref{sec:bose_dress}.

\section{Further discussion and concluding remarks}\label{sec:conclude}

In this paper we have introduced the method of vertex reconstruction as a means of checking the fidelity of various tensor representations of lattice vertex functions.~We've 
used the method to show that, for general kinematics, the description of lattice gluon propagator and three-gluon vertex in terms of their continuum tensor bases leads to a 
non-negligible loss of information.~On the other hand, we have also demonstrated that there exist special kinematic configurations for which these functions can be respresented
correctly with tensor elements of the continuum theory.~To summarise, these special kinematics include 1) situations with all vertex momenta pointing along the diagonal, see 
e.\,g.~Figure \ref{fig: configs_3d}\,d), 2) configurations with vertex momenta having components equal to either zero, or to the same non-zero value $s$, arranged such that one
of the momenta is diagonal, see e.\,g.~\er{eqn:quasy_symm} and first line of \er{eqn:3d_symm}, and 3) kinematic choices with one vanishing vertex momentum, wherein the last
example requires a momentum substitution $p \rightarrow \hat{p} = 2\sin(p/2)$ in order to work, see Figure \ref{fig:configs_3g_2d}\,a).~We have shown analytically why the
aforementioned  kinematic configurations are special concerning the continuum tensor representations, and provided some arguments on the applicability of these ideas to other
primitively divergent vertices of lattice QCD.

In addition to the above, we attempted to provide quantitative estimates of the impact that the use of continuum tensor bases has on evaluations of vertex form factors on the 
lattice.~Our, somewhat conservative, estimate is that the resulting uncertainties for vertex dressings do not exceed fifty percent.~While this is a significant figure, it is 
not greater than typical statistical errors encountered in lattice evaluations of vertex dressing functions.~This means that the tensor-related discrepancies will only become
important once the signal for vertex form factors has improved significantly, compared to the current state.  
  
Here we would also like to make some additional comments on topics which are indirectly related to our results.~Let us start with the diagonal kinematic configurations.~One
of our main conclusions in this paper is that the evaluation of vertex functions near the lattice diagonal can be advantageous due to a reduction of discretisation artifacts
in vertex tensor elements.~But there is yet another reason to favour the near-diagonal configurations over some generic lattice kinematics:~close to the diagonal, there is
also a reduction in hypercubic artifacts inherent to lattice vertex form factors.~This constitutes the basis of the so-called cylindrical kinematic cut \cite{Leinweber:1998uu},
where one only considers momentum configurations which are a certain (short) distance away from the diagonal.~Thus, computations of lattice propagators/vertices for near-diagonal 
momenta can be doubly useful, as they reduce the discretisation effects in both the tensor structures and dressing functions of a given lattice correlator. 

\newpage

Another issue which merits a further discussion is the zero-momentum discrepancy in Figure \ref{fig:gluon_diag}.~It is very likely that most of this effect comes from finite volume
artifacts.~But it should be mentioned that there might exist other factors which contribute at $p=0$.~One of the possible ``culprits'' is the appearance of additional tensor terms
proportional to a Dirac delta function $\delta(p)$:~it is argued in \cite{Lowdon:2017uqe}, by means of axiomatic field theory, that such terms may arise in tensor decompositions of
gauge field propagators in both continuum and lattice theories.~Besides this, on the lattice there are also tensor structures that contribute to gluon correlators at zero momentum,
which have no continuum analogue and which vanish as $a \rightarrow 0$ \cite{Rothe:1992nt}.~In order to truly assess the influence of either of these structures at vanishing lattice
momentum, one would need to conduct a dedicated study with a careful consideration of finite volume and Gribov copy effects \cite{Ilgenfritz:2006he, Bakeev:2003rr, Silva:2004bv, 
Sternbeck:2005tk, Sternbeck:2008mv, Maas:2009ph, Maas:2011se}.~We are looking forward to contributing to some of these endeavours in future studies.                    
       
\section*{ACKNOWLEDGMENTS} 

We are grateful to R.~Williams, A.~Maas, A.~Cucchieri, G.~Eichmann and R.~Alkofer for helpful discussions and a careful reading of this manuscript.~TM acknowledges the
partial support by CNPq.~MV gratefully acknowledges the support of the Austrian science fund FWF, under Schr\"odinger grant J3854-N36.~We would also like to thank our 
referee, whose comments have helped us produce a significantly better paper, compared to the original version. 

\appendix


\section{Vertex tensor bases in Landau gauge}

\subsection{Orthonormal transverse basis}\label{sec:ortho_basis}   

Each of the gluon legs that comprise the three-gluon interaction comes with its own momentum variable:~in lattice literature, these three momenta are often denoted as
$p,\,q$ and $r$.~Due to momentum conservation at the vertex, only two of the momenta are independent.~A construction principle for the three-gluon vertex basis proposed
in \cite{Eichmann:2014xya} starts from the following combinations 
\begin{equation}\label{eqn:kV_def}
k = \frac{q - p}{2}, \quad V = -\,r
\end{equation}

The first step is to orthonormalise $k$ and $V$ with respect to each other:~this is done in a standard way as 
\begin{equation}\label{eqn:momenta_sd}
d_\mu = \widetilde{V}_\mu\,, \quad s_\mu = \widetilde{k}^{\,tr}_\mu\,,
\end{equation}
where $k_\mu^{\,tr}= T^{\,V}_{\,\mu\nu} \, k_\nu$ is a component of $k$ transverse to $V$,\, and $T^{\,V}_{\,\mu\nu} = \delta_{\,\mu\nu} -  V_\mu V_\nu/V^2$.~Tilde 
(\,\,$\widetilde{}$\,\,) in these expressions denotes a normalised vector.~For purposes of later discussion we introduce the auxiliary tensors
\begin{align}\label{eqn:basis_prelim_3}
&\mathsf{T}^{\,1}_{\mu\nu} = \delta_{\,\mu\nu}\,, \quad \,\,\,\mathsf{T}^{\,4}_{\mu\nu} = s_\mu d_\nu + d_\mu s_\nu \,, \nonumber \\
&\mathsf{T}^{\,2}_{\mu\nu} = s_\mu s_\nu\,, \quad \mathsf{T}^{\,5}_{\mu\nu} = s_\mu d_\nu - d_\mu s_\nu\,. \nonumber \\
&\mathsf{T}^{\,3}_{\mu\nu} = d_\mu d_\nu\,,
\end{align}

In Landau gauge the vertex is transverse with respect to $p, \, q$ and $r$, or explicitly
\begin{align}\label{eqn:trans_cond}
p_\mu \, \Gamma_{\mu\nu\rho}(p,q,r) \, = \, q_\nu \, \Gamma_{\mu\nu\rho}(p,q,r) \, = \, r_\rho \, \Gamma_{\mu\nu\rho}(p,q,r) \, = \, 0 \, .
\end{align} 

Thus, in Landau gauge it is sufficient to retain those linear combinations of elements in \eqref{eqn:basis_prelim_3} which are transverse to all of the vectors $p,\,q$
and $r$.~It turns out that there are only 4 of them.~To shorten the upcoming equations, we will use the following notation for kinematic variables  
\begin{align}\label{eqn:ab_define}
&t = \frac{V^2}{4}\,, \qquad \eta = \frac{4 \, k^2}{3 \, V^{\, 2}}\,, \qquad z = \widetilde{k}\cdot\widetilde{V}\,, \nonumber \\[1.4mm] 
&a = \sqrt{3\eta}\,z, \qquad b = \sqrt{3\eta} \, \sqrt{1 - z^2} \, .
\end{align} 

The above quantities are all dimensionless, except for $t$.~The momenta $p, \, q, \, r$ can now be rewritten as 
\begin{align}\label{eqn:sd_to_pqr}
p_\mu = -\sqrt{t}\,(b \, s_\mu + (a & - 1) \, d_\mu)\,, \qquad  q_\mu = \sqrt{t}\,(b \, s_\mu + (a + 1) \, d_\mu) \, ,\nonumber \\ 
&V = - \, r = 2\sqrt{t} \, d \, .
\end{align} 

We now need linear combinations of quantities in \eqref{eqn:basis_prelim_3} which have definitive transversality properties with respect to $p$ and $q$.~These have
been constructed in \cite{Eichmann:2012mp}.~Here we only provide the elements relevant for the vertex in Landau gauge:~for more general cases, consult \cite{Eichmann:2012mp}
or \cite{Eichmann:2014xya}.~The important objects are
\begin{align}\label{eqn:transver_Y}
\mathsf{Y}^{\,1}_{\mu\nu} &= \frac{1}{\sqrt{D - 2}}\left( \mathsf{T}^{\,1}_{\mu\nu} - \mathsf{T}^{\,2}_{\mu\nu} - \mathsf{T}^{\,3}_{\mu\nu} \right), \nonumber \\
\mathsf{Y}^{\,2}_{\mu\nu} &= \frac{1}{\sqrt{n_1 n_2}}\left[ (1-a^2)\, \mathsf{T}^{\,2}_{\mu\nu} - b^2\,\mathsf{T}^{\,3}_{\mu\nu} + ab\,\mathsf{T}^{\,4}_{\mu\nu} - 
b\,\mathsf{T}^{\,5}_{\mu\nu}  \right]\,.
\end{align}

In the above expression, $D$ denotes the number of dimensions, and we used the abbreviations
\begin{equation}\label{eqn:n1n2_def} 
 n_1 = 1+a^2+b^2\,, \qquad n_2 = n_1 - \frac{4a^2}{n_1}\,.
\end{equation}

From equations \eqref{eqn:sd_to_pqr} and \eqref{eqn:transver_Y}, and using the properties of $s$ and $d$ ($s^2 = d^2 = 1$ and $s\cdot d = 0$), it is
straightforward to show that the objects $\mathsf{Y}^{\,1}_{\mu\nu}$ and $\mathsf{Y}^{\,2}_{\mu\nu}$ have the following transversality properties: 
\begin{align}\label{eqn:bu!}
&\mathsf{Y}^{\,1}_{\mu\nu}\,s_\mu \,=\,\mathsf{Y}^{\,1}_{\mu\nu} \, s_\nu \,=\mathsf{Y}^{\,1}_{\mu\nu} \, d_\mu \,=\,\mathsf{Y}^{\,1}_{\mu\nu}\,d_\nu = \,  0 \,, \nonumber \\ 
&\mathsf{Y}^{\,2}_{\mu\nu} \, p_\mu \, = \, \mathsf{Y}^{\,2}_{\mu\nu} \, q_\nu \, = \, 0 \,.
\end{align}

From equations \eqref{eqn:sd_to_pqr} and \eqref{eqn:bu!}, one can see that $\mathsf{Y}^{\,1}_{\mu\nu}$ is transverse to all momenta $p, \, q$ and $r$, in both of its
indices.~Now, vector $s_\mu$ is, by construction, orthogonal to $r_\mu$, and one thus immediately gets two fully transverse elements, $\mathsf{Y}^{\,1}_{\mu\nu}\,s_\rho$
and $\mathsf{Y}^{\,2}_{\mu\nu}\,s_\rho$.~The remaining objects can be obtained by taking the vector $s$ and transversely projecting it with respect to momenta $p$ and $q$,
i.\,e.~$s_\mu^{\,p,\,q} = T_{\mu\alpha}^{\,p,\,q}s_\alpha$.~The resulting normalised momenta are 
\begin{align}\label{eqn:sp_sq_def}
&\widetilde{s}^{\,p}_{\,\mu} = \frac{1}{\sqrt{n_1 - 2a}}\left[ (a - 1)\,s_\mu - b \, d_\mu  \right] \, ,\nonumber \\ 
&\widetilde{s}^{\,q}_{\,\mu} = \frac{1}{\sqrt{n_1 + 2a}}\left[ (a + 1)\,s_\mu - b \, d_\mu  \right] \, .
\end{align} 

From equations \eqref{eqn:sp_sq_def} and \eqref{eqn:sd_to_pqr} one can see that $\widetilde{s}^{\,p}_{\,\mu}$ and $\widetilde{s}^{\,q}_{\,\mu}$ are orthogonal to $p$ and
$q$, respectively.~We now have all the ingredients to write down a complete, orthonormal and transverse basis for the Landau gauge three-gluon vertex:
\begin{align}\label{eqn:on_basis} 
&\rho^{\,1}_{\mu\nu\sigma} =  \mathsf{Y}^{\,1}_{\mu\nu}\,s_\sigma , \,  \qquad \rho^{\,2}_{\mu\nu\sigma} =  \mathsf{Y}^{\,2}_{\mu\nu}\,s_\sigma, \, \nonumber \\
&\rho^{\,3}_{\mu\nu\sigma} =  \mathsf{Y}^{\,1}_{\sigma\nu} \, \widetilde{s}^{\,p}_\mu, \, \qquad \rho^{\,4}_{\mu\nu\sigma} = \mathsf{Y}^{\,1}_{\rho\mu}\,\widetilde{s}^{\,q}_\nu \,.
\end{align}

Now let us discuss the case of two dimensions.~One first notes that in 2D the vectors $s$ and $d$, defined in equation \eqref{eqn:momenta_sd}, take the form 
\begin{align}\label{eqn:sd_2d}
s =  \left(\begin{array}{c} e \\ f \end{array}\right), \quad d =  \left(\begin{array}{c} f \\ -e \end{array}\right),  
\end{align}

\noindent
where $f^2 + e^2 = 1$.~In two dimensions, this is the only combination that satisfies the defining characteristics of $s$ and $d$, namely that $s^2 = d^2 = 1$ and $s 
\cdot d = 0$.~By plugging in the expressions of \eqref{eqn:sd_2d} into the definition of $\mathsf{Y}^{\,1}_{\mu\nu}$, component by component, one can see that the
(non-normalised) version of this tensor vanishes:
\begin{align}
&\mathsf{Y}^{\,1}_{11} \, = \, \delta_{11} - s_1 s_1 - d_1 d_1 \, = \, 1 - e^2 - f^2 = \, 0, \, \nonumber\\
&\mathsf{Y}^{\,1}_{21} \, = \, \delta_{21} - s_2 s_1 - d_2 d_1 \, = \, f e - e f = \, 0 \, ,  
\end{align}

\noindent
and similarly for $\mathsf{Y}^{\,1}_{12}$ and $\mathsf{Y}^{\,1}_{22}$.~The fact that $\mathsf{Y}^{\,1}_{\mu\nu}$ identically equals zero in 2D means that three out of
four basis elements in equation \eqref{eqn:on_basis} also vanish, and the only surviving tensor in Landau gauge is $\rho^{\,2}_{\mu\nu\sigma}$.~With the same kind
of calculation one can show that for special kinematics in three dimensions, e.\,g.~$p=(m,\,n,\,0)$ and $q=(g,\,l,\,0)$ (with arbitrary numbers $m,\,n,\,g,\,l$), the 
structure $\rho^{\,2}_{\mu\nu\sigma}$ will be the dominant one, as all the other elements of \eqref{eqn:on_basis} vanish for all but a few values of their indices
$(\mu\nu\sigma)$.~For such special 3D kinematics, the contributions of tensors $\rho^{\,j}_{\mu\nu\sigma}$ ($j=1,3,4$) to the index average of \eqref{eqn: gamma_aver}
will be negligible, rendering the calculations essentially two-dimensional.~In connection to this, one may look up the results in Figure \ref{fig: configs_3d} and compare
the ``quality'' of data between the upper and lower panels of the Figure. 

\subsection{\large  Simple Landau gauge basis}\label{sec:simple_basis}

Due to the properties of orthonormality and manifest transversality, the basis given in \eqref{eqn:on_basis} is very useful for numerics.~However, a somewhat convoluted
construction can make the transverse orthonormal (ON) basis difficult to manage for analytic manipulations.~In this section we describe another tensor basis for the 
three-gluon vertex, arguably the simplest one (in a certain sense) which one can use in Landau gauge.~The upcoming analytic proofs, relevant for our study, will be carried
out in full only for the Simple elements.~Here we will also establish a connection between the ON and Simple bases, and it will be used to argue that all of the forthcoming
results are equally well applicable to the ON structures, or indeed to any other tensor representation that one might choose to describe the three-gluon interaction.      

We start the basis construction with an observation that the three-gluon coupling has two independent momentum variables (say, $p$ and $q$) and three 
Lorentz indices, meaning that the following 14 tensor elements should suffice to parameterise the vertex (compare Appendix A of \cite{Davydychev:1996pb}):
\begin{align}\label{eqn:simple_full}
&  \delta_{\mu\nu} \times \big\{ p_\rho, \,q_\rho  \big\} , \quad \, \delta_{\mu\rho} \times \big\{ p_\nu, \, q_\nu  \big\} , \quad \, \delta_{\nu\rho}
\times \big\{ p_\mu, \, q_\mu  \big\} \, , \nonumber \\ 
&  p_\mu \times \big\{ p_\rho \, p_\nu , \,\,\,\, p_\rho \, q_\nu , \,\,\,\, q_\rho \, q_\nu \big\}, \quad \, p_\nu \, q_\mu \, p_\rho \, , \nonumber \\
&  q_\mu \times \big\{ q_\rho \, q_\nu , \,\,\,\, q_\rho \, p_\nu , \,\,\,\, p_\rho \, p_\nu \big\}, \quad \, q_\nu \, p_\mu \, q_\rho \, . 
\end{align} 

Full transversality, as expressed in \eqref{eqn:trans_cond} (with $r=-\,p-q$) means that among the elements of \eqref{eqn:simple_full}, one can ignore those which are
proportional to components $p_\mu$ and $q_\nu$.~Even more than that, from momentum conservation ($r_\rho =  - \, p_\rho - q_\rho$) and the fact that the tensor $r_\rho$
is eliminated in Landau gauge, one gets that $p_\rho$ and $q_\rho$ are degenerate, with~$p_\rho = - \, q_\rho$.~Taking into account this degeneracy and neglecting the
elements proportional to $p_\mu$ and $q_\nu$ in \eqref{eqn:simple_full}, one ends up with only four fully transverse elements in Landau gauge.~In terms of dimensionless
quantities, these tensors are
\begin{align}\label{eqn:simple_landau}
\sqrt{p^2} \, & S_{\mu\nu\rho}^{\,1} = T_{\alpha\mu}^{\,p} \, T_{\beta\nu}^{\,q} \, T_{\gamma\rho}^{\,r} \cdot  \delta_{\alpha\beta} \, p_\gamma \nonumber \, , \\
\sqrt{q^2} \, & S_{\mu\nu\rho}^{\,2} = T_{\alpha\mu}^{\,p} \, T_{\beta\nu}^{\,q} \, T_{\gamma\rho}^{\,r} \cdot  \delta_{\beta\gamma} \, q_\alpha \nonumber \, , \\
\sqrt{p^2} \, & S_{\mu\nu\rho}^{\,3} = T_{\alpha\mu}^{\,p} \, T_{\beta\nu}^{\,q} \, T_{\gamma\rho}^{\,r} \cdot  \delta_{\gamma\alpha} \, p_\beta \nonumber \, , \\
p^2\,\sqrt{q^2} \, & S_{\mu\nu\rho}^{\,4} = T_{\alpha\mu}^{\,p} \, T_{\beta\nu}^{\,q} \, T_{\gamma\rho}^{\,r} \cdot q_\alpha \, p_\beta \, p_\gamma \, ,
\end{align}            

\noindent
with a transverse projector $T_{\mu\alpha}^{p} = \delta_{\mu\alpha} - p_\mu \, p_\alpha /p^2$, and similarly for others.~The above structures are deceptively simple,
since we have refrained from writing out the full expressions, with transverse projections carried out.~For most of the upcoming analytic arguments, the forms given in
\eqref{eqn:simple_landau} are perfectly adequate.

The bases of \eqref{eqn:on_basis} and \eqref{eqn:simple_landau} describe the same object, and thus there has to be a connection between them.~In other words, there
should exist a rotation operator $R$ that effects the transformation  
\begin{align}\label{eqn:rotate!}
\rho^{\,j}_{\mu\nu\sigma} = \sum_{k=1}^4 \, R_{jk} \, S_{\mu\nu\sigma}^{\,k} \, , \quad j = 1\ldots 4 \, .
\end{align}

The procedure to obtain the components of $R$ can be broken down into a few simple steps, but we shall not provide the details here.~We will just write down the 
non-vanishing elements of $R$, for a three-dimensional theory.~In terms of kinematic variables $a,\, b, \, t$ and $n_1$ defined in equations \eqref{eqn:ab_define} and
\eqref{eqn:n1n2_def}, the non-zero entries of $R$ are 
\begin{align}
& R_{\,11} = -4 \sqrt{p^2} , \qquad R_{\,14} = \frac{p^2 \sqrt{q^2} \, (a^2 + b^2 - 1)}{b^2 \, t}, \qquad R_{\,24} = \frac{p^2 \sqrt{q^2}}{b^2 \, t \, n_+ \, n_-} , 
\qquad R_{\,32} = \frac{-2 \sqrt{q^2}}{n_-} ,
\nonumber \\[0.4cm] 
& R_{\,34} = \frac{p^2 \sqrt{q^2} \, (a + 1)}{b^2 \, t \, n_-}, \qquad R_{\,43} = \frac{-2 \sqrt{p^2}}{n_+} , \qquad R_{\,44} = \frac{p^2 \sqrt{q^2} \, (a - 1)}{b^2 \, t \, n_+} \, , 
\end{align}

\noindent
where $n_{\pm} = 1/\sqrt{n_1 \pm 2a}$.~Additionaly, all of the elements of $R$ are to be divided by $4 \, b \,\sqrt{t}$.~The matrix transpose of $R$ can be used to effect 
a different kind of transformation, namely to rotate the dressing functions of the ON basis (denoted $\mathcal{B}^{\,j}$ and calculated via \eqref{eqn:project_on}) into the 
dressings of the Simple basis.~We've used this fact to simultaneously perform vertex reconstructions with both ON and Simple bases, and we've checked that the two methods 
give the same results.~With the connection between the ON and Simple elements established via \eqref{eqn:rotate!}, we are ready to move on with our analytic proofs.    

\section{Special kinematic configurations}

\subsection{Generalised diagonal kinematics}\label{sec: diagonal}

In the following we wish to show that for certain kinematic configurations, the three-gluon vertex of lattice Monte Carlo calculations can be represented as a linear 
combination of tensor elements from the continuum theory.~The proof, which will unfortunately turn out to be rather lengthy, will consist of two parts.~In the first 
part, we will demonstrate that there exist kinematics for which the continuum tensors do not change under a transformation  
\begin{align}
p_\mu \rightarrow \hat{p}_\mu \, ,
\end{align}

\noindent
with $p$ being the continuum momentum and $\hat{p}$ being its lattice-adjusted version.~Following \eqref{eqn:landau_mom}, which is valid for standard lattice Landau gauge
implementations, we will look at a concrete example where 
\begin{align}\label{eqn:latt_adjust}
\hat{p}_\mu = 2 \sin\left(\frac{p_\mu}{2}\right) \, .
\end{align}

An invariance under the above adjustment, for certain kinematics, is a necessary requirement for continuum tensor bases to ``work'' on a lattice, since momenta which enter the
construction of vertex tensor elements should all be modified according to \eqref{eqn:latt_adjust}, see equation \er{eqn:3g_landau_latt}.~However, the above invariance 
condition alone does not guarantee an exact representation of lattice vertices in terms of continuum tensor structures.~We will discuss the reasons for this in the second part of
our proof:~there we will take a look at the perturbative lattice correlator and see what additional prerequisites have to be met, to make this function fully describable 
by tensor bases of the continuum theory.~Let us begin the arguments by looking at a specific momentum configuration, namely the fully symmetric case in three dimensions:
\begin{align}\label{eqn:fully_symm}
 \quad &p \, = \, (-n, \, 0, \,n) \, , \quad q \, = (n, \, -n, \, 0) \, , \quad  r \, = \, (0,\,n,\,-n) \, ,
\end{align}   
\noindent
with $n$ being a number consistent with lattice momentum discretisation.~The first part of our proof will hold not just for \eqref{eqn:fully_symm}, but for any
configurations which satisfy the demands that 1) the components of $p$ and $q$ are either $n$ or 0, and 2) the components of $p$ and $q$ are organised in such a way that
the lattice adjustment does not break momentum conservation.~In other words, from $r = -\, p - q$ it should follow that $\hat{r} = -\, \hat{p} - \hat{q}$.~Besides \eqref{eqn:fully_symm},
this class of configurations would also include the 2D example of \eqref{eqn:quasy_symm}, and others.  

One first notes that, without any loss of generality, the lattice transformation on momenta of \eqref{eqn:fully_symm} can be written as a multiplicative factor, i.\,e.
\begin{align}\label{eqn:scale!}
\hat{p}_\mu = \xi \cdot p_\mu, \qquad \hat{q}_\mu = \xi \cdot q_\mu , \qquad  \hat{r}_\mu = \xi \cdot r_\mu ,  
\end{align} 

\noindent
with $\xi$ being some number.~The above relation follows from the special form of the vectors in \eqref{eqn:fully_symm}:~since the components of all the vectors are either 
$\pm \,n$ or 0, they all get modified in the same way, i.\,e.~$\sin(n/2) = \xi\cdot n, \, \sin(0) =\xi\cdot 0, \, \sin(-n/2) = -\,\xi\cdot n $.~We shall temporarily assume
that the factor $\xi$ is strictly positive, and the possibility $\xi < 0$ will be discussed in detail later.~Now, one can easily see that the transverse projectors that enter
the construction of Simple elements in \eqref{eqn:simple_landau}, are invariant under the scaling transformation of \eqref{eqn:scale!}.~As an example, 
\begin{align}\label{eqn:transver_scaling}
\frac{\hat{p}_\alpha \hat{p}_\mu}{\hat{p}^2} \, = \, \frac{\xi^{\,2} \, p_\alpha p_\mu}{\xi^{\,2} \, p^2} \, = \, \frac{ p_\alpha p_\mu}{p^2} \, ,
\end{align}

\noindent
and similarly for the non-trivial parts of $T_{\beta\nu}^{\,q}$ and  $T_{\gamma\rho}^{\,r}$.~Since these operators are unaffected by the lattice adjustment, we will drop
them from the definition of the Simple basis, for the derivation of the following expression.~Let us see what happens with the remaining parts of the $S^{\,k}_{\mu\nu\rho}$
structures under \eqref{eqn:scale!}.~Concretely, let us look only at $S^{\,1}_{\mu\nu\rho}$ and $S^{\,4}_{\mu\nu\rho}$, as it should be fairly obvious that the same thing
happens with the other elements as well:  
\begin{align}\label{eqn:no_change}
&\widehat{S}_{\mu\nu\rho}^{\,1} \,\, = \,\, \frac{\delta_{\mu\nu} \, \hat{p}_\rho}{\sqrt{\hat{p}^{\,2}}} \,\, = \,\, \frac{\xi \, \delta_{\mu\nu} \, p_\rho}{\xi
\, \sqrt{p^2}} \,\, = \,\, \frac{\delta_{\mu\nu} \, p_\rho}{\sqrt{p^2}} \, , \nonumber \\[2mm] 
&\widehat{S}_{\mu\nu\rho}^{\,4} \,\, = \,\, \frac{\hat{q}_\mu \, \hat{p}_\nu \, \hat{p}_\rho}{ \hat{p}^{\,2} \sqrt{\hat{q}^{\,2}}} \,\, = \,\,
\frac{\xi^3 \, q_\mu \, p_\nu \, p_\rho}{\xi^3 \, p^2 \sqrt{q^2}} \,\, = \,\, \frac{\,q_\mu \, p_\nu \, p_\rho}{p^2 \sqrt{q^2}} \, .
\end{align}

Thus, even with lattice-adjusted momenta, the vertex tensor structures remain the same as in the continuum.~The same can be shown for other tensor representations, like
the orthonormal one.~From the invariance of elements $S^{\,k}_{\mu\nu\sigma}$ and equation \eqref{eqn:rotate!} one can see that, to establish an absence of change for the
basis $\rho^{\,j}_{\mu\nu\sigma}$ under \eqref{eqn:scale!}, it should be proven that the operator $R$ remains unaffected by lattice momentum modifications.~We will not go
into a detailed demonstration of this, but will outline the main steps.~Combining the transformation of \eqref{eqn:scale!} with definitions of \eqref{eqn:kV_def} and 
\eqref{eqn:ab_define} one gets:
\begin{align}
& \hat{t} = \xi^2 \, t, \quad \hat{\eta} = \eta, \quad \hat{z} = z , \nonumber \\ 
& \hat{a} = a , \quad \hat{b} = b \, .
\end{align}

From the above results and the definition of \eqref{eqn:n1n2_def}, it also follows that $\hat{n}_1 = n_1$ and (consequently) $\hat{n}_\pm = n_\pm$.~With this, one has all
the necessary ingredients to prove the invariance of $R$.~For instance, the lattice version of the element $R_{\,44}$ would be 
\begin{align}
R^{\mathcal{\,L}}_{\,44} \, = \, \hat{p}^{\,2} \sqrt{\hat{q}^{\,2}} \, (\hat{a} - & 1)/(4 \,\hat{b}^{3} \, \hat{t}^{\,\frac{3}{2}} \, \hat{n}_+) \, = \, \xi^3 \, p^2 \sqrt{q^2} 
\, (a - 1)/(\xi^3 \, 4 \, b^3 \, t^{\,\frac{3}{2}} \, n_+) \, = \nonumber \\ 
& =\, p^2 \sqrt{q^2} \, (a - 1)/(4 \, b^3 \, t^{\,\frac{3}{2}} \, n_+) = R_{\,44} \, .
\end{align}
 
Similar relations hold for other entries in $R$.~To conclude this part of our argument, we wish also to comment on the case $\xi < 0$.~While such a scenario can happen in principle,
for relatively general $\hat{p}\,(p)$ dependencies, it is in fact not possible in our current framework, with a periodic lattice and $\hat{p}\,(p) = 2 \, \sin(p/2)$.~The function
$\sin (x/2)$ has the same sign as $x$, for all $x \in [\,0,2\pi\,]$, with $[\,0,2\pi\,]$ being the relevant interval of values in standard lattice formulations.~Thus, in our present 
numerical setup, $\xi$ will always be a strictly positive factor.

This conlcudes the first part of our proof, where we have shown that continuum tensor bases do not change under a modification akin to \eqref{eqn:latt_adjust}, for certain 
kinematics.~As already noted, the above scaling invariance does not guarantee that the lattice three-gluon vertex can be described exactly by continuum tensor bases, with a 
counterexample provided by the configuration \eqref{eqn:fully_symm}, cf.~Figure \ref{fig:3d_symmetric}.~To understand why some lattice kinematic choices are special, in terms
of continuum tensor representations, we shall take a look at the perturbative lattice vertex of equation \eqref{eqn: lattice_vertex}.~We will show that, under some additional
assumptions which we have not addressed so far, this vertex can be represented exactly with the Simple basis elements of \eqref{eqn:simple_landau}.~This kind of argument directly 
pertains only to the perturbative lattice correlator, and not to the three-gluon vertex of Monte Carlo simulations:~however, the latter seems to behave similarly to the former,
regarding the continuum tensor descriptions, see Figure \ref{fig:3d_symmetric}.~Such similarities are somewhat expected, since the Monte Carlo correlator calculated with the
Wilson gauge action should approximately reduce to \eqref{eqn: lattice_vertex} in the high-momentum region.
  
As an example of kinematics where the perturbative lattice vertex can be completely described with elements \eqref{eqn:simple_landau}, we shall look at the upper configuration in
\eqref{eqn:3d_symm}, i.\,e.~the situation
\begin{align}\label{eqn:buci_bu}
 p \, = \, (n, \, 0, \,n) \, , \quad q \, = (0, \, n, \, 0) \, , \quad  r \, = \, -(n,\,n,\,n) \, .
\end{align}  

Let us begin by pointing out some special features of the above momentum configuration, which will become important in the following.~First, at least one momentum (in the case 
of \eqref{eqn:buci_bu}, it is vector $r$) is diagonal, meaning that it has all the components equal to each other.~In some sense, it makes the vector $r$ behave as if it were a 
constant, since $r_\mu = - \, n$ for all possible values of index $\mu$.~The second important characteristic of the kinematic choice \eqref{eqn:buci_bu} is that (\textit{no}
summation implied): 
\begin{align}\label{eqn:p_q_vanish}
 p_\mu \, q_\mu = 0 \, .
\end{align}  

In other words, whenever there is an expression where momenta $p$ and $q$ appear with the same index $\mu$, the said expression can be set to zero.~The above is a consequence of 
the arrangement of zeros and non-zero entries in vectors $p$ and $q$.~Namely, since we have $p_1 = p_3 = n$ and $q_2 = n$, with all the other components of $p$ and $q$ vanishing.         
 
One last trick which we shall require for the upcoming proofs, is the manipulation of the cosine functions of momenta \eqref{eqn:buci_bu}, so as to get continuum-like 
momentum factors.~In other words, we wish to see how the terms like $\cos(p_\mu/2)$ can be transformed to give expressions similar to \eqref{eqn:scale!}.~Using the fact that
$\cos(0) = 1$, one can write [\,here, the vector $p$ is taken from \eqref{eqn:buci_bu}\,]
\begin{align}\label{eqn:scale_cos}
\cos\left(\frac{p}{2}\right) =  \left(\begin{array}{c} \cos(n/2) \\ 1 \\ \cos(n/2) \end{array}\right) \, = \, 1 - \left(\begin{array}{c} 1 - \cos(n/2) \\ 0 \\ 1 - \cos(n/2)
\end{array}\right) \, = \, 1 - \eta\cdot p \, ,
\end{align}  
where $\eta = (1 - \cos(n/2))/n$.~One can thus equate $\cos(p_\mu/2)$ with $1 - \eta\cdot p_\mu$, with the aforegiven factor $\eta$.~The same kind of transformation works for the 
momentum $q$ of \eqref{eqn:buci_bu}, so that one has $\cos(q_\nu/2) = 1 - \eta\cdot q_\nu$.~With these relations at our disposal, we are finally ready to demonstrate that the vertex
\eqref{eqn: lattice_vertex} reduces to a linear combination of Simple basis elements $S_{\mu\nu\rho}^{\,k} \, (k = 1\ldots4) $, for the kinematics \eqref{eqn:buci_bu}.~For convenience, 
we repeat here the definition of the lattice perturbative three-gluon vertex in Landau gauge and for Wilson gauge action (an overall factor of 2 for the perturbative vertex will be 
ignored in the following):
\begin{align}\label{eqn:latt_again}
&\Gamma^{\,\text{latt,\,Landau}}_{\mu\nu\rho}(p,q,r) \, = \, T_{\alpha\mu}^{\,p} \, T_{\beta\nu}^{\,q} \, T_{\gamma\rho}^{\,r} \cdot 
\Gamma^{\,\text{latt}}_{\alpha\beta\gamma}(p,q,r), \,\, \text{where} \nonumber \\[0.13cm]
& \Gamma^{\,\text{latt}}_{\alpha\beta\gamma}(p, q, r) \, = \, \delta_{\alpha\beta}\sin\left(\frac{p_\gamma - q_\gamma}{2}\right)\cos\left(\frac{r_\alpha}{2}\right)
\, + \, \delta_{\beta\gamma}\sin\left(\frac{q_\alpha - r_\alpha}{2}\right)\cos\left(\frac{p_\beta}{2}\right) \nonumber \\[0.1cm]
& \qquad \qquad \quad \,\,\, + \, \delta_{\gamma\alpha}\sin\left(\frac{r_\beta - p_\beta}{2}\right)\cos\left(\frac{q_\gamma}{2}\right) \,\, .
\end{align}  

Note that, unlike in equation \eqref{eqn: lattice_vertex}, in the above expression we use the continuum definitions for transverse projectors.~This is justified by the fact that,
for the kinematics of \eqref{eqn:buci_bu}, the continuum and lattice-adjusted versions (with $T_{\alpha\mu}^{\,p,\,l} = \delta_{\mu\alpha} - \, \hat{p}_\alpha\hat{p}_\mu/\hat{p}^2$\,)
become equivalent, see equation \eqref{eqn:transver_scaling} and the accompanying text.~We now need to analyse all of the terms in \eqref{eqn:latt_again} carefully, to see how they may
be brought into the form of tensor elements \eqref{eqn:simple_landau}.~We shall start with the quantity $\sin[(q_\alpha - r_\alpha)/2]$.~By means of a trigonometric identity
\begin{align}
\sin(\alpha - \beta) = \sin(\alpha)\cos(\beta) - \sin(\beta)\cos(\alpha) \, , 
\end{align}  

\noindent
the said expression can be recast into a new form
\begin{align}\label{eqn:first_sin}
\sin\left(\frac{q_\alpha - r_\alpha}{2}\right) \,\, = \,\, \sin\left(\frac{q_\alpha}{2}\right)&\cos\left(\frac{r_\alpha}{2}\right) \, - \, \sin\left(\frac{r_\alpha}{2}\right)
\cos\left(\frac{q_\alpha}{2}\right) \,\, =  \,\, \mathcal{C}_1 \sin\left(\frac{q_\alpha}{2}\right) \, - \, \mathcal{C}_2 \cos\left(\frac{q_\alpha}{2}\right) \, . 
\end{align} 

In the last step of \eqref{eqn:first_sin}, we have introduced the constants 
\begin{align}\label{eqn:c1_c2_def}
\mathcal{C}_1 = \cos\left(\frac{n}{2}\right) \, , \quad  \mathcal{C}_2 = - \, \sin\left(\frac{n}{2}\right) \, .
\end{align} 

The final equality in \eqref{eqn:first_sin} follows from the fact that the vector $r$ is diagonal, for kinematics \eqref{eqn:buci_bu}.~In other words, the index `$\alpha$'
on this vector does not really matter, since one has
\begin{align}
\cos\left(\frac{r_\alpha}{2}\right) =  \cos\left(\frac{-n}{2}\right) = \mathcal{C}_1 \, , 
\end{align} 

\noindent
for \textit{all} values of the index $\alpha$.~The same argument holds for the constant $\mathcal{C}_2$ in \eqref{eqn:first_sin}.~Now, by using the identities \eqref{eqn:scale!}
and \eqref{eqn:scale_cos}, we can further transform the final result of \eqref{eqn:first_sin} as follows:
\begin{align}\label{eqn:until_later}
\mathcal{C}_1 \sin\left(\frac{q_\alpha}{2}\right) \, - \, \mathcal{C}_2 \cos\left(\frac{q_\alpha}{2}\right) \,\, = \,\,  (\mathcal{C}_1\cdot\xi + \mathcal{C}_2\cdot\eta) 
\cdot q_\alpha \, - \, \mathcal{C}_2 \,\, = \,\, (\mathcal{C}_1\cdot\xi + \mathcal{C}_2\cdot\eta)\cdot q_\alpha \, - \, \xi\cdot r_\alpha \, .  
\end{align} 

For convenience, we collect here all of the difinitions we have used to arrive at \eqref{eqn:until_later} ($\xi$ and $\eta$ were defined in \eqref{eqn:scale!} and \eqref{eqn:scale_cos},
respectively):
\begin{align}\label{eqn:subs_conts_one}
&\xi = \sin(n/2)/n \, , \, \quad \eta = (1 - \cos(n/2))/n \, , \nonumber \\ 
&\mathcal{C}_1 = \cos(n/2) \, , \, \quad \mathcal{C}_2 = - \sin(n/2)  \, .
\end{align}

In the last step in \eqref{eqn:until_later}, we rewrote the constant $\mathcal{C}_2$ as $\mathcal{C}_2 = \xi\cdot r_\alpha = - \, \xi\cdot n$:\,we repeat that for vector $r$ of
\er{eqn:buci_bu}, it holds that $r_\alpha = -\, n$ for all values of the index $\alpha$.~The same steps which were used to arrive at the final expression in \er{eqn:until_later}, can 
be applied to the remaining two sine functions in \eqref{eqn:latt_again}.~In particular, the term $\sin[(r_\beta - p_\beta)/2]$ can be transformed into 
\begin{align}\label{eqn:sin_subs_two}
\sin\left(\frac{r_\beta - p_\beta}{2}\right) \,\, = \,\, \xi\cdot r_\beta \, - \, (\mathcal{C}_1\cdot\xi + \mathcal{C}_2\cdot\eta)\cdot p_\beta \, .
\end{align}

The remaining sine factor in the perturbative lattice vertex \eqref{eqn:latt_again} does not manifestly depend on the vector $r$, whose ``diagonality'' we have used to obtain
relations like \eqref{eqn:until_later} and \eqref{eqn:sin_subs_two}.~This is easily remedied by employing momentum conservation $r = - \, p - q$ to get 
\begin{align}\label{eqn:sin_subs_three}
\sin\left(\frac{p_\gamma - q_\gamma}{2}\right) \,\, = \,\, \sin\left(\frac{2 p_\gamma + r_\gamma}{2}\right) \,\, = \,\, (\tau\cdot\mathcal{C}_1 - 
\lambda\cdot\mathcal{C}_2)\cdot p_\gamma + \xi\cdot r_\gamma \, .
\end{align} 

In arriving at the final form in \eqref{eqn:sin_subs_three}, we have introduced new constants 
\begin{align}\label{eqn:subs_conts_two}
\tau = \sin(n)/n \, , \, \quad \lambda = (1 - \cos(n))/n \, .
\end{align}

The final expressions which we have derived in \eqref{eqn:until_later}, \eqref{eqn:sin_subs_two} and \eqref{eqn:sin_subs_three} are quite ungainly, but the main point is that the
sine terms of the perturbative lattice correlator can be recast into a form which features continuum-like tensor structures (e.\,g.~\eqref{eqn:until_later} is linear in $q_\alpha$
and $r_\alpha$), with all of the non-linear dependence on $n$ ``stored away'' in coefficients which multiply the vectors $p, \,q$ and $r$.

We now wish to (arguably) simplify the above continuum-like expressions, by using the fact that we work in Landau gauge.~To see how this may help with simplifying things, note that
from momentum conservation it follows that (say) $r_\alpha = - \, q_\alpha - p_\alpha$.~Now, the tensorial term $p_\alpha$ is projected out by the operator $T_{\alpha\mu}^{\,p}$ in
\er{eqn:latt_again}, from which it follows $r_\alpha = - q_\alpha$.~This last identity is to be understood to hold formally, when constructing vertex tensor elements, and not as a
literal equality between vectors $r$ and $q$.~By exploting the formal relations like $r_\alpha = - q_\alpha$, some of the above equations can be recast, in Landau gauge, as follows:    
\begin{align}\label{eqn:subs_final}
&\sin\left(\frac{q_\alpha - r_\alpha}{2}\right) \,\, = \,\, (\mathcal{C}_1\cdot\xi + \mathcal{C}_2\cdot\eta + \xi)\cdot q_\alpha \, , \nonumber \\ 
&\sin\left(\frac{r_\beta - p_\beta}{2}\right) \,\, = \,\, - \, (\mathcal{C}_1\cdot\xi + \mathcal{C}_2\cdot\eta + \xi)\cdot p_\beta \, , \nonumber \\ 
&\sin\left(\frac{p_\gamma - q_\gamma}{2}\right) \,\, = \,\, (\tau\cdot\mathcal{C}_1 - \lambda\cdot\mathcal{C}_2)\cdot p_\gamma \, .
\end{align}

As one of the final transformations of the perturbative lattice vertex, we apply the trick \eqref{eqn:scale_cos} to cosine factors in \eqref{eqn:latt_again}, and combine this with the
relations \eqref{eqn:subs_final} to get
\begin{align}\label{eqn:almost_vert_subs}
&\sin\left(\frac{q_\alpha - r_\alpha}{2}\right)\cos\left(\frac{p_\beta}{2}\right) \,\, = \,\, \sin\left(\frac{q_\alpha - r_\alpha}{2}\right)(1 - \eta\cdot p_\beta) \,\, = \,\, 
(\mathcal{C}_1\cdot\xi + \mathcal{C}_2\cdot\eta + \xi)\cdot(q_\alpha - \eta\cdot q_\alpha \, p_\beta) \, , \nonumber \\
&\sin\left(\frac{r_\beta - p_\beta}{2}\right)\cos\left(\frac{q_\gamma}{2}\right) \,\, = \,\, \sin\left(\frac{r_\beta - p_\beta}{2}\right)(1 - \eta\cdot q_\gamma) \,\, = \,\, 
- \, (\mathcal{C}_1\cdot\xi + \mathcal{C}_2\cdot\eta + \xi)\cdot(p_\beta  - \eta \cdot p_\beta \, q_\gamma) \, , \nonumber \\ 
&\sin\left(\frac{p_\gamma - q_\gamma}{2}\right)\cos\left(\frac{r_\alpha}{2}\right) \,\, = \,\, \mathcal{C}_1 \cdot (\tau\cdot\mathcal{C}_1 - \lambda\cdot\mathcal{C}_2)\cdot p_\gamma \, .
\end{align}

For the last equality in \eqref{eqn:almost_vert_subs}, we used the definition of the constant $\mathcal{C}_1$, given in \eqref{eqn:c1_c2_def}.~With the above identities, all of the 
individual terms which comprise the vertex \eqref{eqn:latt_again} can be formulated as (almost) continuum-like tensor structures, i.\,e.
\begin{align}\label{eqn:vert_subs_final}
&\left[ \delta_{\beta\gamma}\sin\left(\frac{q_\alpha - r_\alpha}{2}\right)\cos\left(\frac{p_\beta}{2}\right) \right]^\text{tr} \,\, = \,\, (\mathcal{C}_1\cdot\xi + \mathcal{C}_2\cdot\eta + \xi)
\cdot \left[ (\,\delta_{\beta\gamma} \, q_\alpha - \eta\cdot \delta_{\beta\gamma} \, q_\alpha \, p_\beta\,) \right]^\text{\,tr}\, , \nonumber \\
&\left[ \delta_{\gamma\alpha}\sin\left(\frac{r_\beta - p_\beta}{2}\right)\cos\left(\frac{q_\gamma}{2}\right) \right]^\text{tr}\,\, = \,\, - \, (\mathcal{C}_1\cdot\xi + \mathcal{C}_2\cdot\eta + \xi)
\cdot \left[ (\,\delta_{\gamma\alpha} \, p_\beta  - \eta \cdot \delta_{\gamma\alpha} \, p_\beta \, q_\gamma\,) \right]^\text{\,tr} \, , \nonumber \\ 
&\left[ \delta_{\alpha\beta}\sin\left(\frac{p_\gamma - q_\gamma}{2}\right)\cos\left(\frac{r_\alpha}{2}\right) \right]^\text{tr} \,\, = \,\, \mathcal{C}_1 \cdot (\tau\cdot\mathcal{C}_1 - \lambda\cdot
\mathcal{C}_2)\cdot \left[ \, \delta_{\alpha\beta} \, p_\gamma \, \right]^\text{\,tr} \, .
\end{align}

The superscript ``\,tr\,'' in \eqref{eqn:vert_subs_final} indicates that objects inside the square brackets are to be transversely projected.~What remains to be done is to demonstrate 
that all of the tensorial terms in \eqref{eqn:vert_subs_final} [\,i.\,e.~pieces which carry the indices $(\alpha\beta\gamma)$\,] are proportional to some of the Simple basis elements in 
\eqref{eqn:simple_landau}, upon tranverse projection.~For certain quantities in \eqref{eqn:vert_subs_final}, the connection with the Simple basis is rather obvious.~As an example, for
the tensor in the third line of \eqref{eqn:vert_subs_final}, one has 
\begin{align}\label{eqn:whatever}
\mathcal{C}_1 \cdot (\tau\cdot\mathcal{C}_1 - \lambda\cdot\mathcal{C}_2)\cdot \sum_{\alpha\beta\gamma} T_{\alpha\mu}^{\,p} \, T_{\beta\nu}^{\,q} \, T_{\gamma\rho}^{\,r} 
\cdot \delta_{\alpha\beta} \, p_\gamma \,\, = \,\, \mathcal{C}_1 \cdot (\tau\cdot\mathcal{C}_1 - \lambda\cdot\mathcal{C}_2)\cdot \sqrt{p^2} \, S_{\mu\nu\rho}^{\,1} \, ,
\end{align}
\noindent
with $S_{\mu\nu\rho}^{\,1}$ being the first element of the basis \eqref{eqn:simple_landau}.~Note that in the above relation, we explicitly wrote the sum over indices 
$(\alpha\beta\gamma)$:~for the remainder of this section, we shall \textit{not} be using the Einstein summation convention, and our motivation for this will become clear
shortly.~Similarly to \eqref{eqn:whatever}, the transversely projected versions of $\delta_{\beta\gamma} q_\alpha$ and $\delta_{\gamma\alpha} p_\beta$ structures in
\eqref{eqn:vert_subs_final} are proportional to elements $S^{\,2}_{\mu\nu\rho}$ and $S^{\,3}_{\mu\nu\rho}$, respectively.~However, the equation \eqref{eqn:vert_subs_final} also
contains two problematic terms, namely the tensors 
\begin{align}\label{eqn:fus}
&\mathcal{F}^{\,1}_{\mu\nu\rho} = \sum_{\alpha\beta\gamma} T_{\alpha\mu}^{\,p} \, T_{\beta\nu}^{\,q} \, T_{\gamma\rho}^{\,r} \cdot \delta_{\beta\gamma} \, q_\alpha \, p_\beta \, , \nonumber \\
&\mathcal{F}^{\,2}_{\mu\nu\rho} = \sum_{\alpha\beta\gamma} T_{\alpha\mu}^{\,p} \, T_{\beta\nu}^{\,q} \, T_{\gamma\rho}^{\,r} \cdot \delta_{\gamma\alpha} \, p_\beta \, q_\gamma \, .
\end{align}

The elements on the right-hand side of the above relations are not Lorentz-covariant, since e.\,g.~the expression $\delta_{\beta\gamma} q_\alpha \, p_\beta$ contains 
two instances of index $\beta$, without an implied contraction:~it is because of such quantities that we will avoid using the summation convention for the rest of this section, 
and it should be understood that an index is summed over only if the appropriate sum symbol is present.~Due to absence of manifest Lorentz-covariance, it is not obvious that the
structures $\mathcal{F}^{\,1}_{\mu\nu\rho}$ and $\mathcal{F}^{\,2}_{\mu\nu\rho}$ can be reduced to linear combinations of Simple basis elements.~Nonetheless, one can show that, 
for kinematics \eqref{eqn:buci_bu}, it holds that
\begin{align}\label{eqn:f_s_relate}
&\mathcal{F}^{\,1}_{\mu\nu\rho} = n^2 \, S_{\mu\nu\rho}^{\,2} \, , \nonumber \\
&\mathcal{F}^{\,2}_{\mu\nu\rho} = -2 n^2 \,  S_{\mu\nu\rho}^{\,4} \, , 
\end{align}
 
\noindent
with number $n$ coming from \eqref{eqn:buci_bu}.~We will now go through a detailed proof for the first of the above equations, and simply remark that the second identity can be
shown with similar steps.~The proof will consist of a direct comparison, component-wise, between structures $\mathcal{F}^{\,1}_{\mu\nu\rho}$ and $\mathcal{S}^{\,2}_{\mu\nu\rho}$.~For 
this we first need to evaluate each of the tensors separately.~From the definition of $\mathcal{F}^{\,1}_{\mu\nu\rho}$ in equation \eqref{eqn:fus}, one gets 
\begin{align}\label{eqn:f_one}
\mathcal{F}^{\,1}_{\mu\nu\rho} \,\, = \,\, \sum_{\alpha\beta\gamma} \left( \delta_{\alpha\mu} - \frac{p_\alpha p_\mu}{p^2} \right) \left(\delta_{\beta\nu} - \frac{q_\beta q_\nu}{q^2}
\right) \left(\delta_{\gamma\rho} - \frac{r_\gamma r_\rho}{r^2} \right) \cdot \delta_{\beta\gamma} \, q_\alpha \, p_\beta \,\, = \,\, q_\mu \, p_\nu \, \delta_{\rho\nu} \, - \, \frac{q_\mu
\, p_\nu \, r_\rho \, r_\nu}{r^2} \, .
\end{align}  

When evaluating the sums over $\alpha$ and $\beta$ in \eqref{eqn:f_one}, we've used the property \eqref{eqn:p_q_vanish}, which eliminates most of the terms.~We now employ the 
fact that vector $r$ of \eqref{eqn:buci_bu} is diagonal, to make a substitution $r_\nu = - n$ in the above equation, and obtain 
\begin{align}\label{eqn:f_two}
\mathcal{F}^{\,1}_{\mu\nu\rho} =  q_\mu \, p_\nu \, \delta_{\rho\nu} \, + \, \frac{ n \, q_\mu\, p_\nu \, r_\rho}{r^2} \, .
\end{align}  

This concludes our manipulations with the structure $\mathcal{F}^{\,1}_{\mu\nu\rho}$, for now, and we turn our attention to the tensor $S_{\mu\nu\rho}^{\,2}$.~We consider 
the definition of this object in \eqref{eqn:simple_landau}, and we get for kinematics \eqref{eqn:buci_bu} the relation (we also use $\sqrt{q^2} = \sqrt{n^2} = n$) 
\begin{align}\label{eqn:s_one}
\mathcal{S}^{\,2}_{\mu\nu\rho} \,\, & = \,\, \sum_{\alpha\beta\gamma} \left( \delta_{\alpha\mu} - \frac{p_\alpha p_\mu}{p^2} \right) \left(\delta_{\beta\nu} - \frac{q_\beta q_\nu}{q^2}
\right) \left(\delta_{\gamma\rho} - \frac{r_\gamma r_\rho}{r^2} \right) \cdot \frac{\delta_{\beta\gamma} \, q_\alpha}{n} \,\, = \nonumber \\ 
& = \,\, \frac{\delta_{\nu\rho} \, q_\mu}{n} \, - \, \frac{q_\mu \, q_\nu \, q_\rho}{n \, q^2} \, - \, \frac{q_\mu \, r_\nu \, r_\rho}{n \, r^2} + \frac{q_\mu \, q_\nu \, r_\rho \, q\cdot r}{ n\, 
q^2 \, r^2} \, .
\end{align} 

When computing the sum over $\alpha$ in the above equation, we've used the relation \eqref{eqn:p_q_vanish} to eliminate some of the contributions.~Now, one can employ the relations $q\cdot
r = - n^2$ and $q^2 = n^2$, valid for kinematics \eqref{eqn:buci_bu}, together with momentum conservation $p = - \, r - q$ to combine the last two terms of \eqref{eqn:s_one} into 
\begin{align}\label{eqn:s_two}
 - \, \frac{q_\mu \, r_\nu \, r_\rho}{n \, r^2} + \frac{q_\mu \, q_\nu \, r_\rho \, q\cdot r}{ n \, q^2 \, r^2} \,\, = \,\, \frac{q_\mu \, p_\nu \, r_\rho}{n \, r^2} \, .
\end{align}

Taking the results of \eqref{eqn:f_two}, \eqref{eqn:s_one} and \er{eqn:s_two} we see that the first equation in \eqref{eqn:f_s_relate} reduces to a claim that 
\begin{align}\label{eqn:almost_there}
q_\mu \, p_\nu \, \delta_{\rho\nu} \, + \, \frac{ n \, q_\mu\, p_\nu \, r_\rho}{r^2} \,\, \overset{?}{=} \,\, n \, \delta_{\nu\rho} \, q_\mu \, - \, \frac{n \, q_\mu \, q_\nu \, q_\rho}{q^2}
\, + \, \frac{n \, q_\mu \, p_\nu \, r_\rho}{r^2} \, .
\end{align} 

The second term on the left-hand side of the above equation obviously agrees with the third term on the right-hand side, and we shall drop these from further comparison.~Also, we shall 
drop the factor $q_\mu$, which multiplies all of the contributions on both sides of \eqref{eqn:almost_there}.~This leaves us with a presumed identity
\begin{align}
p_\nu \, \delta_{\rho\nu}  \,\, \overset{?}{=} \,\, n \, \delta_{\nu\rho} \, - \, \frac{n \, q_\nu \, q_\rho}{q^2} \, .
\end{align} 

We now transform the left-hand side by substituting $p_\nu = - \, r_\nu - q_\nu$, and using $r_\nu = -n$ to get 
\begin{align}
n \, \delta_{\rho\nu} - q_\nu \, \delta_{\rho\nu}  \,\, \overset{?}{=} \,\, n \, \delta_{\nu\rho} \, - \, \frac{n \, q_\nu \, q_\rho}{q^2} \, .
\end{align} 

Neglecting the factors $n \, \delta_{\rho\nu}$ on both sides of the equation, we finally end up with a comparison 
\begin{align}
q_\nu \, \delta_{\rho\nu}  \,\, \overset{?}{=} \,\, \frac{n \, q_\nu \, q_\rho}{q^2} \, .
\end{align} 

By looking at the momentum $q$ in \eqref{eqn:buci_bu}, it is clear that both sides of the above assumed equality are non-vanishing only when $\nu = \rho = 2$.~For the 
particular case of $\nu = \rho = 2$, one obtains $(\text{using} \, q_2 = n)$
\begin{align}
n \, \delta_{22}  \,\, = \,\, \frac{n \, n^2}{n^2} \, ,
\end{align} 

\noindent
which is an obviously true statement.~This concludes our proof of the first relation in \eqref{eqn:f_s_relate}, and the second one can be demonstrated in exactly the same way.~At
this point we want to make a comment regarding the generality of the above results.~Expressions \eqref{eqn:f_s_relate} become incorrect if vectors $p$ and $q$ of \eqref{eqn:buci_bu}
are changed in any way, even if the said change does not affect the validity of essential arguments used in the above proofs.~As an example of such an alteration, one may consider
the same kinematics as in \eqref{eqn:buci_bu}, but with swapped momenta $p$ and $q$, i.\,e.~$p \leftrightarrow q$.~Another example would be a four-dimensional generalisation of
\eqref{eqn:buci_bu}, with 
\begin{align}\label{eqn:buci_4d}
 p \, = \, (n, \, 0, \,n, \, 0) \, , \quad q \, = (0, \, n, \, 0, \, n) \, , \quad  r \, = \, -(n,\,n,\,n,\, n) \, .
\end{align}

For both of these cases [\,swapped momenta $p$ and $q$, 4D version of \eqref{eqn:buci_bu}\,], the relations \eqref{eqn:f_s_relate} no longer hold, but for both configurations it turns
out that tensors $\mathcal{F}_{\mu\nu\rho}^{\,1}$  and $\mathcal{F}_{\mu\nu\rho}^{\,2}$ can still be expressed as linear combinations of Simple basis elements.~Here we shall 
state without proof, that for kinematics \eqref{eqn:buci_4d}, the tensors \eqref{eqn:fus} can be written as 
\begin{align}
&\mathcal{F}^{\,1}_{\mu\nu\rho} = \sqrt{2} \, n^2 \left(\, S_{\mu\nu\rho}^{\,2} +  S_{\mu\nu\rho}^{\,4} \right) \, , \nonumber \\
&\mathcal{F}^{\,2}_{\mu\nu\rho} = \sqrt{2} \, n^2 \left(\, S_{\mu\nu\rho}^{\,3} -  S_{\mu\nu\rho}^{\,4} \right) \, . 
\end{align}

It thus ``seems'' that it is always possible, for kinematic configurations akin to \eqref{eqn:buci_bu}, to (re)write the Lorentz-non covariant structures \eqref{eqn:fus} in terms of continuum
tensor elements.~We are not aware if these facts are a consequence of some deeper mechanism at play, or merely a coincidence resulting when kinematics like \eqref{eqn:buci_bu} are combined
with a Landau gauge condition for the gauge fields. 

In either case, returning to the kinematics \eqref{eqn:buci_bu}, we combine the relations \eqref{eqn:f_s_relate}, \eqref{eqn:simple_landau} and \eqref{eqn:vert_subs_final}, to (finally) 
obtain a representation of the perturbative lattice vertex \eqref{eqn:latt_again} in terms of continuum tensor structures (in the following, we also use $\sqrt{p^2} = \sqrt{2}n, \, \sqrt{q^2}
 = n$):
\begin{align}\label{eqn:ultimate}
&\left[ \delta_{\beta\gamma}\sin\left(\frac{q_\alpha - r_\alpha}{2}\right)\cos\left(\frac{p_\beta}{2}\right) \right]^\text{tr} \,\, = \,\, (\mathcal{C}_1\cdot\xi + \mathcal{C}_2\cdot\eta + \xi)
\cdot (1 \, - \, \eta\cdot n) \cdot n \, S_{\mu\nu\rho}^{\,2} \, , \nonumber \\
&\left[ \delta_{\gamma\alpha}\sin\left(\frac{r_\beta - p_\beta}{2}\right)\cos\left(\frac{q_\gamma}{2}\right) \right]^\text{tr}\,\, = \,\, - \, (\mathcal{C}_1\cdot\xi + \mathcal{C}_2\cdot\eta + \xi)
\cdot (\sqrt{2}\,n \, S_{\mu\nu\rho}^{\,3} + 2 \,\eta \cdot n^2 \, S_{\mu\nu\rho}^{\,4}) \, , \nonumber \\ 
&\left[ \delta_{\alpha\beta}\sin\left(\frac{p_\gamma - q_\gamma}{2}\right)\cos\left(\frac{r_\alpha}{2}\right) \right]^\text{tr} \,\, = \,\, \mathcal{C}_1 \cdot (\tau\cdot\mathcal{C}_1 - \lambda\cdot
\mathcal{C}_2)\cdot \sqrt{2} \, n \, S_{\mu\nu\rho}^{\,1} \, .
\end{align}

\begin{figure}[!t]
\begin{center}
\graph[width = 0.44\tew]{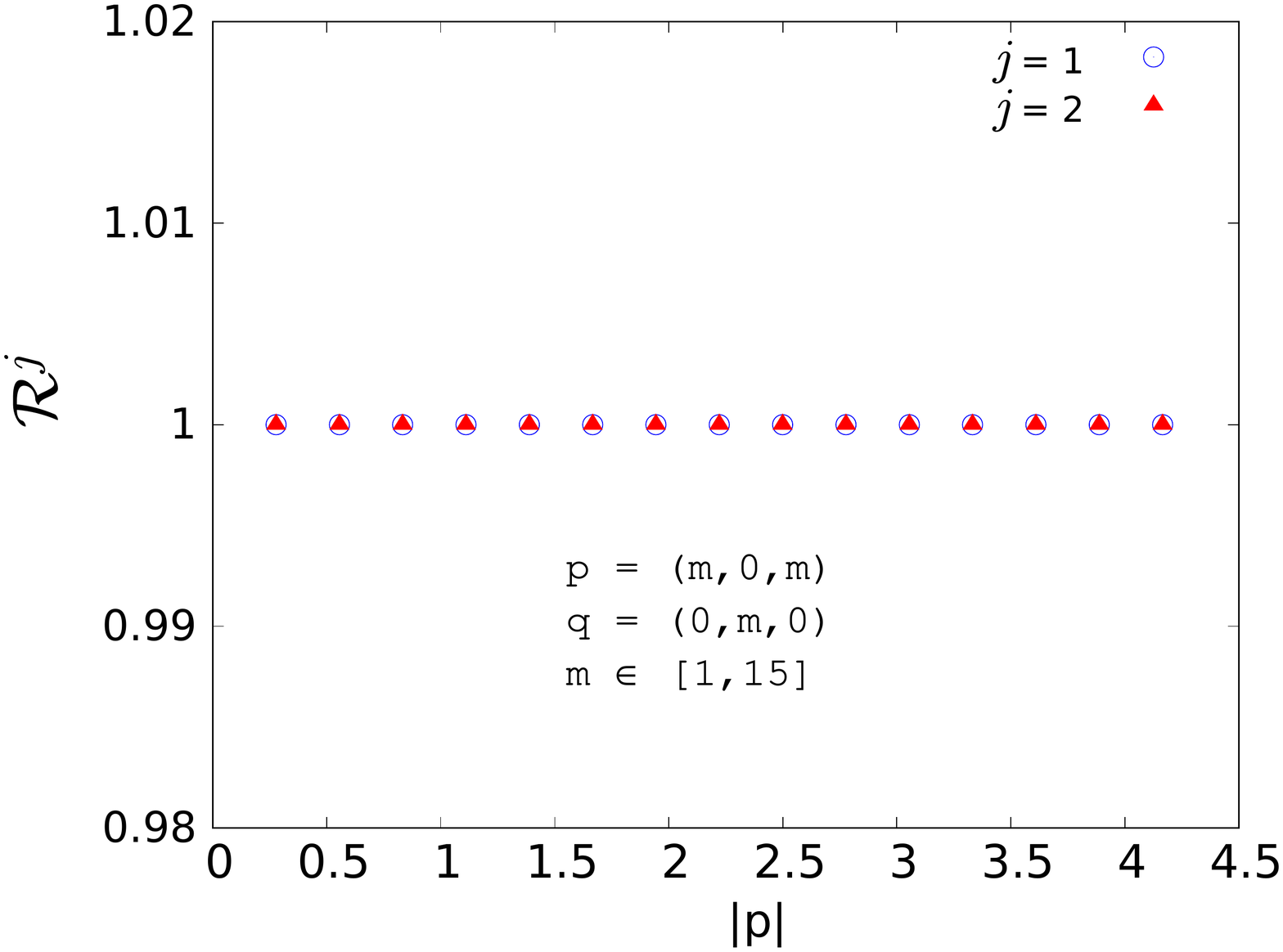}\\
\caption{Ratio of equations \eqref{eqn:append_ratio} and \eqref{eqn:num_tens_latt} on a 32$^3$ lattice, for kinematics \eqref{eqn:buci_bu}.~Results are in lattice units, with momenta
given in terms of vector $n_\mu$ of \eqref{eqn:fourier_glue}.} 
\label{fig:append_ratio}
\end{center}
\end{figure}

We repeat that in the above relations, ``\,tr\,'' denotes the full transverse projection, number $n$ is defined in \eqref{eqn:buci_bu}, and the quantities $\xi,\, \eta, \, \mathcal{C}_1,
\, \mathcal{C}_2, \, \tau$ and $\lambda$ all depend on $n$, via relations \eqref{eqn:c1_c2_def}, \eqref{eqn:subs_conts_one} and \eqref{eqn:subs_conts_two}.~Despite the cumbersomeness of the
above expressions, they unambiguously show that the perturbative lattice vertex can be represented via a continuum tensor basis.~As a check on the validity of the above relations, we shall  
evaluate numerically the following ratio
\begin{align}\label{eqn:append_ratio}
\mathcal{R} = \frac{\sum_{\mu\nu\rho} | \tau_{\mu\nu\rho}^{\, j, \, \text{latt}} |}{ \sum_{\mu\nu\rho} | \tau_{\mu\nu\rho}^{\, j, \, \text{cont}} |} \, , \quad j = 1, \, 2 \, ,  
\end{align}
\noindent
with $|.|$ denoting an absolute value, and tensors $\tau_{\mu\nu\rho}^{\, j, \, \text{latt}}$ and $\tau_{\mu\nu\rho}^{\, j, \, \text{cont}} \, (j = 1, 2)$ standing for certain objects on the
left- and right-hand sides of \eqref{eqn:ultimate}, respectively.~More precisely, we have 
\begin{align}\label{eqn:num_tens_latt}
&\tau_{\mu\nu\rho}^{\, 1, \, \text{latt}} \,\, = \,\, \left[ \delta_{\beta\gamma}\sin\left(\frac{q_\alpha - r_\alpha}{2}\right)\cos\left(\frac{p_\beta}{2}\right)
\right]^\text{tr} \, , \quad \tau_{\mu\nu\rho}^{\, 1, \, \text{cont}} \,\, = \,\, C_n \cdot (1 \, - \, \eta\cdot n) \cdot n \,
S_{\mu\nu\rho}^{\,2} \,\, , \nonumber \\ 
&\tau_{\mu\nu\rho}^{\, 2, \, \text{latt}} \,\, = \,\, \left[ \delta_{\gamma\alpha}\sin\left(\frac{r_\beta - p_\beta}{2}\right)\cos\left(\frac{q_\gamma}{2}\right)
\right]^\text{tr} \, , \quad \tau_{\mu\nu\rho}^{\, 2, \, \text{cont}} \,\, = \,\,  - \, C_n \cdot (\sqrt{2}\,n \, S_{\mu\nu\rho}^{\,3}
+ 2 \,\eta \cdot n^2 \,\,  S_{\mu\nu\rho}^{\,4}) \,\, ,
\end{align}
with an overall factor $C_n = (\mathcal{C}_1\cdot\xi + \mathcal{C}_2\cdot\eta + \xi)$.~Numerical results for the ratio are shown in Figure \ref{fig:append_ratio}, and they are equal to
unity (within numerical precision), as they should be.~We do not show the data for the third tensor(s) in \eqref{eqn:ultimate}, but simply remark that they are virtually indistinguishable
from the points in Figure \ref{fig:append_ratio}. 

With the equalities \eqref{eqn:ultimate} thus confirmed numerically, we immediately tackle the question on why the same kind of correspondence should not be expected to hold for some
other kinematics, like the symmetric one of \eqref{eqn:fully_symm}.~Put simply, the symmetric momentum partitioning has almost none of the essential features that made it possible
to establish the relations \eqref{eqn:ultimate}.~For instance, none of the momenta $p, \, q$ or $r$ of \eqref{eqn:fully_symm} are diagonal, making the manipulations like \er{eqn:first_sin} inapplicable.~Additionally, the cosine trick \er{eqn:scale_cos} does not result in continuum-like tensor structures for either vertex momentum, since one would have [\,we take the vector $p$
of \er{eqn:fully_symm} for illustration\,]
\begin{align}
\cos\left(\frac{p}{2}\right) =  \left(\begin{array}{c} \cos(-n/2) \\ 1 \\ \cos(+n/2) \end{array}\right) \, = \, 1 - \left(\begin{array}{c} 1 - \cos(n/2) \\ 0 \\ 1 - \cos(n/2)
\end{array}\right) \, \neq \, 1 - \eta\cdot p \, .
\end{align}  

With the last (in)\,equality above, we emphasised that the factors like $\cos(p_\mu/2)$ cannot be recast into a form $1 - \eta \cdot p_\mu$, for symmetric kinematics.~Finally, the
property \er{eqn:p_q_vanish} does not hold for kinematic choices akin to \er{eqn:fully_symm}, and this identity played an important role in deriving the relations \er{eqn:ultimate}.~Of
course, these arguments do not constitute an actual proof that equalities like \er{eqn:ultimate} are impossible for symmetric momentum arrangements, but they should make it clear that 
it is at least very unlikely that the perturbative lattice vertex can be fully described in terms of continuum tensor structures.~It should also be clear that it is almost a small
miracle (or at least it seems to be) that there exist any kinematic configurations, where the above correspondences between lattice and continuum tensor elements can be established.

\subsection{Collinear kinematics}\label{sec:collinear}

We now wish to show that for collinear kinematic configurations, all of the tensor elements of the three-gluon vertex vanish in Landau gauge.~This is the reason that, in our 
plots which include vertex reconstruction, we leave out the points corresponding to $p=q$:~such a scenario is simply a special case of collinear configurations.~Collinear
kinematics are defined by 
\begin{align}\label{eqn:pq_collin}
q =  \mathcal{C}_q \cdot p \, , \quad \mathcal{C}_q = \text{const}.
\end{align}
 
From the above relation and momentum conservation it follows that all of the vertex vectors are multiples of a single momentum variable, which we shall take to be $p$\,:
\begin{align}\label{eqn:all_collin}
p = p \ , \quad q = \mathcal{C}_q \cdot p , \quad r = \mathcal{C}_r \cdot p \, , 
\end{align} 

\noindent
where $\mathcal{C}_r = - \,1 -\, \mathcal{C}_q$.~When the relation \eqref{eqn:all_collin} is satisfied, the different transverse projectors of \eqref{eqn:simple_landau}
become equivalent, in a sense.~As an example,
\begin{align}
\frac{q_\nu q_\beta}{q^2} \, = \frac{\mathcal{C}_q^{\,2} \, p_\nu p_\beta}{\mathcal{C}_q^{\,2} \, p^2} \, = \, \frac{ p_\nu p_\beta}{p^2} \, . 
\end{align}

The same kind of equality holds for the operator $T_{\rho\gamma}^{\,r}$, which becomes equal to $T_{\rho\gamma}^{\,p}$.~Going back to the definitions of equation \eqref{eqn:simple_landau},
these facts entail the vanishing of all of the involved tensor structures, for instance 
\begin{align}
& S_{\mu\nu\rho}^{\,1} \, = \, T_{\alpha\mu}^{\,p} \, T_{\beta\nu}^{\,p} \, T_{\gamma\rho}^{\,p} \cdot \frac{\delta_{\alpha\beta} \, p_\gamma} {\sqrt{\smash[b] p^2}}
 \, \sim \, \bigg(\delta_{\gamma\rho} - \frac{p_\gamma \, p_\rho}{p^2}\bigg) \cdot \frac{\delta_{\alpha\beta} \, p_\gamma} {\sqrt{\smash[b] p^2}} \, = \, 0 \, , \nonumber \\[2mm] 
& S_{\mu\nu\rho}^{\,4} \, = \, T_{\alpha\mu}^{\,p} \, T_{\beta\nu}^{\,p} \, T_{\gamma\rho}^{\,p} \cdot \frac{\mathcal{C}_q \, p_\alpha \, p_\beta \, p_\gamma}{\mathcal{C}_q \,p^2
\sqrt{\smash[b] p^2} } \, \sim \, \bigg(\delta_{\gamma\rho} - \frac{p_\gamma p_\rho}{p^2}\bigg) \cdot \frac{p_\alpha \, p_\beta \, p_\gamma}{p^2 \sqrt{\smash[b] p^2} } \,= \, 0 \, . 
\end{align}
 
Similar relations hold for the elements $S_{\mu\nu\rho}^{\,2}$ and $S_{\mu\nu\rho}^{\,3}$.~In the above expressions, we've assumed that $\mathcal{C}_q$ is strictly positive, 
since its sign makes no difference for the end result.~We point out that none of these considerations apply to the case of one vanishing momentum, with either $\mathcal{C}_q$
or $\mathcal{C}_r$ being equal to zero.~This is because in such a scenario (say, $\mathcal{C}_r = 0$), the corresponding transverse projector is reduced to a Kronecker delta, 
i.\,e.~$T_{\rho\gamma}^{\,r} \rightarrow\delta_{\rho\gamma}$, meaning that some of the vertex tensor structures will survive the transverse projection.

With the vanishing of continuum Landau gauge three-gluon vertex thus established, for collinear configurations, we turn briefly to the case of the lattice correlation 
function.~From non-linearity of the transformation \eqref{eqn:latt_adjust}, it should be fairly easy to see that the conditions of \eqref{eqn:all_collin} will in general
not survive the lattice momentum adjustment.~In other words, from equation \eqref{eqn:all_collin} it does not follow that   
\begin{align}
\hat{q} = \mathcal{D}_q \cdot \hat{p} , \quad \hat{r} = \mathcal{D}_r \cdot \hat{p} \, , 
\end{align}
  
\noindent
with some constants $\mathcal{D}_q$ and $\mathcal{D}_r$.~Even if one chooses a configuration with $p=q$, so that necessarily $\hat{p}=\hat{q}$, one cannot keep both the 
condition of collinearity and the momentum conservation condition.~In general, then, the lattice three-gluon correlator would not be expected to vanish, for collinear 
kinematic configurations.  

\section{Bose symmetric tensor basis and vertex dressing functions}\label{sec:bose_dress}

In this section we will briefly discuss the dressing functions of the three-gluon vertex of lattice Monte Carlo simulations.~For our final results regarding the form factors,
we shall employ neither the orhonormal tensor structures \er{eqn:on_basis}, nor the Simple ones of \er{eqn:simple_landau}.~This is because we want to have a clear separation between
the continuum tree-level term and the beyond tree-level tensor structures, and this is a property that none of the aforementioned bases possesses.~Here we thus chose to work with
manifestly Bose-symmetric tensor elements, whose explicit construction is provided in \cite{Eichmann:2014xya}.~The basis elements are (equation (60) of \cite{Eichmann:2014xya}\,)
\begin{align}\label{eqn: bose_landau}
&\tau^1_{\mu\nu\rho}(p,q,r) \, = \, (p_\rho - q_\rho)\,\delta_{\mu\nu} \, + \, (q_\mu - r_\mu)\,\delta_{\nu\rho} \, + \, (r_\nu - p_\nu)\,\delta_{\rho\mu}\, , \nonumber \\
\mathcal{S}_0 \, &\tau^2_{\mu\nu\rho}(p,q,r) \, = \, (p_\rho - q_\rho) \, (q_\mu - r_\mu) \, (r_\nu - p_\nu) \, , \nonumber \\ 
\mathcal{S}_0 \, &\tau^3_{\mu\nu\rho}(p,q,r) \, = \, r^2 \, (p_\rho - q_\rho)\,\delta_{\mu\nu} \, + \, p^2 \, (q_\mu - r_\mu)\,\delta_{\nu\rho} \, + \, q^2 \, (r_\nu - p_\nu)
\,\delta_{\rho\mu} \, , \nonumber \\
\mathcal{S}_0 \, &\tau^4_{\mu\nu\rho}(p,q,r) \, = \, \omega_3 \, (p_\rho - q_\rho)\,\delta_{\mu\nu} \, + \, \omega_1 \, (q_\mu - r_\mu)\,\delta_{\nu\rho} \, + \, \omega_2 \, 
(r_\nu - p_\nu)\,\delta_{\rho\mu} \, , 
\end{align}

where 
\begin{align}\label{eqn: omegas}
\omega_1 \, = \, -\, q^2 + r^2 \, , \quad \omega_2 \, &= \, - \, r^2 + p^2 \, , \quad \omega_3 \, = \, - \, p^2 + q^2 \, , \quad \mathcal{S}_0 = p^2 + q^2 + r^2 \, .
\end{align}

To be more specific, we work with transversely projected versions of the above tensors, in Landau gauge.~The element $\tau_{\mu\nu\rho}^1$ in \er{eqn: bose_landau} represents
the continuum tree-level vertex structure.~The factor $\mathcal{S}_0$ was introduced in the above relations to make all the basis elements have the same mass dimension.~To get the
dressing functions pertaining to the tensor description \er{eqn: bose_landau}, we use the orthonormal basis \er{eqn:on_basis} in intermidiary steps.~Namely, we first obtain the form
factors of the orthonormal elements, denoted $\mathcal{B}_{\, j}$ and calculated via contractions \er{eqn:project_on}, and then rotate the results into dressing functions of the basis
\er{eqn: bose_landau}.~Denoting the coefficient functions of the Bose symmetric elements as $F_k$, one has 
\begin{align}\label{eqn:rot_dress}
F_k \, = \, \sum_{m = 1}^4 \, R_{\,km} \, \mathcal{B}_m \, , \quad k = 1 \ldots 4 \, .
\end{align}  

The rotation matrix $R$ corresponding to the transformation \er{eqn:rot_dress} is rather complicated, and here we shall provide its components in a somewhat condensed notation.~Writing
out the rows of $R$ as vectors, one has
\begin{align}\label{eqn: r_glory_one}
&R(1,:) \, = \, 2\,t \cdot [ \, m_+ \, n_-^2 - m_-\,n_+^2 \, , \quad 0 , \quad -2 n_-\,(m_- n_+^2 + a) \, , \quad -2 n_+ \,(m_+ n_-^2 + a) \, ] \, , \nonumber \\
&R(3,:) \, = \, n_-n_+ \, \mathcal{S}_0 \cdot [ \, 4 \, (m - 6) \, n_-n_+ \, , \quad 0 \, , \quad n_+\,(6a + 6 - m) \, , \quad n_-\,(6a - 6 + m) \, ] \, , \nonumber \\ 
&R(4,:) \, = \, \mathcal{S}_0 \cdot [ \, 2 \, (n_+^2 - n_-^2) \, , \quad 0 \, , \quad n_-\,(1 - 4n_+^2) \, , \quad n_+\,(1 - 4n_-^2) \, ] \, .
\end{align}   

The remaining elements of $R$, which are too long to fit neatly into the above vector notation, are 
\begin{align}\label{eqn: r_glory_two}
&R(2,1) =  - (2 \, (a - 3) \, m + m^2) \, n_-^2 \, + \, (16 \, b^2 \, (m - 6) \, n_-^2 \, + \, 2\,(a + 3) \, m \, - \, m^2) \, n_+^2   \, , \nonumber \\ 
&R(2,2) = - \frac{l}{(n_- \, n_+)} \, , \nonumber \\ 
&R(2,3) = 2n_- \cdot ( \, (12 \, (a + 1) \, b^2 \, - \, 2 \, (b^2 + a + 3)\,m + m^2) \, n_+^2 \, - \, a\,m \, ) \, , \nonumber \\ 
&R(2,4) = 2n_+ \cdot ( \, (12 \, (a - 1) \, b^2 \, + \, 2 \, (b^2 - a + 3) \, m - m^2) \, n_-^2 \, - \, a \, m)  \, .
\end{align}     

\begin{figure}[!t]
\begin{center}
\graph[width = 0.39\tew]{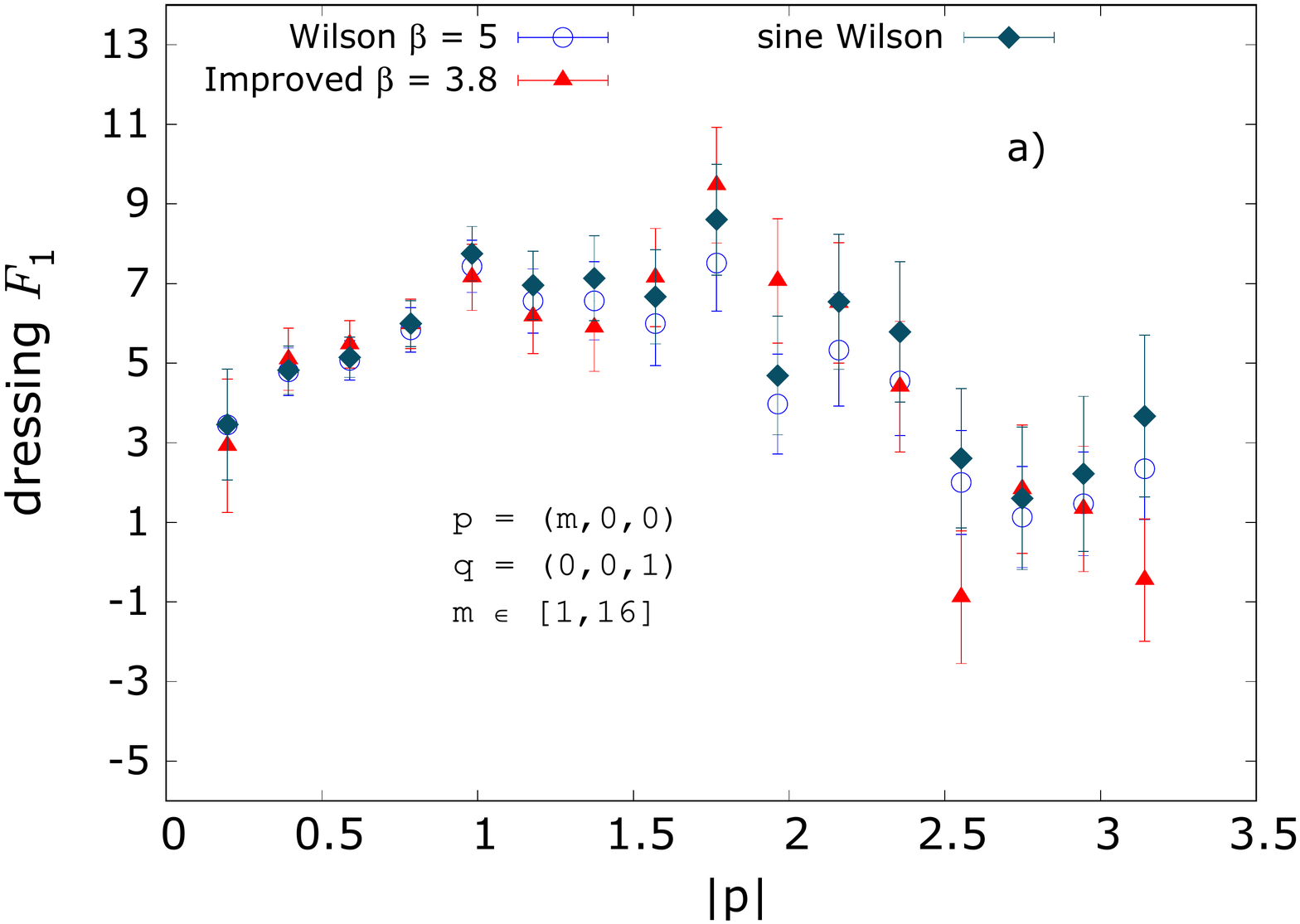}\graph[width = 0.39\tew]{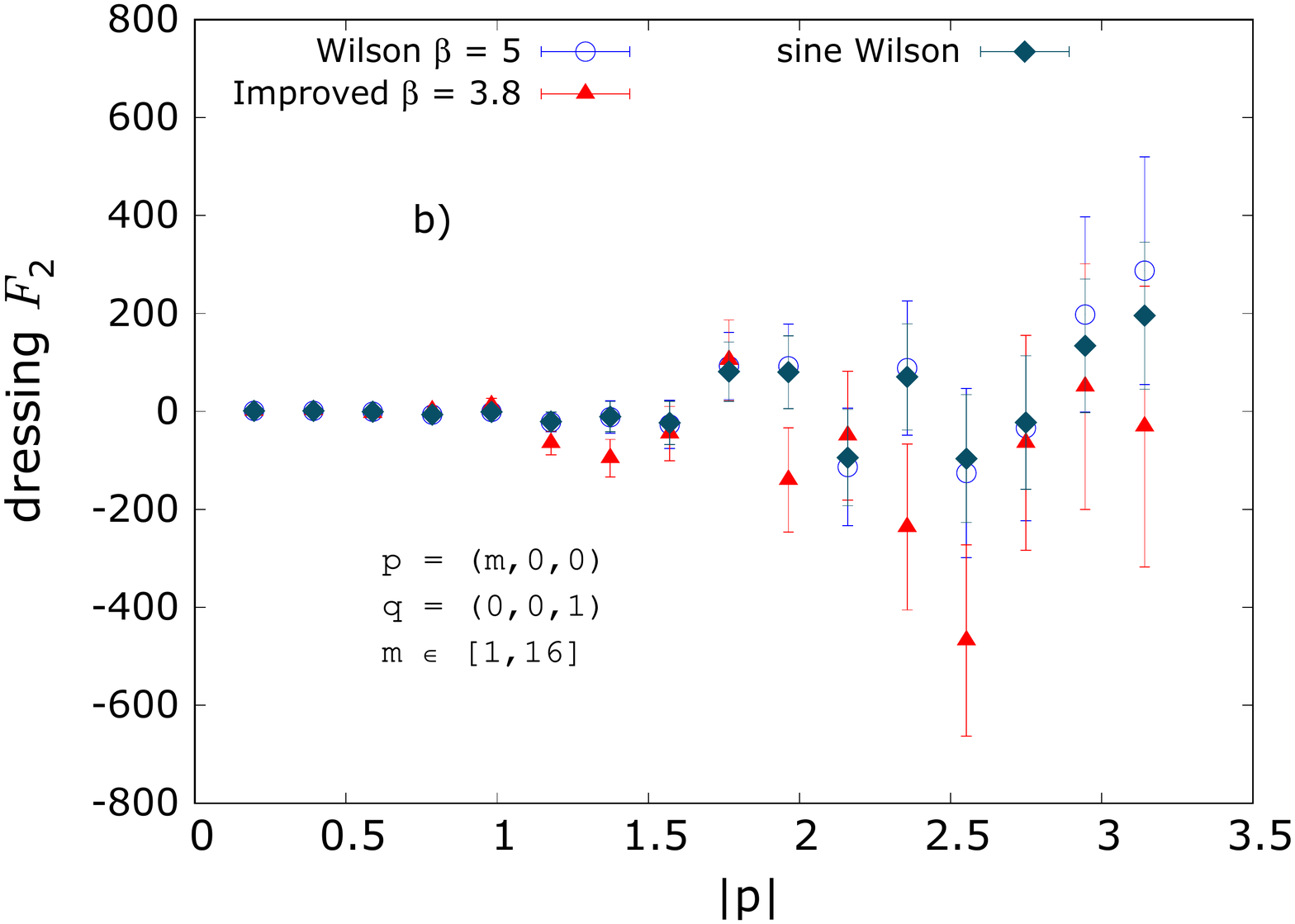}
\graph[width = 0.39\tew]{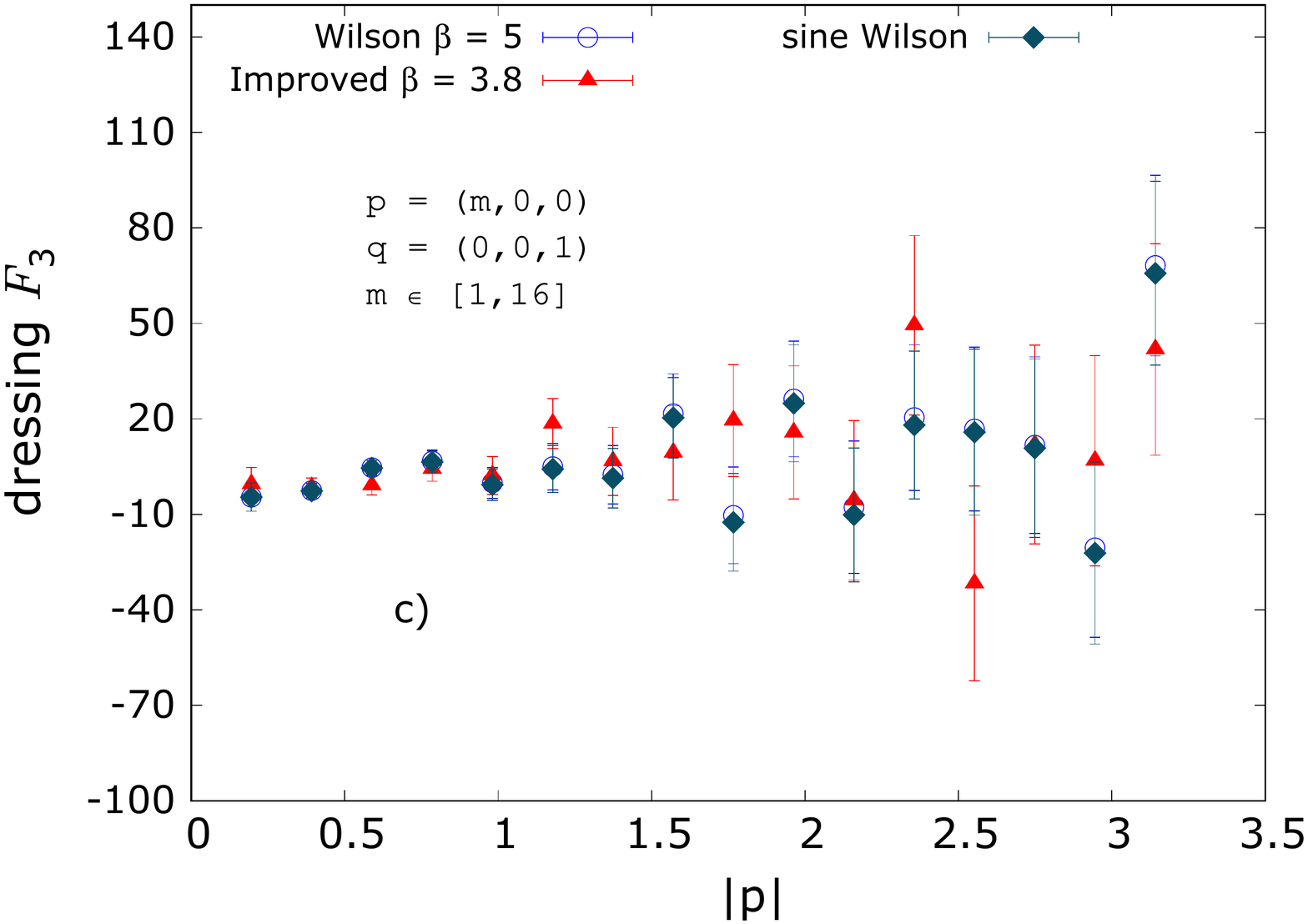}\graph[width = 0.39\tew]{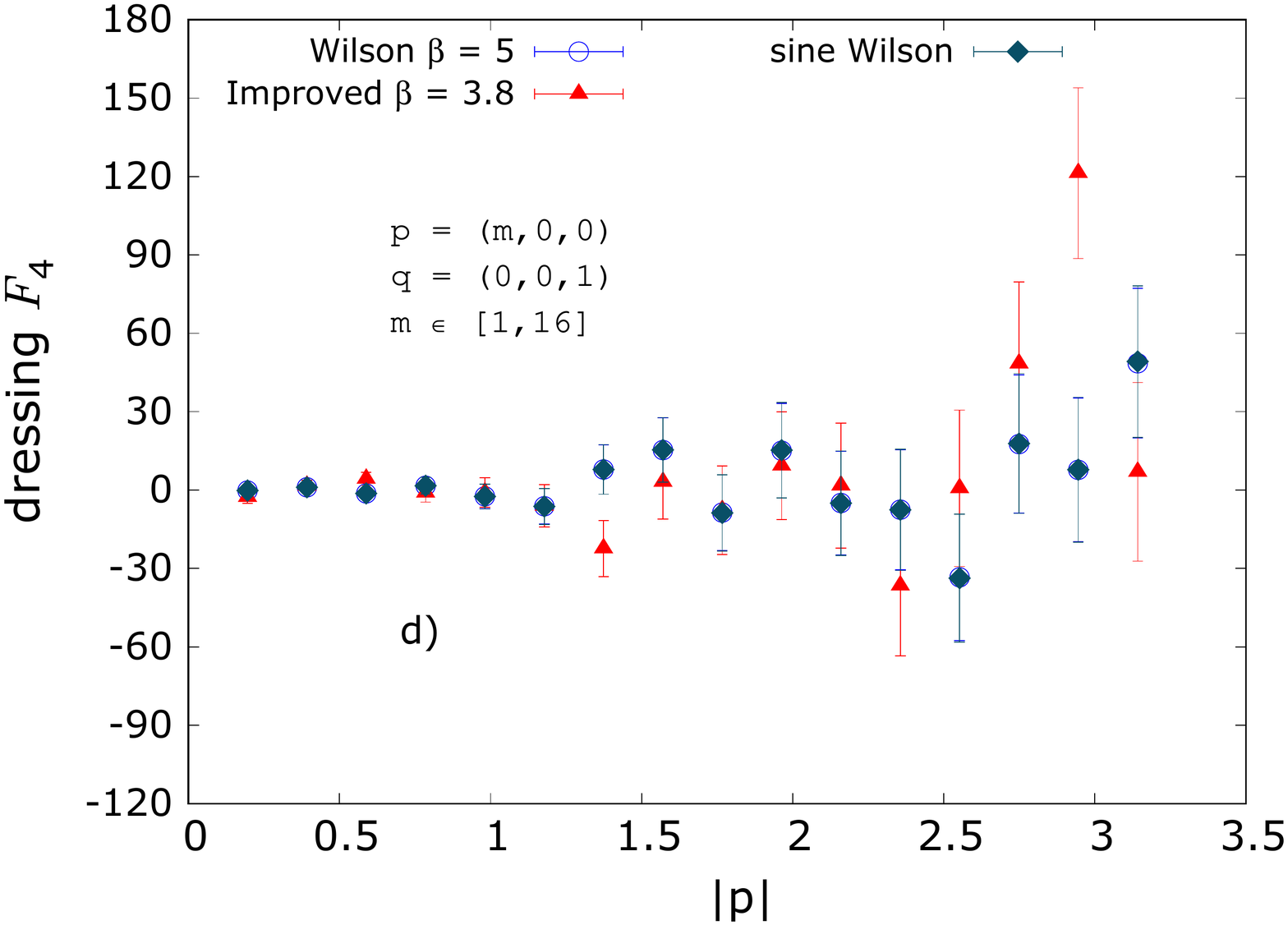}\\
\caption{Dressing functions of the vertex \er{eqn:g3_lattice}, for particular kinematics on a 32$^3$ lattice, as functions of $|\, p\,| = \sqrt{\smash[b]p^2}$.~The form factors
correspond to the transversely projected version of the basis decomposition \er{eqn: bose_landau}.~~Results are in lattice units, with momenta given in terms of vector $n_\mu$
of \er{eqn:fourier_glue}.~``Sine'' data refers to calculations with momenta $(\hat{p},\,\hat{q})$, where e.\,g.~$\hat{p} = 2\sin(p/2)$.} 
\label{fig:3d_dress}
\end{center}
\end{figure}

Additionally, all of matrix elements of \er{eqn: r_glory_one} are to be divided out with the factor $C_d = 4 \, l \cdot \, b \cdot \, t^{3/2}$, while the components \er{eqn: r_glory_two}
are to be multiplied with $ \mathcal{S}_0/(8 \, b^2 \cdot C_d)$.~In the above expressions, we used the shorthand notations 
\begin{align}\label{eqn: holy_maccaroni}
& n_{\pm} = 1/\sqrt{n_1 \pm 2a} \, , \quad m = a^2 + b^2 + 3 \, , \quad m_{\pm} = 2 a \pm m \mp 6 \, , \nonumber \\  
& l \, = \, ((6 \, a + m - 6) \, n_-^2 - (8 \, (m - 6) \, n_-^2 + 6 \, a - m + 6) \, n_+^2) \, ,
\end{align} 

\noindent
with quantities $a, \, b $ and $n_1$ being defined in equations \er{eqn:ab_define} and \er{eqn:n1n2_def}.~With the details of our calculation thus specified, we can turn to the actual
results for the dressing functions of the basis \er{eqn: bose_landau}.~These are provided in Figure \ref{fig:3d_dress}.~We note that the displayed data corresponds to non-renormalised 
functions, i.\,e.~we have imposed no renormalisation conditions when evaluating $F_k \, (k = 1\ldots 4)$.  

One thing that one may immediately note about the results in Figure \ref{fig:3d_dress}, is a rather poor signal quality for most of the displayed functions.~The dressing of the tree-level 
term [\,i.\,e.~the function $F_1$ in \ref{fig:3d_dress}\,] is the only one which clearly does not vanish, within the statistical uncertainties.~On the other hand, almost all of the values
for the beyond tree-level form factors are consistent with zero, within very large error bars.~Clearly, much better statistics would be needed before any definitive statements can be made
about the functions $F_k$, with $k = 2, 3, 4$.~In connection to this, we would like to repeat that the results of Fig.~\ref{fig:3d_dress} were obtained with 9600 gauge field configurations,
which is significantly more than what most other lattice studies of the three-gluon vertex use.~It thus seems that future Monte Carlo investigations will either have to employ many more 
configurations than what is typically considered today, or some algorithmic improvements will have to be made, to refine the signal for the beyond tree-level tensor structures and their
dressings.

Concerning the tree-level form factor $F_1$, with our present data we can neither confirm nor definitely refute the presence of the so-called zero crossing in the infrared (IR) region,
which was observed in certain continuum studies of the vertex \cite{Pelaez:2013cpa, Aguilar:2013vaa, Blum:2014gna, Eichmann:2014xya}.~There are also some lattice results which support
the existence of this sign change for the function $F_1$ \cite{Cucchieri:2008qm, Duarte:2016jhj, Athenodorou:2016oyh, Boucaud:2017obn}, but the available data is still not fully 
conclusive.~Our own results seem to be consistent with the value of $F_1$ slowly going down, as one approaches the region of small momentum $p$, but at the moment we cannot make any 
stronger statement in this regard.~The presence of a zero crossing, and even signs of a possible collinear divergence in $F_1$, are topics which will have to wait for future dedicated
studies, where better statistics and larger lattice volumes are considered.~We are currently working on some of these endeavours, and hope to get some decisive results in the near future.   

\vspace{1.84cm}  

\begin{table}[!h]
\begin{center}
\bgroup
\def\arraystretch{1.4}%
\begin{tabular}{||c|c|c|c|c|c|c|c|c|}
\hline\hline
$ V $    &  Action  & $ \beta $   & $a$ [GeV$^{-1}$] & $\langle W_{1,1}\rangle$ & $(\sqrt{\sigma} \, a)^\text{\,exp}$ & $(\sqrt{\sigma} \, a)^\text{\,calc}$ & $\alpha^\text{APE}$ & $\alpha^\text{gauge}$\\
\hline\hline
 $32^2$  &  W  & 10  & 0.93(2) & 0.85432(10) &   0.396   & 0.411(9) & 0.7 &  0.495  \\  
\hline  
 $32^2$  &  I  & 8  & 0.95(2)  & 0.87441(9)  &    $-$    &   $-$    & 0.7 &  0.495  \\  
\hline 
 $32^3$  &  W  & 5  & 0.74(2)  & 0.78694(9)  & 0.313(27) & 0.327(8) & 0.3 &  0.348  \\     
\hline 
 $32^3$  &  I & 3.8 & 0.72(2)  & 0.81195(9)  &    $-$    &   $-$    & 0.3 &  0.346  \\
\hline
 $32^3$  &  W & 12  & 0.35(1)  & 0.91481(8)  & 0.119(14) & 0.142(5) & 0.3 &  0.324 \\ 
\hline\hline
\end{tabular}
\egroup
\caption{Some details for our gauge field configurations.~``W'' stands for Wilson gauge action, ``I'' for the improved one.~The value of the spacing $a$ in GeV
was set via a static $q\bar{q}$ potential $U_\text{pot}$, with $\sqrt{\sigma}$ = 0.44 GeV.~For Wilson gauge action, we provide the expected (superscript ``exp'') and
calculated (superscript ``calc'') values for the quantity $\sqrt{\sigma} a$, with the expected ones coming from the analytic results of \cite{Dosch:1978jt} (for 2D),
and from a fit of equation (67) of \cite{Teper:1998te} (for 3D).~$\langle W_{1,1}\rangle$ is the expectation value of the $1 \times 1$ Wilson loop, needed for the fit
in \cite{Teper:1998te}.~$\alpha^\text{APE}$ denotes the APE smearing parameter \cite{Albanese:1987ds}, used in measurements of $U_\text{pot}$.~$\alpha^\text{gauge}$ is
the parameter for the gauge fixing procedure, the Cornell method \cite{Davies:1987vs}.}
\label{tab:config_details}
\end{center}
\end{table}

\end{document}